\documentclass[12pt]{article}
\usepackage{natbib}
\newcommand{\indep}{\rotatebox[origin=c]{90}{$\models$}}
\usepackage{graphicx}
\usepackage{amsmath}
\usepackage{titlesec}
\usepackage{color}
\usepackage{algorithm}
\usepackage{algpseudocode}
\usepackage{mathrsfs}
\usepackage{appendix}
\usepackage{soul}
\usepackage{subcaption}
\usepackage{booktabs}
\usepackage{lipsum}
\usepackage[shortlabels]{enumitem}
\usepackage[english]{babel} 
\usepackage{blindtext}
\usepackage{amsmath, amsthm, amssymb,nccmath}
\usepackage[margin=2.5cm]{geometry}
\usepackage[T1]{fontenc}
\def\bSig\mathbf{\Sigma}

\def \bZ {\boldsymbol{Z}}
\def \bz {\boldsymbol{z}}
\usepackage{chngcntr}
\usepackage{setspace}
\usepackage{caption}
\newtheorem{prop}{Proposition}[section]
\makeatletter

\title{A sensitivity analysis approach for the causal hazard ratio in randomized and observational studies}

\author
{Rachel Axelrod and Daniel Nevo\\
Department of Statistics and Operations Research, Tel Aviv University
}

\begin{document}
\begin{onehalfspacing}

\label{firstpage}

\maketitle

\begin{abstract}
The Hazard Ratio (HR) is often reported as the main causal effect when studying survival data. Despite its popularity, the HR suffers from an unclear causal interpretation. As already pointed out in the literature, there is a built-in selection bias in the HR, because similarly to the truncation by death problem, the HR conditions on post-treatment survival. A recently proposed alternative, inspired by the Survivor Average Causal Effect (SACE), is the causal HR, defined as the ratio between hazards across treatment groups among the study participants that would have survived regardless of their treatment assignment.
We discuss the challenge in identifying the causal HR and present a sensitivity analysis identification approach in randomized controlled trials utilizing a working frailty model. We further extend our framework to adjust for potential confounders using inverse probability of treatment weighting. We present a Cox-based and a flexible non-parametric kernel-based estimation under right censoring.  We study the finite-sample properties of the proposed estimation method through simulations. We illustrate the utility of our framework using two real-data examples.
\end{abstract}



\section{Introduction}
\label{sec:Introduction}

The analysis of time-to-event data, also known as survival analysis, is one of the backbones of clinical and epidemiological research \citep{andersen2012statistical,kragh2021analysis}. When studying the causal effect of a treatment or exposure on a time-to-event outcome, arguably the most ubiquitously reported measure is the hazard ratio (HR). At each time $t$, the HR compares the outcome rate between the treated and untreated, among the event-free individuals. To make a fair comparison, researchers seek to conduct randomized controlled trials (RCTs) within which treatment is assigned randomly. Alternatively, in observational studies, when the treatment or exposure assignment is not randomized, researchers collect rich data and spend considerable time selecting the appropriate covariates and models for their analyses.

A commonly-employed tool for HR estimation is Cox proportional hazard (PH) model  \citep{cox1972regression}. The standard Cox PH model assumes that the HR is constant through time.
Nevertheless, because the true HR may change over time, researchers often consider specifications of the Cox model that allow for time-varying HR \citep{zucker1990nonparametric, tian2005cox}. 

In the past decade, it has become increasingly clear that a causal interpretation of the HR is unlikely \citep{hernan2010hazards,aalen2015does,martinussen2020subtleties}. As firstly discussed by \cite{hernan2010hazards}, there is a built-in selection bias in the HR as a parameter which prevents a causal interpretation. While the treated and untreated are comparable at baseline, either by design or conditionally on covariates, among those who survived until time $t$, if the treatment is beneficial, then at time $t$ the treated are expected to be on average more frail than the untreated as a result of lower mortality among the treated up to time $t$. 

Although alternatives to the HR were suggested, the HR remains the most popularly reported measure even in studies targeting causal effects implicitly or explicitly \citep{lang2015basic,assel2019guidelines}. 
Therefore, having an alternative HR-like measure that can be endowed with a causal interpretation will be useful in practice. As part of teasing out the complexity of the interpretation of HRs, \cite{martinussen2020subtleties} have presented a causal HR defined within the sub-population of those who will potentially survive until a given time $t$ regardless of the treatment they will receive. This definition of causal effect is similar to the survivor average causal effect (SACE) \citep{rubin2006causal}. The SACE is the effect of treatment on a non-survival outcome among the sub-population that would have survived under either treatment arm. The SACE is not point-identified under standard assumptions, prompting researchers to often center their efforts around sensitivity analysis methods \citep{hayden2005estimator,egleston2007causal,zehavi2021matching}.

Although the casual HR has already been defined by \cite{martinussen2020subtleties}, the framework by which a researcher needs to follow to identify and estimate the casual HR from real data has not been fully established. In this paper, we present a detailed identification strategy for the casual HR in randomized and observational studies. Then, building on this approach, we develop sensitivity analysis techniques coupled with a flexible non-parametric estimation of the hazard function at each treatment arm. We also extend our approach to account for observed confounders and propose inverse probability of treatment weighting (IPTW) estimators.
Our sensitivity analysis approach is built upon the idea of \textit{frailty models}, accounting for the imbalance between the treatment groups due to unobserved factors that affect survival and cannot be adjusted for. 

The rest of the paper is organized as follows. First, in Section \ref{sec:Motivating example}, we review two motivating examples, one is an RCT and the other is an observational study. In Section \ref{sec:not_chr_assm}, we present the notations, assumptions, and clarify the motivation for targeting the causal HR. In Section \ref{sec:A sensitivity analysis approach}, we provide identification formulas and propose a sensitivity analysis approach stemming from these formulas. In Section \ref{sec:estimation}, we present a kernel-based and a Cox-based estimation methods in randomized studies and extend the approach to observational studies using IPTW. Evaluation of this estimators using a simulation study is presented in Section \ref{sec:simulation}. Finally, in Section \ref{sec:real_data}, we return to the two motivating examples and apply our proposed methodology. The \textbf{R} package \texttt{CausalHR} implements our proposed methodology. The package and fully reproducible simulations and data analyses are available from the first author's Github account. 

\section{Two motivating examples}
\label{sec:Motivating example}

Our first motivating example is the Urothelial Carcinoma data \citep{powles2018atezolizumab}. These data are from the IMvigor211 RCT study, which aimed to evaluate the effectiveness of immunotherapy treatment, Atezolizumab (Atezo), in comparison to standard care of chemotherapy (Chemo) in patients with locally advanced or metastatic urothelial carcinoma. The study outcome was the survival time defined as the time between patient's study enrollment and death. The data available for us \citep{gorfine2020k} included the 625 patients from the IC1/2/3 subgroup \citep{powles2018atezolizumab}, out of which 316 (51\%) patients were treated with Atezo and 309 (49\%) with Chemo. We focus on the first 12 months due to low number of events after this time in each treatment arm. 

\begin{figure}[h]
	\centering
\includegraphics[scale=0.5]{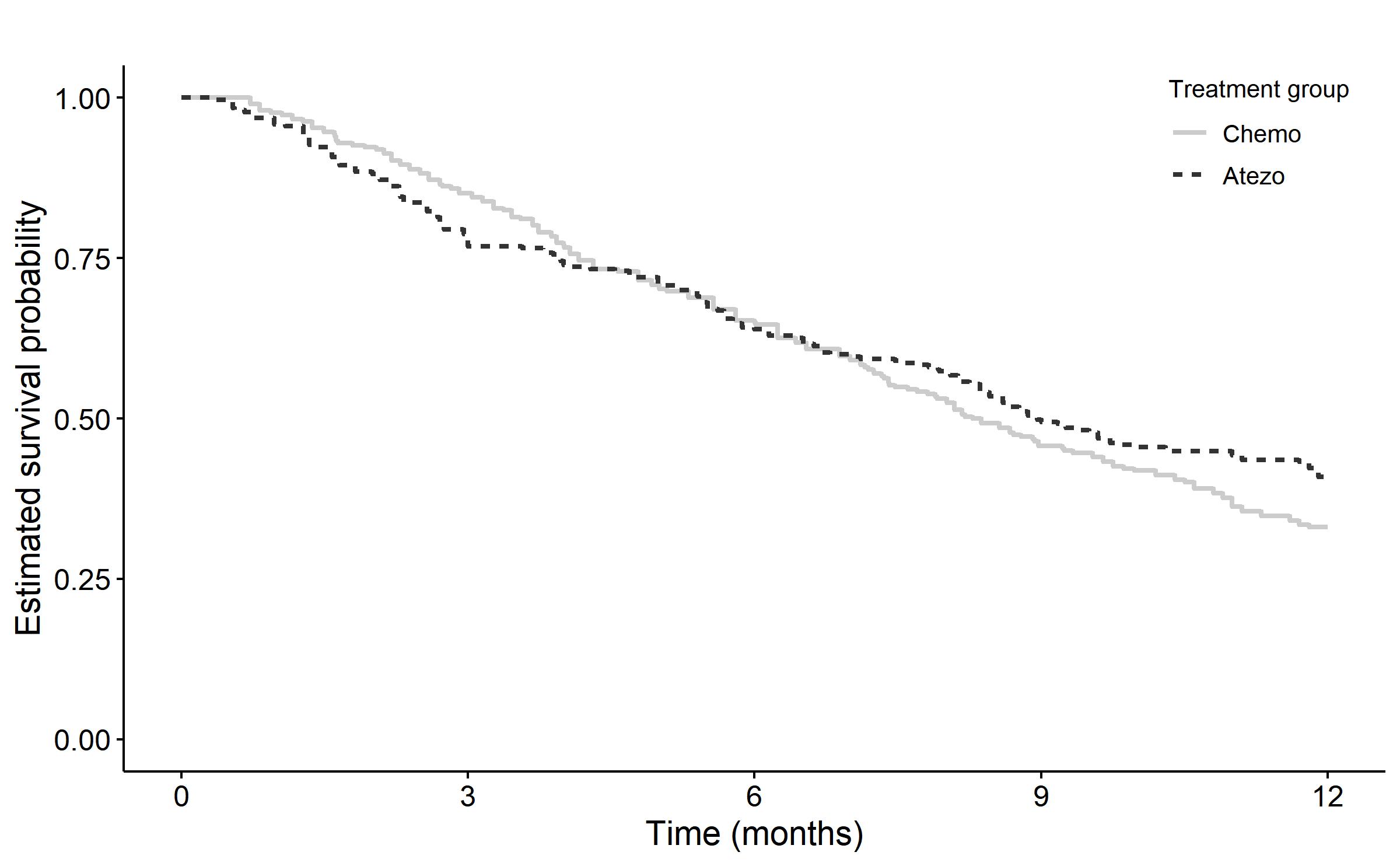}
\caption{Survival curves estimated by Kaplan–Meier estimator for Urothelial Carcinoma patients according to the treatment arm: Atezolizumab (Atezo) versus standard care of chemotherapy (Chemo).}	
\centering
\label{fig1:KM_carcinoma}
\end{figure}

From the Kaplan-Meier curves in Figure \ref{fig1:KM_carcinoma}, it seems that while at early times the Atezo patients are at larger risk than the Chemo patients, at later times the opposite is true. Furthermore, consider a Cox model analysis that partitioned the time axis into three segments: 0--4 months; 4--8 months; and 8--12 months, and estimated a constant HR within each segment. The estimated HR for the 0--4  and 4--8 months periods is larger than one, although insignificant ($\widehat{HR}$=1.34, 95\% confidence interval (CI): 0.96--1.88 and $\widehat{HR}$=1.09, 95\% CI: 0.76--1.57, respectively). However, the estimated HR for the 8--12 months period is significantly lower than one ($\widehat{HR}$=0.71, 95\% CI: 0.53--0.95), indicating that Atezo treatment may have a long-term preventive effect. Nevertheless, reporting three HRs instead of a single HR does not alleviate the problem of using HRs as a measure of causal effect;  The period-specific HRs are also prone to the selection bias presented in Section \ref{sec:Introduction}.

Our second motivating example is the Données Informatisées et VAlidées en Transplantation (DIVAT), an ongoing prospective cohort study following kidney transplant recipients. Analyzing these data, we want to conclude whether donor's characteristics (expanded versus standard donor criteria; see Section \ref{subsec:kidney}) influence the kidney transplant recipient's survival time. As this is an observational study, the two treatment groups (defined by the two donor types) are not balanced even at the study's start. 

\section{Notations, assumptions and causal estimands}\label{sec:not_chr_assm}
\subsection{Notations and assumptions}
\label{subsec:Notations}

We employ the potential outcomes framework \citep{imbens2015causal}. Let $A$ denote the binary treatment and $T^{A=a,C=\infty}$ denote the potential event time had a patient was assigned to treatment group  $a, a=0,1$, and there was no loss to follow up or administrative censoring, namely that the  censoring time was $C=\infty$. This notion of the outcome had there was no censoring aligns with the survival analysis literature that often defines the event time had there was no censoring, although typically without using the potential outcomes framework explicitly. We  assume that $C^{A=a}$, the potential censoring time under treatment level $A=a$, is independent of $T^{A=a, C=\infty}$ for $a=0,1$. This assumption can be thought of the usual \textit{independent censoring assumption}, assumed in each of the two counterfactual worlds defined by hypothetically setting the treatment to $a=0,1$. Let $X^{A=a}=\min\{T^{A=a, C=\infty},C^{A=a}\}$ and $\delta^{A=a}=I\{T^{A=a, C=\infty}\leq C^{A=a}\}$ represent the observed time and the event indicator, respectively, had we intervened and set $A=a$. We also define the potential outcomes' event counting processes $N^{A=a}(t)=\delta^{A=a}I\{X^{A=a}\leq t\}$, and let 
$$
\lambda^{A=a,C=\infty}(t):=\lambda^{A=a}(t)=\lim\limits_{dt \to 0}(dt)^{-1}\Pr\left(t\leq T^{A=a,C=\infty} <t+dt | T^{A=a,C=\infty}\geq t\right)
$$
be the hazard rate in terms of the potential outcomes, and
$$
\lambda^{A=a,C=\infty}(t|Q):=\lambda^{A=a}(t|Q)=\lim\limits_{dt \to 0}(dt)^{-1}\Pr\left(t\leq T^{A=a,C=\infty}<t+dt | T^{A=a,C=\infty}\geq t,Q\right)
$$
be the  hazard rate in terms of the potential outcomes given an event $Q$.

Throughout the paper, we assume the \textit{Stable Unit Treatment Value Assumption (SUTVA)} \citep{imbens2015causal}, which means that the potential outcome $X^{A=a}$ for each participant does not depend on the treatments received by other participants, and that there are no multiple versions of each treatment value $a$ leading to a different outcome. Consequently, for a participant with actual treatment $A$ we have $X=(1-A)X^{A=0}+AX^{A=1}$,  $\delta=(1-A)\delta^{A=0}+A\delta^{A=1}$ and $N(t)=(1-A)N^{A=0}(t)+AN^{A=1}(t)$. 

We first assume \textit{randomization} $A \indep X^{A=a}$, for $a=0,1$. That is, the actual treatment value is not informative about the two potential outcomes. Later, we relax this assumption and replace it with the \textit{conditional exchangeability assumption}, meaning that  $A\indep X^{A=a}\|\bZ$, for $a=0,1$, where $\bZ$ is a vector of observed confounders.

Under the randomized trial setting, the observed data for each patient $i, i=1,...,n$, is $\left(A_i, X_i, \delta_i\right)$. In the presence of confounders, we assume that we additionally observe $\bZ_i$. From $X_i$ and $\delta_i$ one can obtain the risk indicator $Y_i(t)=I\{X_i\geq t\}$; the counting process $N_i(t)$; and the indicator of the jump in the process $N_i(t)$ during the interval $[t, t + dt)$, denoted by $dN_i(t)$. Finally, let $N_i(t|A=a)=N_i(t)I\{A_i=a\}$, $dN_i(t|A=a)=dN_i(t)I\{A_i=a\}$, and $Y_i(t|A=a)=Y_i(t)I\{A_i=a\}$ be the analogous quantities in each treatment arm.

\subsection{The causal and the non-causal hazard ratio}
\label{subsec:the.causal.hr}

Let $HR^{PO}(t)=\frac{\lambda^{A=1}(t)}{\lambda^{A=0}(t)}$ to be the HR in terms of the potential outcomes. As recently discussed \citep{hernan2010hazards,aalen2015does,martinussen2020subtleties},  $HR^{PO}(t)$ does not have a natural causal interpretation because it contrasts hazard rates between two non-identical sub-populations of patients, those who would have survived at a given time $t$ when treated $\left(T^{A=1,C=\infty}\geq t\right)$ and those who would have survived at the same time $t$  when untreated $\left(T^{A=0,C=\infty}\geq t\right)$. 
Intuitively, while the treated and untreated patients are comparable at time $t=0$ due to randomization, as time progresses, the two sub-populations may become less and less balanced with respect to covariates other than the treatment that determine survival. Weaker or more ``frail'' patients may have survived because they received the treatment, and they would not have survived had they been placed in the untreated group. As a result, the treated are generally more frail than the untreated, and $HR^{PO}(t)$ will contrast two sub-populations with different frailty levels. 

The aforementioned selection bias resembles the issue of \textit{truncation by death} \citep{rubin2006causal,hayden2005estimator,egleston2007causal,zehavi2021matching}. One solution employs the principal stratification approach \citep{frangakis2002principal} to focus instead on the SACE, which compares the outcome under different treatment levels among the sub-population who would have survived regardless of treatment assignment. As recently discussed \citep{martinussen2020subtleties}, an approach similar to the SACE can be used to remedy the selection problem in HR.  The \textit{causal HR},
\begin{equation}
\label{eq:HRCdef}
HR^{C}(t)=\frac{\lim\limits_{dt \to 0}dt^{-1}\Pr\left[t\leq T^{A=1,C=\infty} <t+dt| T^{A=1,C=\infty}\geq t, T^{A=0,C=\infty}\geq t\right]}{\lim\limits_{dt \to 0}dt^{-1}\Pr\left[t\leq T^{A=0,C=\infty}< t+dt| T^{A=1,C=\infty}\geq t, T^{A=0,C=\infty}\geq t\right]},
\end{equation}
contrasts the instantaneous risk at time $t$ of the
treatment versus no treatment, in the sub-population containing participants who would have survived up to time $t$ regardless of their treatment assignment. Because $HR^{C}(t)$ is a contrast between hazard rates defined on the same sub-population, it a well-defined causal effect. 

Analogous to challenges with identification of the SACE, even had we were able to avoid censoring, for each participant we would have known the survival status at time $t$ under one treatment value only, the actual treatment. Therefore, we do not know which participants would have survived until time $t$ regardless of their treatment assignment and cannot identify $HR^{C}(t)$ from the observed data using standard identification assumptions (e.g., randomization and SUTVA). Usually, point-identification of the SACE entails strong assumptions \citep{hayden2005estimator,zehavi2021matching} that are unlikely to hold in our setup \citep{martinussen2020subtleties}. Therefore, to avoid making strong and implausible assumptions, we focus on a sensitivity analysis approach.

\section{A sensitivity analysis approach}
\label{sec:A sensitivity analysis approach}
When a causal effect is non-parametrically unidentifiable, sensitivity analysis is often implemented. The idea is to identify the causal effect up to an unknown and unidentifiable sensitivity parameter(s). Then, varying the value of the sensitivity parameter, an estimator of the causal effect as a function of the unknown sensitivity parameter is obtained.

\subsection{Identification and sensitivity model}\label{sec:ident_sen}
We adopt a frailty point of view to model the unidentifiable cross-world dependence between event times under different treatment values ($T^{A=1,C=\infty}$ and $T^{A=0,C=\infty}$). Frailty variables are commonly used in the survival analysis literature to model unobserved heterogeneity with known source, for example, to account for and study dependence in clustered survival data \citep{hougaard2000analysis,clayton1985multivariate}. More recently, a frailty approach was taken to model cross-world dependence between event times under different treatments
\citep{aalen2015does,stensrud2017exploring,martinussen2020subtleties,nevo2021causal}.

We take three working assumptions on the frailty variable $V$, henceforth referred to as the \textit{frailty assumptions}: (a) $V$ is an unmeasured time-fixed variable from a known parametric family with mean $1$ and variance $\theta$; (b) conditionally on the frailty, the potential event times are independent $T^{A=1,C=\infty}\indep T^{A=0,C=\infty}|V$; and (c) the \textit{multiplicity assumption} $\lambda^{A=a}(t|V=v)=v\psi^{A=a}(t)$, where $\psi^{A=a}(t)$ is a function that does not depend on $V$. 

Importantly, we stress that our frailty-based approach is a working sensitivity analysis model. One my argue that a variable like $V$ leading to independence between the potential survival times does not exist. Here, we use the frailty approach simply to account for the cross-world dependence.  

In the following proposition, we build on the frailty assumptions to identify $HR^{C}(t)$ in a randomized trial setting under independent right-censoring, as a function of terms estimable from the observed data and of the unknown and unidentifiable variance of the frailty distribution denoted by $\theta$. 
\begin{prop} 
\label{prop:RCT}
\normalfont Under SUTVA, randomization, independent censoring, and the frailty assumptions, the $HR^{C}(t)$ is identified by
\begin{equation}
\label{eq:ident_no_conf}
HR^{C}(t)=\frac{\lambda(t|A=1)}{\lambda(t|A=0)}\varphi\big[\Lambda(t|A=1),\Lambda(t|A=0),\theta\big],
\end{equation}
where $\lambda(t|A=a)=\frac{E\left[dN(t|A=a) \right]}{E\left[Y(t|A=a)\right]}$, $\Lambda(t|A=a)=\int_{0}^{t}\lambda(u|A=a)du$, and $\varphi\big[\Lambda^{A=1}(t),\Lambda^{A=0}(t),\theta\big]$ is a function depending on $\Lambda^{A=1}(t)$, $\Lambda^{A=0}(t)$ and $\theta$ and may take a closed form depending on the specific parametric family distribution for $V$. The proof is given in Section \ref{SM:prop1} of the supplementary materials (SM).
\end{prop}

Below, we describe how Proposition \ref{prop:RCT} serves as the basis for our proposed sensitivity analysis. But first, we consider the analogous identification result in observational studies. 
We relax the randomization assumption and instead assume conditional exchangeability assumption. Under this assumption, we present an IPTW-based identification formula for the $HR^C(t)$. Let $\pi(\bZ)=\Pr(A=1|\bZ)$ be the \textit{propensity score}, namely the probability of being treated given $\bZ$, and define the weights $w(\bZ)=\frac{A}{\pi(\bZ)}+\frac{1-A}{1-\pi(\bZ)}$.

We also replace assumption (b) of the \textit{frailty assumptions} with the assumption that conditionally on the frailty and $\bZ$, the potential event times are independent, $T^{A=1,C=\infty}\indep T^{A=0,C=\infty}|$
$(V,\bZ)$. Under this modified assumption, our approach does not lead to a closed identification formula for $HR^C(t)$. However, we suggest an identification formula which approximates the true $HR^C(t)$, that should work well when the event of interest is rare, as is often the case, for example, in cancer epidemiology studies, and/or when the association between $T^{A=a,C=\infty}$ and $\bZ$ is low. We further investigate this issue in Section \ref{sec:simulation}, and illustrate that under high censoring rate and/or low association between $T^{A=a,C=\infty}$ and $\bZ$, an estimator built upon our approximated identification formula for $HR^C(t)$ can be robust even in the case the multiplicity assumption on the frailty (assumption (c)) does not hold exactly. 

In Section \ref{SM:prop2} of the SM we show that when the event is rare and/or the association between $T^{A=a,C=\infty}$ and $\bZ$ is weak, then under SUTVA, conditional exchangeability, independent censoring, and the frailty assumptions, $HR^{C}(t)$ can be approximated by
\begin{equation}\label{eq:ident_with_conf}
HR_{IPTW}^{C}(t)=\frac{\lambda_{IPTW}(t|A=1)}{\lambda_{IPTW}(t|A=0)}\varphi\big[\Lambda_{IPTW}(t|A=1),\Lambda_{IPTW}(t|A=0),\theta\big],
\end{equation}
where $\lambda_{IPTW}(t|A=a)=\frac{E\left[w(\bZ)dN(t|A=a)\right]}{E\left[w(\bZ)Y(t|A=a)\right]}$ and $\Lambda_{IPTW}(t|A=a)=\int_{0}^{t}\lambda_{IPTW}(u|A=a)du$, and $\varphi(\cdot)$ is identical to the one appearing on \eqref{eq:ident_no_conf}. 

We can now  propose a general sensitivity analysis approach for $HR^C(t)$ based on Equations \eqref{eq:ident_no_conf} and \eqref{eq:ident_with_conf}. The first step is choosing the parametric family distribution for $V$ and a range of possible $\theta$ values. There are several options for the frailty distribution choice; a natural common choice for frailty distributions is the Gamma distribution. In Section \ref{SM:formulas_for_varphi} of the SM we present closed-form formulas of $\varphi$ and the identification of $HR^C(t)$ under different distribution choices. Although the identification formulas differ between frailty distribution specifications, one might expect the resulting differences in the $HR^C(t)$ estimator to be small, even for high values of $\theta$, given previous empirical evidence in frailty modeling \citep{gorfine2012conditional}. We demonstrate this point in our analyses of the motivating examples (Section \ref{sec:real_data}). In both of the examples, the differences in the $HR^C(t)$ estimator under different frailty distribution specifications were mild and did not change the analysis conclusions. The range of $\theta$ values can be selected, for example, by inverting a corresponding range of Kendall's $\tau$ correlation values between $T^{A=0,C=\infty}$ and $T^{A=1,C=\infty}$ \citep{oakes1989bivariate}. For example, under the Gamma frailty, Kendall's $\tau$ between $T^{A=0,C=\infty}$ and $T^{A=1,C=\infty}$ is $\frac{\theta}{\theta + 2}$.

The next step is estimating the identifiable but unknown quantities. Under randomization, these are the functions $\lambda(t|A=a)$ and $\Lambda(t|A=a)$  in  \eqref{eq:ident_no_conf}. In the presence of confounders, we need to estimate $w(\bZ)$, as well as $\lambda_{IPTW}(t|A=a)$ and $\Lambda_{IPTW}(t|A=a)$ in \eqref{eq:ident_with_conf}. In the final step of the sensitivity analysis, the obtained estimators are plugged in \eqref{eq:ident_no_conf} or \eqref{eq:ident_with_conf} to obtain an estimator for the $HR^C(t)$ function, $\widehat{HR}^{C}(t)$ for a range of $\theta$ values chosen by the researcher.

\section{Estimation}
\label{sec:estimation}
We present two estimation methods, a semi-parametric Cox-based approach and a non-parametric kernel-based approach. For each estimation method we start with the scenario of randomized study and then discuss how to modify the analysis in the presence of confounders. For the Cox-based approach under randomization see also \cite{martinussen2020subtleties}. 

\subsection {Cox-based estimation}\label{sec:cox.estimation}
The Cox-based approach assumes that the hazard function of the potential event times  $\lambda^{A=a}(t)$ follows the marginal structural Cox PH model \citep{hernan2000marginal}
\begin{equation}\label{eq:Cox model}
\lambda^{A=a}(t)=\lambda_0(t)\exp(\beta a).
\end{equation}
Because under randomization $\lambda^{A=a}(t)=\lambda(t|A=a)$, then  \eqref{eq:ident_no_conf} becomes a function of $\Lambda_0(t)=\int_{0}^{t}\lambda_0(u)du$ and $\beta$. For example, in Section \ref{app:proof_gamma_frailty} of SM we show that under Gamma frailty, by substituting \eqref{eq:Cox model} back into \eqref{eq:ident_no_conf}, we get that $HR^{C}(t)=\exp\big\{\beta\exp \big\{\theta \Lambda_0(t) \big[\exp(\beta)-1\big]\big\}$,
and the proposed estimator for $HR^{C}$ for each $\theta$ value 
is 
$$
\widehat{HR}^{C}(t)=\exp\big\{\hat{\beta}\exp \big\{\theta \hat{\Lambda}_0(t) \big[\exp(\hat{\beta})-1\big]\}\big\},
$$
where $\hat{\beta}$ and $\hat{\Lambda}_0(t)$ are obtained by fitting a standard Cox model. 

In the presence of confounders, we propose to estimate $\beta$ and $\Lambda_0(t)$ by first estimating the weights $w(\bZ)$ and then fitting a weighted Cox model \citep{cole2004adjusted}. To this end, one can estimate the propensity score $\pi(\bZ)$, for example, using a logistic regression model, and then obtain the weights
$\hat{w}(\bZ)=\frac{A}{\hat{\pi}(\bZ)}+\frac{1-A}{1-\hat{\pi}(\bZ)}$. Alternatively, the stabilized weights $\hat{w}^{st}(\bZ) = \frac{\widehat{\Pr}(A=1)A}{\hat{\pi}(\bZ)} + \frac{\widehat{\Pr}(A=0)(1-A)}{1-\hat{\pi}(\bZ)}$ can be used to gain reduced variance \citep{robins2000marginal}. 

\subsection{Kernel-based estimation}
As a flexible non-parametric alternative to the Cox-based estimation, we propose to estimate $\lambda(t|A=a)$ or $\lambda_{IPTW}(t|A=a)$ at each treatment arm separately without imposing any relationship between hazards under different treatment values (other than the frailty). Furthermore, we do not impose a model for $\lambda^{A=a}(t)$ and specifically we avoid the PH assumption.

\subsubsection{Pre-smoothed estimators}
We first look at a pre-smoothed estimator, which is the building block of the kernel estimator. Under randomization, $\lambda(t|A=a)$ can be estimated by
\begin{equation}
\label{eq:est.no.confounders}
\hat{\lambda}^{pre}(t|A=a)=\frac{1}{n}\sum_{i=1}^{n} \frac{dN_i(t|A_i=a)}{\frac{1}{n}\sum_{j=1}^{n}Y_j(t|A_j=a)}
,
\end{equation}
where $\sum_{i=1}^{n}dN_i(t|A_i=a)$ and $\sum_{j=1}^{n}Y_j(t|A_j=a)$ are the number of observed events and at-risk patients at time $t$ in treatment arm $A=a$. The cumulative hazard of the observed data distribution, $\Lambda(t|A=a)$, is then estimated at each treatment arm by the well-known Nelson-Aalen estimator $\hat{\Lambda}^{pre}(t|A=a)=\int_{0}^{t}\hat{\lambda}^{pre}(u|A=a)du$. When confounders are present, the IPTW estimator analogue of 
\eqref{eq:est.no.confounders} is 
\begin{equation}\label{eq:iptw_est.confounders}
\hat{\lambda}^{pre}_{IPTW}(t|A=a)=\frac{1}{n}\sum_{i=1}^{n}\frac{\hat{w}_i\left(z_i\right)dN_i(t|A_i=a)}{\frac{1}{n}\sum_{j=1}^{n}\hat{w}_j\left(z_j\right)Y_j(t|A_j=a)}.
\end{equation}
Then, $\Lambda_{IPTW}(t|A=a)$ is estimated by the IPTW version of the Nelson-Aalen estimator $\hat{\Lambda}^{pre}_{IPTW}(t|A=a)=\int_{0}^{t}\hat{\lambda}^{pre}_{IPTW}(u|A=a)du$.

\subsubsection{Smoothed estimators}
\label{sec:Smoothed estimator}
The estimators $\hat{\lambda}^{pre}(t|A=a)$ and $\hat{\lambda}^{pre}_{IPTW}(t|A=a)$ are not smooth functions of $t$. Moreover, for a finite sample these estimators are expected be zero for some time points $t$, such that the ratios $\frac{\lambda(t|A=1)}{\lambda(t|A=0)}$ in \eqref{eq:ident_no_conf} or $\frac{\lambda_{IPTW}(t|A=1)}{\lambda_{IPTW}(t|A=0)}$  in \eqref{eq:ident_with_conf} may equal zero or $\infty$ and more generally, to heavily fluctuate and eventually to considerably increase the variance of $\widehat{HR}^{C}(t)$. Therefore, we propose to estimate $\lambda(t|A=a)$ or $\lambda_{IPTW}(t|A=a)$ with kernel-smoothed estimators \citep{ramlau1983smoothing} based on the pre-smoothed estimators \eqref{eq:est.no.confounders} and \eqref{eq:iptw_est.confounders}. 

Denote $m_a=\sum_{i=1}^{n}I\{A_i=a,\delta_i=1\}$ and let $T^{(i)}_a, i=1,...,m_a,$ be the ordered event times at treatment arm $A=a$. Let also $K_t(\cdot)$ and $b(t)$ be the chosen kernel function and bandwidth, respectively. We omit the dependence on $a$ for simplicity of presentation, although in practice both the kernel function and the bandwidth can be chosen separately for each treatment arm. Under randomization, $\lambda(t|A=a)$ is estimated by a weighted average of the Nelson-Aalen increments $d\hat{\Lambda}(T^{(i)}_a)$ over the interval $[t - b(t), t + b(t)]$,
\begin{equation}
\label{eq:kernel_estimator}
\hat{\lambda}(t|A=a)=b(t)^{-1}\sum_{i=1}^{m_a}K_{t}\left(\frac{t-T^{(i)}_a}{b(t)}\right) d\hat{\Lambda}(T^{(i)}_a).
\end{equation}
Note that here $K_t(\cdot)$ and $b(t)$ depend on the point $t$. We use a boundary kernel $K_t(\cdot)$ to account for the bias problem that may occur when estimating near the endpoints of the event times' range.
We use a time-varying bandwidth $b(t)$ to improve local performance over the fixed-bandwidth estimator. See \cite{hess1999hazard}, \cite{muller1990locally}, and \cite{muller1994hazard} for further discussion of these issues. 

The local bandwidth $b(t)$ can be chosen by minimizing the estimator of the local mean squared error (MSE), 
\begin{equation*}
\widehat{MSE}[t,b(t)]=\widehat{var}[t,b(t)]+\widehat{bias}^2[t,b(t)],
\end{equation*}
where the exact formulas of the estimated local variance $\widehat{var}[t,b(t)]$ and bias $\widehat{bias}[t,b(t)]$ are given in \cite{muller1994hazard}.

In the presence of confounders, we propose to estimate $\lambda_{IPTW}(t|A=a)$ by a weighted average of the IPTW Nelson-Aalen increments $d\hat{\Lambda}_{IPTW}(T^{(i)}_a)$ over the interval $[t - b(t), t + b(t)]$,
\begin{equation}
\label{eq:kernel_estimator_iptw}
\hat{\lambda}_{IPTW}(t|A=a)=b(t)^{-1}\sum_{i=1}^{m_a}K_{t}\left(\frac{t-T^{(i)}_a}{b(t)}\right) d\hat{\Lambda}_{IPTW}(T^{(i)}_a).
\end{equation} 
As before, the local bandwidth $b(t)$ can be chosen by minimizing the $\widehat{MSE}$, but a modified estimators of the bias and the variance that include the weights are needed. Further technical details about the kernel-based estimation and our proposed MSE estimation for the weighted scenario are given in Section \ref{app:details} of the SM.  

\subsection{Asymptotic properties of $\widehat{HR}^{C}(t)$}\label{Asymptotic properties}
For each time point $t$, the Cox-based estimator for $HR^{C}(t)$ is consistent and asymptotically normal by the continuous mapping theorem and the delta method,  since it is a function of $\hat{\Lambda}_0(t)$ and $\hat{\beta}$ which are consistent and asymptotically normal \citep{andersen2012statistical}. The kernel-based estimator $\hat{\lambda}(t|A=a)$ is asymptotically normal (with a rate typically slower than $\sqrt{n}$ that depends on the smoothness of $\lambda(t|A=a)$), and consistent under certain conditions on the local bandwidth $b(t)$ \citep{muller1990locally}. Therefore, given results on weighted estimators, one may expect similar results on $\hat{\lambda}_{IPTW}(t|A=a)$. Finally, the kernel-based estimators for $HR^{C}(t)$ are also consistent and asymptotically normal, again by the continuous mapping theorem and the delta method. However, if the local bandwidth ,$b(t)$, is chosen to minimize the MSE, one cannot expect the estimator's bias to be zero even in large sample size \citep[Chapter 5.7][]{wasserman2006all}. Thus, if $b(t)$ is chosen to minimize the MSE, the confidence intervals may not be centered around the true value of $HR^C(t)$ and their empirical coverage rate is not guaranteed to be close to the desired level. 

\section{Simulation studies}
\label{sec:simulation}
We conducted simulation studies to assess and compare the finite-sample performance of the kernel-based and the Cox-based estimators in the two discussed settings: (I) under randomization, (II) in the presence of confounders. For each scenario described below, we simulated 1,000 datasets. Technical details accompanying this section and further results are provided in Section \ref{SM:simulations} of the SM. 

\subsection{Data generating mechanism}
\label{sec:DGM}

Under the randomization setting (I), for each unit we first generated a frailty variable, $V$, from Gamma distribution with mean 1 and variance $\theta\in(0.2,0.8,2.0,4.6)$. These values correspond to Kendall's $\tau\in (0.1,0.3,0.5,0.7)$ between $T^{A=0,C=\infty}$ and $T^{A=1,C=\infty}$. Next, we generated the correlated potential event times $T^{A=a,C=\infty}$ according to a specified hazard rate $\lambda^{A=a}(t|V)$, $a=0,1$. We considered two specifications for $\lambda^{A=a}(t|V)$: 
\begin{description}
\item[Scenario (Ia)] $\lambda^{A=a}(t|V)=V\exp\{\log(0.5) a+\theta t \exp[\log(0.5) a]\}$. This model results in the Cox model \eqref{eq:Cox model} for the marginal hazard $\lambda^{A=a}(t)$ with a time-varying $HR^C(t)$ (see Section \ref{SM:sceanrio_Ia} for a proof).
\item[Scenario (Ib)] $\lambda^{A=a}(t|V)=V\exp[1.5t+\log(0.5) a]$. Under this model, $\lambda^{A=a}(t)$ does not follow the Cox model \eqref{eq:Cox model}, and $HR^C(t)=0.5$ for all $t$ (see Section \ref{SM:sceanrio_Ib} for a proof). 
\end{description}
We generated the censoring time $C^{A=1}=C^{A=0}=C$ from an exponential distribution whose rate parameter was chosen to yield the desired censoring rate. We considered low to moderate censoring rates  $(10\%, 20\%, 40\%)$. We then calculated $X^{A=a}$ and $\delta^{A=a}$. The actual treatment $A$ was generated with probability $\Pr(A=1)=0.5$. Finally, we determined the observed time $X$ and the event indicator $\delta$ according to the actual treatment value $A$. We considered sample sizes $n \in (500, 1000, 5000)$. 

For the observational studies setting (Setting (II)), we used a data generating mechanism similar to Setting (I) with the following modifications. For each unit, we simulated a confounder $Z$ from a Normal distribution $N(0,1)$. We simulated the potential event times $T^{A=a,C=\infty}$, $a=0,1$ according to following the hazard rate:
\begin{description}
\item[Scenario (II)]$\lambda^{A=a}(t|V,Z)=V\gamma \exp[\log(0.5) a+\beta_z Z]$. That is, the conditional hazard (conditional on $V$ and $Z$) followed a Cox model. The exact value of $\gamma$ was selected to yield the desired event rates. We considered different $\beta_z$ values $\left(\log(0.1),\log(0.5),\log(0.9)\right)$. Under this model, the multiplicity assumption does not hold exactly, and $HR^C(t)=0.5$ for all $t$ (see Section \ref{SM:sceanrio_II} for a proof). 
\end{description}
As opposed to Setting (I), Administrative (fixed) censoring was imposed at $t=10$. We considered a relatively rare outcome by taking low event rate $(30\%, 10\%, 5\%, 3\%, 1\%)$ and thus we considered a larger sample size of $n=50000$, mimicking a large epidemiological cohort scenario. The treatment was generated with probability of $\pi(Z)=\frac{\exp[\log(0.5)Z]}{1+\exp(\log(0.5) Z}$. The rest of the process was identical to the one used for Setting (I).

\subsection{Analyses} 
For each scenario, we estimated $HR^{C}(t)$ using both the kernel-based and Cox-based estimators using the \texttt{CausalHR} \textbf{R} package (which was written based on the \texttt{muhaz} \textbf{R} package). 
In Scenarios (Ia) and (Ib), we estimated at each arm the hazard over a grid of 51 time points, taking the minimal time at which the hazard was estimated to be the first event time, and the maximal time to be the time at which 10 patients remained at risk, both calculated from the first simulated dataset.

In Scenario (II), we first estimated the weights $w(Z)$ via a logistic regression of $A$ on $Z$. 
Due to the administrative censoring, we chose the fixed minimal and maximal times to be $t=0$ and $t=10$. We also compared the results from this scenario to the HR estimated by a conditional Cox model that included the treatment and confounder as covariates (but not the unobserved frailty).  
 
For all methods and scenarios, standard errors were estimated by the bootstrap with 500 repetitions, and 95\% pointwise CIs were calculated by the percentile method.

\subsection{Results}

We present here the main results, under Kendall's $\tau=0.7$. Additional results under different Kendall's $\tau$'s, censoring rates, and sample sizes are presented in Section \ref{SM:additional_results} of the SM.

\begin{figure}[h]
\centering
\includegraphics[scale=0.3]{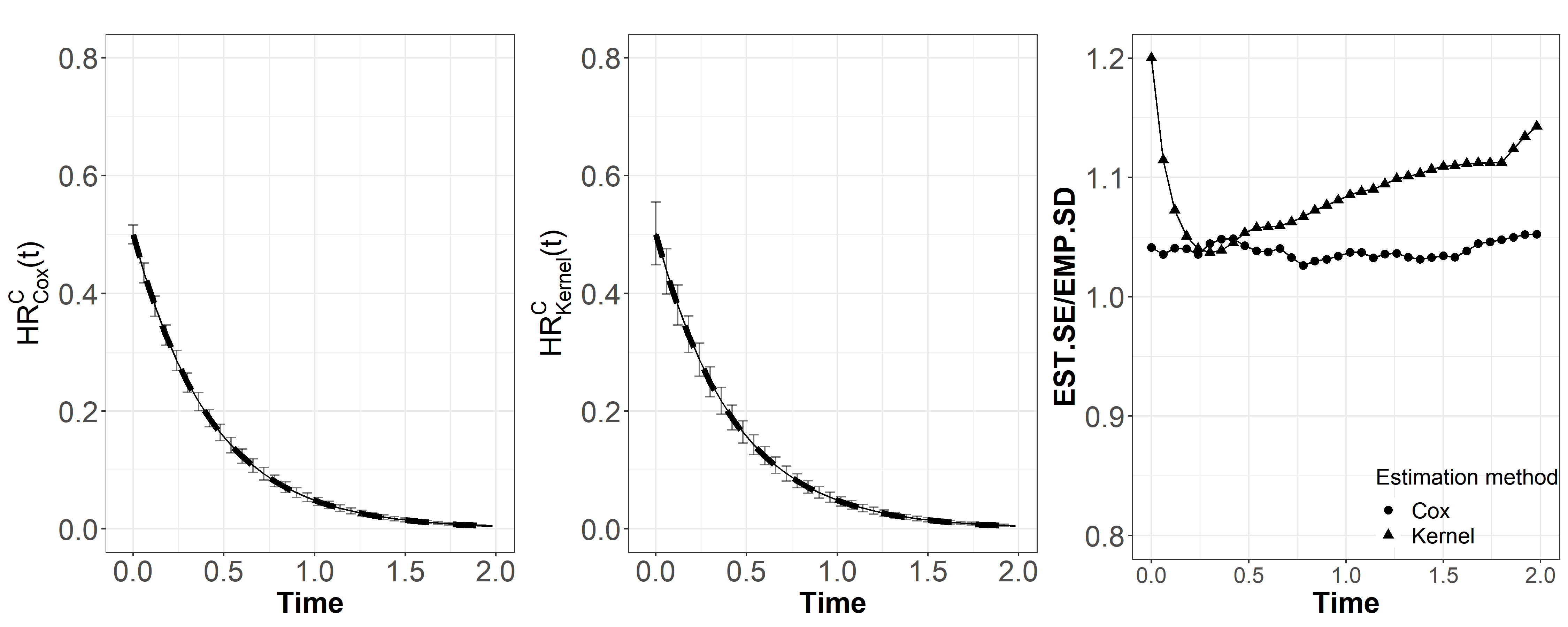}
\caption{Performance of the Cox-based and kernel-based estimators in Scenario (Ia) under sample size of $n=5000$, 20\% censoring rate, and a Gamma frailty distribution with Kendall's $\tau=0.7$. Results are presented only until $t=2$. The two left plots present the mean estimated $HR^C(t)$ across the simulations, plus/minus one empirical standard deviation. The dashed line represents the true $HR^C(t)$. The right plot presents the ratio between the mean estimated standard error (EST.SE) and empirical standard deviation of the estimates (EST.SE). $\widehat{HR}_{Cox}^C(t)$: Cox-based estimator $\widehat{HR}_{kernel}^C(t)$: Kernel-based estimator.}
\label{fig:sim.scen.Ia.results}
\end{figure}

In Scenario (Ia), under which assumption \eqref{eq:Cox model} was met, both estimation methods preformed well (Figure \ref{fig:sim.scen.Ia.results}). The bias and the empirical standard deviation of the estimates were small for all $t$. When the true $HR^C(t)$ were very close to zero (after $t=1$), the standard errors were overestimated especially for the kernel-based estimator. The empirical coverage rates of the 95\% CIs for both methods were close to the desired level (Figure \ref{fig:IA_tau07_n5000}). As expected, for smaller sample sizes ($n=500,1000$) the empirical standard deviations of the estimates were larger and the standard errors of both estimators were overestimated for all $t$ (Figure \ref{fig:IA_07_smaller_sample_sizes}). For higher censoring rate of $40\%$, the standard errors were still overestimated  (Figure \ref{fig:IA_07_different_CR}), but the bias remained less than 10\% for $t\le 1$. For lower Kendall's $\tau$ values, the estimators' performance was generally similar (Figures \ref{fig:IA_07_smaller_different_taus} and \ref{fig:IA_07_smaller_different_taus2}). 

\begin{figure}[h]
\centering
\includegraphics[scale=0.3]{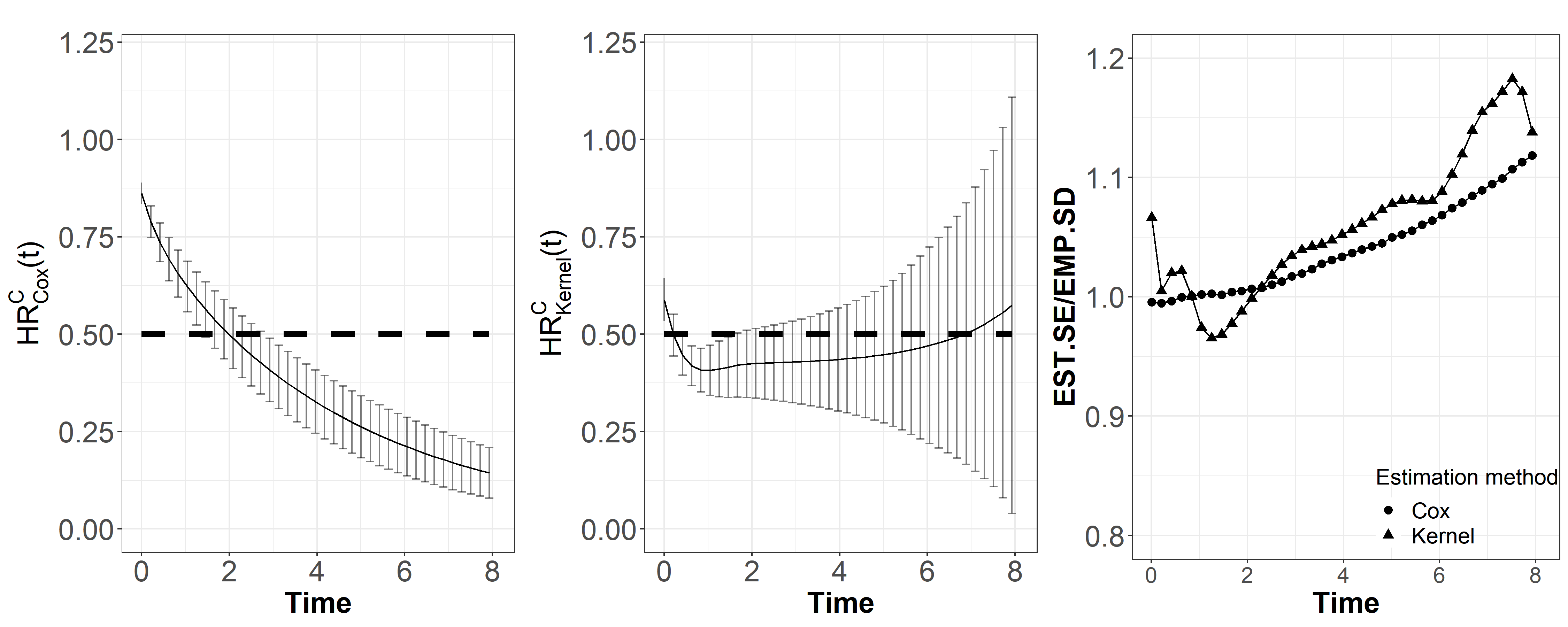}
\caption{Performance of the Cox-based and kernel-based estimators in Scenario (Ib) under sample size of $n=5000$, 20\% censoring rate, and a Gamma frailty distribution with Kendall's $\tau=0.7$. Results are presented only until $t=8$. The two left plots present the mean estimated $HR^C(t)$ across the simulations, plus/minus one empirical standard deviation. The dashed line represents the true $HR^C(t)$. The right plot presents the ratio between the mean estimated standard error (EST.SE) and empirical standard deviation of the estimates (EST.SE). $\widehat{HR}_{Cox}^C(t)$: Cox-based estimator $\widehat{HR}_{kernel}^C(t)$: Kernel-based estimator.}
\label{fig:sim.scen.Ib.results}
\end{figure}

In Scenario (Ib), when assumption \eqref{eq:Cox model} was not met, the kernel-based method preformed better than the Cox-based method in some aspects, especially in terms of bias (Figure \ref{fig:sim.scen.Ib.results}). The bias of the kernel-based was smaller for most $t$ values. The empirical standard deviation of the kernel-based estimates increased with $t$, as the available sample size for estimation was decreased. The standard errors were generally well estimated for both methods, although for later $t$ values they were overestimated. In this scenario,  there was an under-coverage of the confidence intervals for both methods (Figure \ref{fig:IB_tau07_n5000}). For the  kernel-based estimator, this result can be explained by the bias induced by choosing the bandwidth to minimize the MSE (Section \ref{Asymptotic properties}). For the smaller sample sizes  ($n=500,1000$) the empirical standard deviations of the estimates were bigger and the standard errors were overestimated for all $t$ and for both estimation method (Figure \ref{fig:IB_07_smaller_sample_sizes}). For higher censoring rate of $40\%$, the standard errors were larger and overestimated for later time points (Figure \ref{fig:IB_07_different_CR}). For lower Kendall's $\tau$ values, the bias of the kernel-based estimator was smaller and the standard errors were generally better estimated (Figures \ref{fig:IB_07_smaller_different_taus} and \ref{fig:IB_07_smaller_different_taus2}).


\begin{figure}[h]
\centering
\includegraphics[scale=0.3]{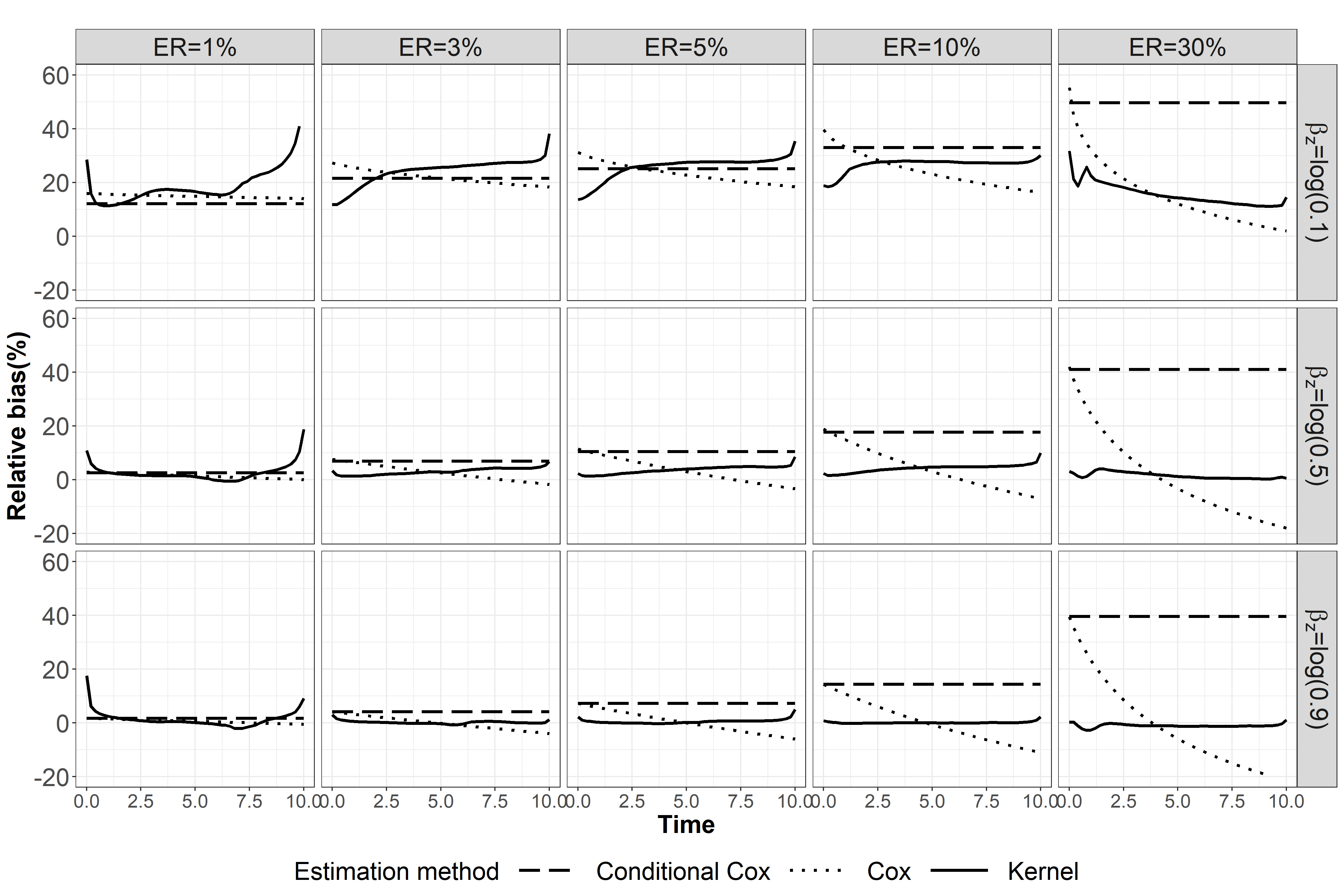}
\caption{Relative bias of Cox-based and kernel-based $HR^C(t)$ estimators in Scenario (II) under sample size of $n=50000$, Gamma frailty distribution and Kendall's $\tau$ of 0.7 between potential event times. ER: Event rate.}

\label{fig:sim.results.IIc}
\end{figure}

In Scenario (II) (Figure \ref{fig:sim.results.IIc}), the bias of the kernel-based method was generally small for nearly all time points whenever the event rate was small and the confounder effect on the event time was not large. Bias observed at early time points could be explained by low cumulative event rates, while bias at late time points is due to small number of events. These phenomenons are expected for non-parametric estimators applied to survival data.  

For non-large effects of $Z$ ($\beta_z = \log(0.5),\log(0.9)$), the relative bias of the kernel-based estimator decreased with the decrease in the event rate. For weak $Z$ effect, the relative bias was small even when the event rate was relatively non-rare (30\%). For a very strong  confounder effect ($\beta_z=\log(0.1$)), the relative bias of the kernel-based estimator was approximately 20\% regardless of the event rate. Similar trends were observed for the Cox-based estimator, although the relative bias was larger when the event rate was large.

For nearly all event rates and $\beta_z$ values considered, the relative bias of the proposed estimators was substantially lower than the bias of the conditional Cox estimator. The relative bias of the conditional Cox estimator reached the minimum when the event rate was the smallest and the confounder's effect was the weakest. 

Generally, standard errors were well-estimated for the proposed methods for all $\beta_z$ values and event rates (Figure \ref{fig:sim.results.II.SD}). Similarly to the picture arising from the relative bias, the empirical coverage rates of the 95\% CIs from the kernel-based approach were close to the desired level for weaker effects of $Z$ ($\beta_z = \log(0.5),\log(0.9)$). The empirical coverage rates of of the 95\% CIs from the Cox-based approach were generally far from the desired level, and near the desired level for low event rate (Figure \ref{fig:sim.results.II.EC}). For a lower Kendall's $\tau=0.5$, the same trends were observed, but the relative bias for all methods was smaller (Figure \ref{fig:sim.results.II.RB.05}).

\section{Real data examples}
\label{sec:real_data}
We applied the proposed sensitivity analysis approach to two real data examples. The first is the RCT of effectiveness of Atezo treatment in Urothelial Carcinoma patients (Section \ref{sec:Motivating example}). The second is the observational study of assessing survival time after kidney transplantation. We estimated the Cox-based (denoted by $\widehat{HR}_{Cox}^C(t)$) and the kernel-based estimator ($\widehat{HR}_{kernel}^C(t)$) for each data over an estimation grid of 51 time points. The minimal and maximal  estimation time points were chosen in similar way to Section \ref{sec:simulation}. We compared between choosing Gamma and Inverse Gaussian (IG) choice for the frailty distribution. Under the Gamma distribution, we considered the Kendall's $\tau\in (0.1,0.3,0.5,0.7)$. Under the IG frailty distribution, we considered Kendall's $\tau\in (0.1,0.3,0.5)$, as the range of the possible Kendall's under the IG distribution is $[0,0.5)$. The corresponding frailty variances, $\theta$, are $(0.2,0.8,2,4.6)$ and $(0.3,2,100)$ for the Gamma and IG distributions, respectively.  The PH assumption was tested by a score test based on scaled Schoenfeld residuals \citep{grambsch1994proportional} and standard errors were computed via the bootstrap with 500 repetitions. Additional results are provided in Sections \ref{SM:Imvigor211} and \ref{SM:DIVAT} of the SM.

\subsection{Effectiveness of Atezo treatment in Urothelial Carcinoma patients}
\label{subsec:atezo}
\begin{figure}[h]
	
\centering
\includegraphics[scale=0.3]{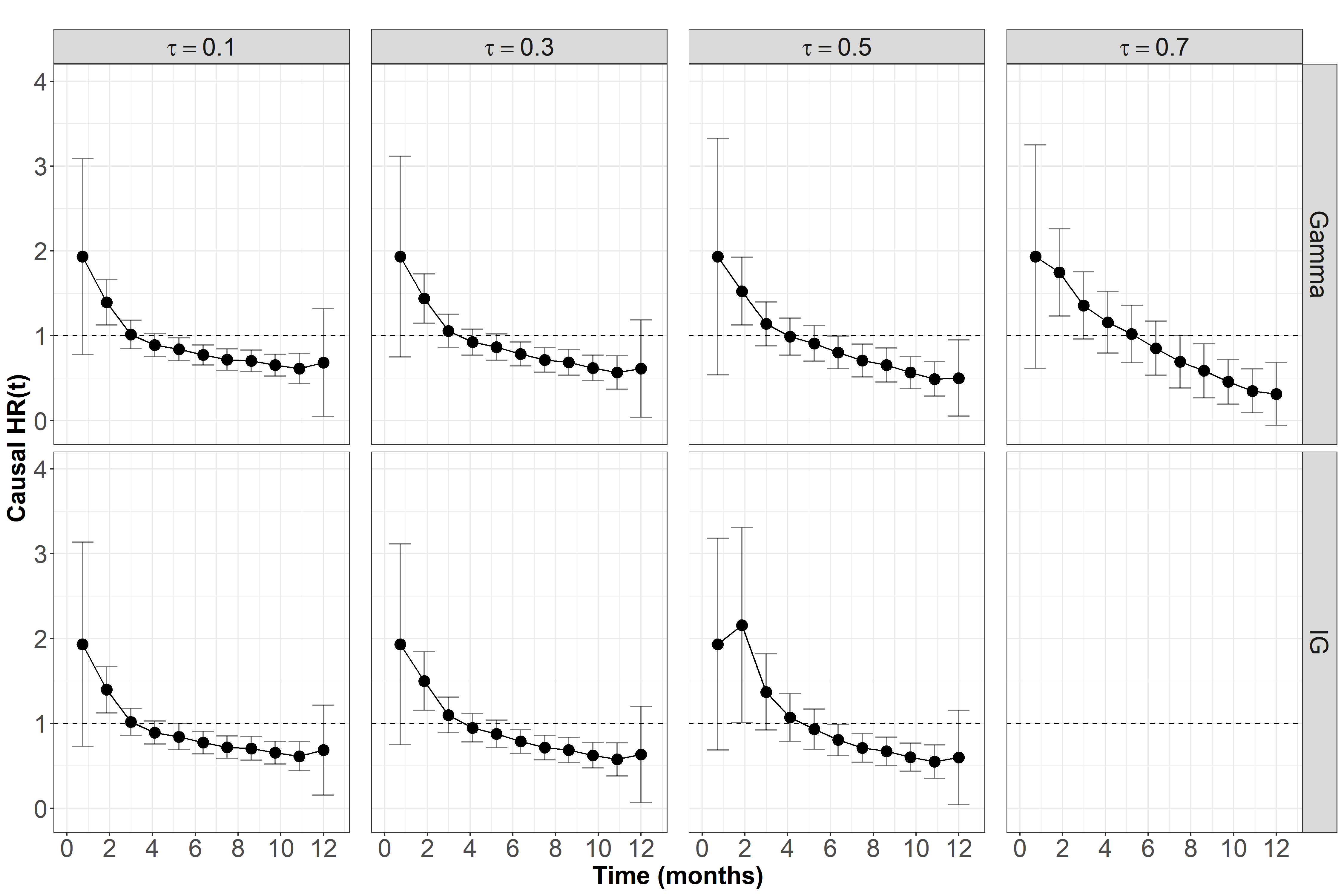}
\caption{Kernel-based estimation along with plus/minus one standard error for the $HR^C(t)$ in the IMvigor211 study, under different frailty distributions and Kendall's $\tau$ values. For IG frailty distribution Kendall's $\tau$ is bounded by 0.5. The horizontal dashed line represents $HR^C(t)=1$. For clarity, we present the estimation every 5 time points.}	
\label{fig:Atezo.HRC}
\end{figure}

In this dataset, there was evidence against the validity of the PH assumption ($p<0.01$). As demonstrated in Section \ref{sec:simulation} (Scenario (Ib)), when the marginal hazard rate $\lambda^{A=a}(t)$ does not follow the Cox model, the Cox-based sensitivity analysis  is expected to be biased. 
Therefore, we focus on our proposed kernel-based analysis. For completeness, the results  the Cox-model sensitivity analysis are presented in Section \ref{SM:Imvigor211-cox}. Generally, under all frailty distributions and Kendall's $\tau$ values, $\widehat{HR}_{Kernel}^C(t)$, at the start of the study was larger than one, indicating that Atezo treatment is inferior to Chemo in terms of short-term survival (Figure \ref{fig:Atezo.HRC}). After approximately four months, $\widehat{HR}_{Kernel}^C(t)$ decreased below one suggesting Atezo treatment prolongs the survival times of patients with Urothelial Carcinoma for those who would have survived up to four months regardless of their treatment assignment. The results were relatively robust to both $\tau$ and the chosen frailty, with the main change being increase in the standard error as $\tau$ increased (Figure \ref{fig:Atezo.HRC}).

The standard (non-causal) HR suggested that during the time interval 4--8 Atezo treatment has an harmful effect (compared with Chemo) and after 8 months has a protective effect, among those surviving beyond 8 months, regardless of their treatment, but weaker than the causal HR estimated by our kernel-based method (Section \ref{sec:Motivating example}). 


\subsection{Survival time after Kidney transplantation}
\label{subsec:kidney}

The DIVAT is an ongoing prospective cohort study following kidney transplant recipients. The data includes medical records for kidney and/or pancreas transplant recipients from 1990 until 2021 from eight different medical centers in France \citep{le2016comparisons}.
 Our analysis included 1281 adults between ages 40 and 70, of which 400 (31\%) had the event of interest, a graft failure, defined as a return to dialysis or death. The rest of  patient event times (881, 69\%) were censored. The treatment variable was donor category, classified into expanded criteria donor (ECD, $A=1$) and standard criteria donor (SCD, $A=0$). ECD is defined as donors older than 60 or aged 50-59 with at least two of the following characteristics: history of hypertension, cerebrovascular accident as the cause of death, or terminal serum creatinine higher than 1.5 mg/dL. ECD transplantations have been described with a substantially worse survival than SCD \citep{metzger2003expanded}. Of the 1281 patients, 562 (56\%) patients received a transplant from ECD donors, and the rest (719, 44\%) from SCD donors. Our analysis included the following confounders: recipient's age at the transplantation, whether there were at least four human leukocyte antigen (HLA) incompatibilities between the donor and the recipient, and whether the transplantation was a re-transplantation.

We fitted a logistic regression for $\pi(\bZ)$, and verified the balance was improved after applying the weights (Tables  \ref{tab:log.table.divat} and \ref{tab:balance.table.divat}, and Figures \ref{fig:age_balance}--\ref{fig:re-transplant_balance}). Those receiving transplant from ECD were generally older (OR=1.20, 95\% CI: 1.18--1.23), and more likely to have HLA incompatibility (OR=0.96, 95\% CI: 0.68--1.34). Additionally, second transplantations were more likely for ECD donors  (OR=1.29, 95\% CI: 0.92--1.83). To improve precision, we used stabilized weights (Section \ref{sec:cox.estimation}). To avoid increased variance in the estimator due to large weights, we truncated the weights at the 99th percentile. Histogram of the obtained weights is given in Figure \ref{fig:histogram of weights}.

Weighted and non-weighted Kaplan-Meier curves \citep{cole2004adjusted} corresponding to the two donor categories (Figure \ref{fig:KP_DIVAT}) showed that recipients from SCD donors survive longer ($p<0.0001$ for  weighted and non-weighted log-rank tests). A naive Cox-model analysis including the donor criterion type and the confounders yielded estimated HR for ECD vs SCD donors of 1.59 (95\% CI: 1.26--2.01, Table \ref{tab:con.cox.table}). A weighted Cox model estimated an HR of 1.56 (95\% CI: 1.19--2.05), and no violation of the PH assumption was detected ($p=0.48$).

\begin{figure}[h]
\centering
\begin{subfigure}[t]{1\textwidth}
 \centering
 \includegraphics[scale=0.3]{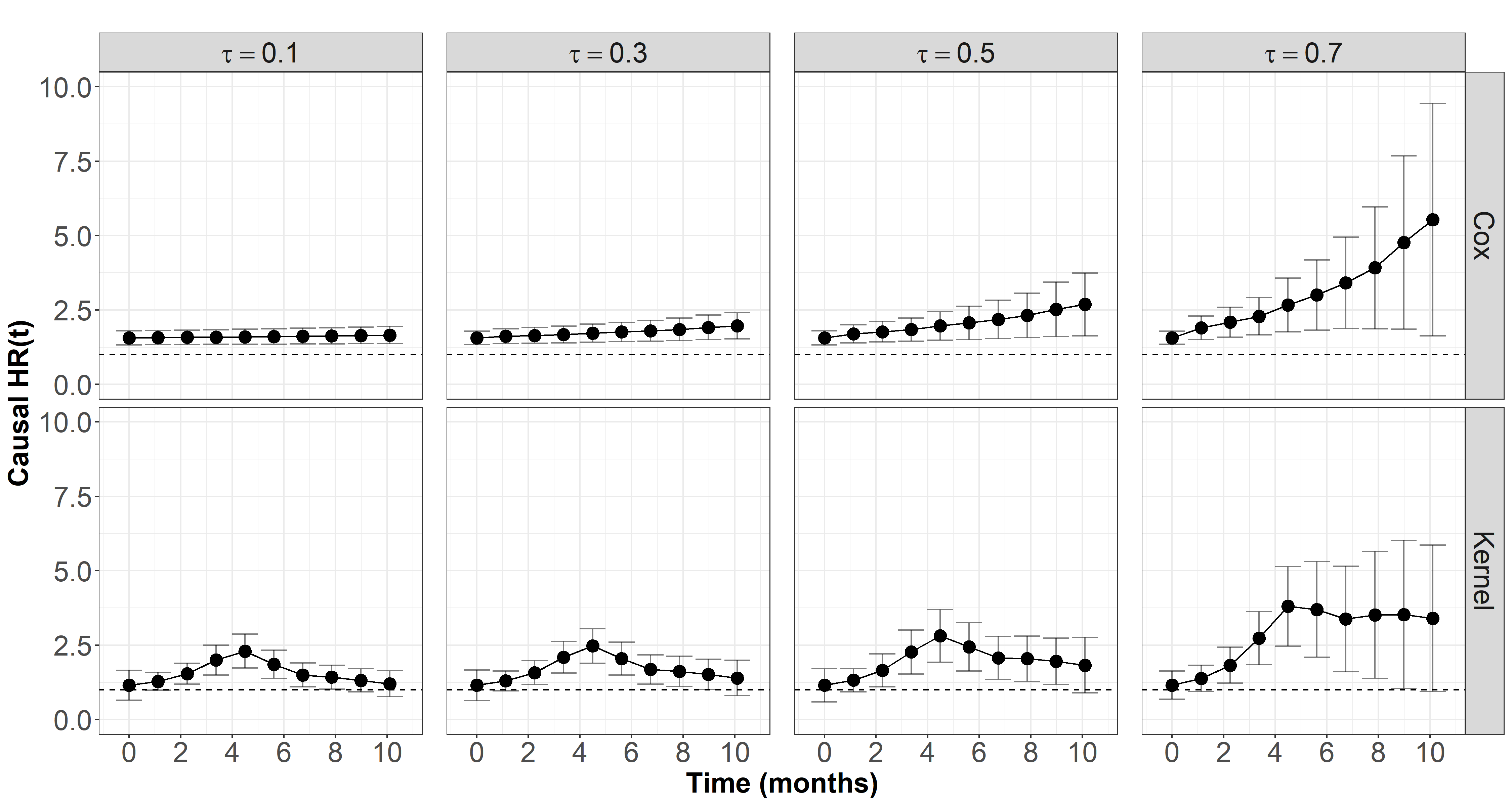}
 \caption{Gamma frailty}
 \label{fig:divat_gamma}
 \end{subfigure}
 \vfill
 \begin{subfigure}[t]{1\textwidth}
 \centering
\includegraphics[scale=0.3]{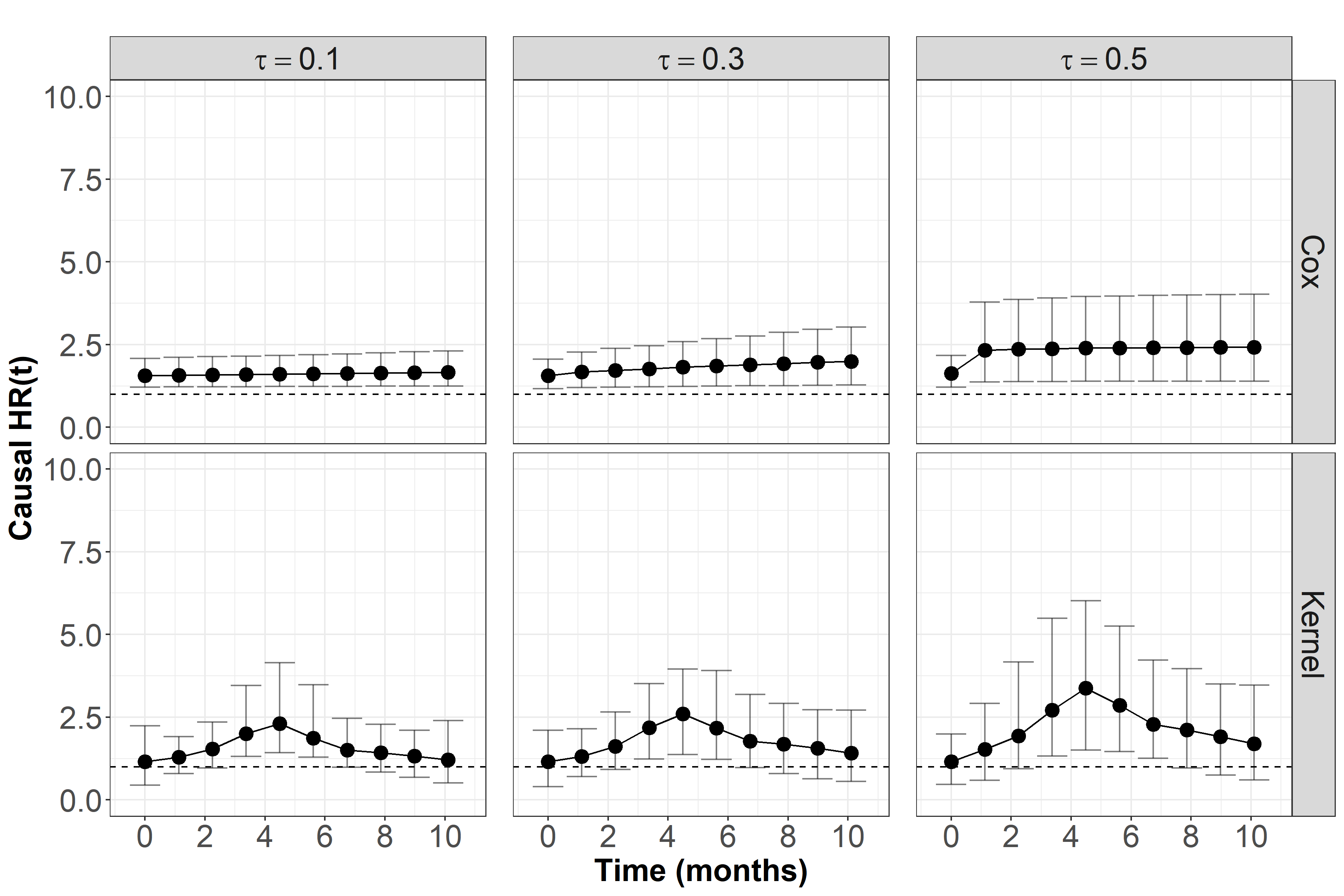}
\caption{IG frailty}
\label{fig:divat_IG}
\end{subfigure}
 \centering
\caption{Cox-based and kernel-based estimation along with plus/minus one standard error of the $HR^C(t)$, under different values of Kendall's $\tau$ and frailty distributions. The horizontal dashed line represents $HR^C(t)=1$. For clarity, we present the estimation every 4 time points and only until $t=10$. Note that the y-axis has been truncated at $HR^C(t)=10$.}
\label{fig:divat.HRC}
\end{figure}

Turning to the proposed sensitivity analysis, since there was no evidence against the validity of the PH assumption, we considered both the Cox-based and kernel-based estimators. The Cox-based estimator of $HR^C(t)$ (Figure \ref{fig:divat.HRC}) was larger than one and monotonically increased over time. This result suggest that transplantation from ECD donors has an harmful effect on graft failure time, and furthermore, that the impact on the hazard becomes more and more substantial with $t$ among those who would have survived beyond time $t$ regardless of the donor type. 
The kernel-based method estimated an increase in the $HR^C(t)$ until approximately $t=5$ years and then a decreasing but larger than one  $HR^C(t)$ at the remaining time points. 

The estimated coefficients of the confounders were relatively small (Table \ref{tab:con.cox.table}) and the event rate was 31\%. In light of the simulation results we might expect the relative bias of the kernel-based estimator to be lower than the naive Cox-model and the Cox-based estimators. 

To summarize the findings, although there is a disagreement between the two methods in the precise $HR^C(t)$ value, there is an agreement in the conclusion that ECD has an harmful effect on the patient and graft survival comparing to transplantations from SCD donors ($HR^C(t)$ above one). Even though the $HR^C(t)$ is above the standard HR estimated from a weighted Cox model, there is an agreement on the conclusion between the the two of them.

\section{Discussion}
\label{sec:discuss}

The HR measure is still one of the most ubiquitous tools for survival analysis although the problematic interpretation it might have. \cite{martinussen2020subtleties} has proposed the causal HR, a well-defined alternative causal effect, similar in its nature to the SACE that have been gradually targeted in studies exposed to truncation by death. However, the strategy of how to apply it to real data was still missing. In this paper, we proposed a sensitivity analysis framework for identification, followed by kernel-based or Cox-based estimation of the $HR^C(t)$. The kernel-based estimator is especially attractive in settings the marginal structural Cox model assumption is unlikely to hold. 
Our simulation results demonstrated that when the marginal structural Cox model is correctly-specified, the Cox-based estimator is preferable over the kernel-based estimator. In all other cases, the kernel-based estimator was less biased even when none of the stated assumptions were exactly met. As with many non-parametric methods, a larger sample size might be needed to obtain satisfactory properties of the estimator. 

The assumption that the potential survival times are independent given the frailty, is a \textit{cross-world  independence assumption}. This assumption cannot be verified but as previously mentioned, we take it as a working sensitivity analysis framework, and not as a belief in the existence of such a variable $V$. In many scenarios, a common alternative strategy constructs bounds for causal effects, e.g., under the monotonicity assumption \citep{zhang2003estimation}. Here, however, a monotonicity-like assumption may constraint the causal HR to be equal to zero (Section A.5 in \cite{martinussen2020subtleties}), and therefore we did not follow this strategy. 

Choosing the bandwidth to minimize the MSE may result in CIs with under-coverage, a well known phenomena in non-parametric inference. A potential solution might be under-smoothing \citep{wasserman2006all}, sacrificing larger standard errors (and wider CIs) to reduce the bias. Another potential extension of the proposed approach is to improve its efficiency by developing an estimation approach that allows for inclusion of more outcome predictors, even in an RCT setting. One way to achieve this can be to use a time-varying coefficient Cox model \citep{zucker1990nonparametric,tian2005cox} rather than the standard Cox model to improve the Cox-based estimator when the PH assumption is unlikely not hold.

\section*{Acknowledgements}
This research was supported by the Israel Science Foundation (ISF grant No. 827/21).
\vspace*{-8pt}
\end{onehalfspacing}


\begin{thebibliography}{}
 	
	\bibitem[Aalen et~al., 2008]{aalen2008survival}
	Aalen, O., Borgan, O., and Gjessing, H. (2008).
	\newblock {\em Survival and event history analysis: a process point of view}.
	\newblock Springer Science \& Business Media.
	\bibitem[Aalen et~al., 2015]{aalen2015does}
	Aalen, O.~O., Cook, R.~J., and R{\o}ysland, K. (2015).
	\newblock Does cox analysis of a randomized survival study yield a causal
	treatment effect?
	\newblock {\em Lifetime data analysis}, 21(4):579--593.
	
	\bibitem[Andersen et~al., 2012]{andersen2012statistical}
	Andersen, P.~K., Borgan, O., Gill, R.~D., and Keiding, N. (2012).
	\newblock {\em Statistical models based on counting processes}.
	\newblock Springer Science \& Business Media.
	
	\bibitem[Andersen et~al., 2021]{kragh2021analysis}
	Andersen, P.~K., Pohar~Perme, M., van Houwelingen, H.~C., Cook, R.~J., Joly,
	P., Martinussen, T., Taylor, J.~M., Abrahamowicz, M., and Therneau, T.~M.
	(2021).
	\newblock Analysis of time-to-event for observational studies: Guidance to the
	use of intensity models.
	\newblock {\em Statistics in Medicine}, 40(1):185--211.
	
	\bibitem[Assel et~al., 2019]{assel2019guidelines}
	Assel, M., Sjoberg, D., Elders, A., Wang, X., Huo, D., Botchway, A., Delfino,
	K., Fan, Y., Zhao, Z., Koyama, T., et~al. (2019).
	\newblock Guidelines for reporting of statistics for clinical research in
	urology.
	\newblock {\em The Journal of urology}, 201(3):595--604.
	
	\bibitem[Clayton and Cuzick, 1985]{clayton1985multivariate}
	Clayton, D. and Cuzick, J. (1985).
	\newblock Multivariate generalizations of the proportional hazards model.
	\newblock {\em Journal of the Royal Statistical Society: Series A},
	148(2):82--108.
	
	\bibitem[Cole and Hern{\'a}n, 2004]{cole2004adjusted}
	Cole, S.~R. and Hern{\'a}n, M.~A. (2004).
	\newblock Adjusted survival curves with inverse probability weights.
	\newblock {\em Computer methods and programs in biomedicine}, 75(1):45--49.
	
	\bibitem[Cox, 1972]{cox1972regression}
	Cox, D.~R. (1972).
	\newblock Regression models and life-tables.
	\newblock {\em Journal of the Royal Statistical Society: Series B
		(Methodological)}, 34(2):187--202.
	
	\bibitem[Egleston et~al., 2007]{egleston2007causal}
	Egleston, B.~L., Scharfstein, D.~O., Freeman, E.~E., and West, S.~K. (2007).
	\newblock Causal inference for non-mortality outcomes in the presence of death.
	\newblock {\em Biostatistics}, 8(3):526--545.
	
	\bibitem[Frangakis and Rubin, 2002]{frangakis2002principal}
	Frangakis, C.~E. and Rubin, D.~B. (2002).
	\newblock Principal stratification in causal inference.
	\newblock {\em Biometrics}, 58(1):21--29.
	
	\bibitem[Gorfine et~al., 2012]{gorfine2012conditional}
	Gorfine, M., De-Picciotto, R., and Hsu, L. (2012).
	\newblock Conditional and marginal estimates in case-control family
	data--extensions and sensitivity analyses.
	\newblock {\em Journal of statistical computation and simulation},
	82(10):1449--1470.
	
	\bibitem[Gorfine et~al., 2020]{gorfine2020k}
	Gorfine, M., Schlesinger, M., and Hsu, L. (2020).
	\newblock K-sample omnibus non-proportional hazards tests based on
	right-censored data.
	\newblock {\em Statistical Methods in Medical Research}, 29(10):2830--2850.
	
	\bibitem[Grambsch and Therneau, 1994]{grambsch1994proportional}
	Grambsch, P.~M. and Therneau, T.~M. (1994).
	\newblock Proportional hazards tests and diagnostics based on weighted
	residuals.
	\newblock {\em Biometrika}, 81(3):515--526.
	
	\bibitem[Hayden et~al., 2005]{hayden2005estimator}
	Hayden, D., Pauler, D.~K., and Schoenfeld, D. (2005).
	\newblock An estimator for treatment comparisons among survivors in randomized
	trials.
	\newblock {\em Biometrics}, 61(1):305--310.
	
	\bibitem[Hern{\'a}n, 2010]{hernan2010hazards}
	Hern{\'a}n, M.~A. (2010).
	\newblock The hazards of hazard ratios.
	\newblock {\em Epidemiology}, 21(1):13.
	
	\bibitem[Hern{\'a}n et~al., 2000]{hernan2000marginal}
	Hern{\'a}n, M.~{\'A}., Brumback, B., and Robins, J.~M. (2000).
	\newblock Marginal structural models to estimate the causal effect of
	zidovudine on the survival of hiv-positive men.
	\newblock {\em Epidemiology}, pages 561--570.
	
	\bibitem[Hess et~al., 1999]{hess1999hazard}
	Hess, K.~R., Serachitopol, D.~M., and Brown, B.~W. (1999).
	\newblock Hazard function estimators: a simulation study.
	\newblock {\em Statistics in medicine}, 18(22):3075--3088.
	
	\bibitem[Hougaard, 2000]{hougaard2000analysis}
	Hougaard, P. (2000).
	\newblock {\em Analysis of multivariate survival data}, volume 564.
	\newblock Springer.
	
	\bibitem[Imbens and Rubin, 2015]{imbens2015causal}
	Imbens, G.~W. and Rubin, D.~B. (2015).
	\newblock {\em Causal inference in statistics, social, and biomedical
		sciences}.
	\newblock Cambridge University Press.
	
	\bibitem[Lang and Altman, 2015]{lang2015basic}
	Lang, T.~A. and Altman, D.~G. (2015).
	\newblock Basic statistical reporting for articles published in biomedical
	journals: the" statistical analyses and methods in the published literature"
	or the sampl guidelines.
	\newblock {\em Int J Nurs Stud}, 52(1):5--9.
	
	\bibitem[Le~Borgne et~al., 2016]{le2016comparisons}
	Le~Borgne, F., Giraudeau, B., Querard, A.~H., Giral, M., and Foucher, Y.
	(2016).
	\newblock Comparisons of the performance of different statistical tests for
	time-to-event analysis with confounding factors: practical illustrations in
	kidney transplantation.
	\newblock {\em Statistics in medicine}, 35(7):1103--1116.
	
	\bibitem[Martinussen et~al., 2020]{martinussen2020subtleties}
	Martinussen, T., Vansteelandt, S., and Andersen, P.~K. (2020).
	\newblock Subtleties in the interpretation of hazard contrasts.
	\newblock {\em Lifetime Data Analysis}, 26(4):833--855.
	
	\bibitem[Metzger et~al., 2003]{metzger2003expanded}
	Metzger, R.~A., Delmonico, F.~L., Feng, S., Port, F.~K., Wynn, J.~J., and
	Merion, R.~M. (2003).
	\newblock Expanded criteria donors for kidney transplantation.
	\newblock {\em American Journal of Transplantation}, 3:114--125.
	
	\bibitem[M{\"u}ller and Wang, 1990]{muller1990locally}
	M{\"u}ller, H.-G. and Wang, J.-L. (1990).
	\newblock Locally adaptive hazard smoothing.
	\newblock {\em Probability Theory and Related Fields}, 85(4):523--538.
	
	\bibitem[Muller and Wang, 1994]{muller1994hazard}
	Muller, H.-G. and Wang, J.-L. (1994).
	\newblock Hazard rate estimation under random censoring with varying kernels
	and bandwidths.
	\newblock {\em Biometrics}, pages 61--76.
	
	\bibitem[Nevo and Gorfine, 2021]{nevo2021causal}
	Nevo, D. and Gorfine, M. (2021).
	\newblock Causal inference for semi-competing risks data.
	\newblock {\em Biostatistics}.
	
	\bibitem[Oakes, 1989]{oakes1989bivariate}
	Oakes, D. (1989).
	\newblock Bivariate survival models induced by frailties.
	\newblock {\em Journal of the American Statistical Association},
	84(406):487--493.
	
	\bibitem[Powles et~al., 2018]{powles2018atezolizumab}
	Powles, T., Dur{\'a}n, I., Van Der~Heijden, M.~S., Loriot, Y., Vogelzang,
	N.~J., De~Giorgi, U., Oudard, S., Retz, M.~M., Castellano, D., Bamias, A.,
	et~al. (2018).
	\newblock Atezolizumab versus chemotherapy in patients with platinum-treated
	locally advanced or metastatic urothelial carcinoma (imvigor211): a
	multicentre, open-label, phase 3 randomised controlled trial.
	\newblock {\em The Lancet}, 391(10122):748--757.
	
	\bibitem[Ramlau-Hansen, 1983]{ramlau1983smoothing}
	Ramlau-Hansen, H. (1983).
	\newblock Smoothing counting process intensities by means of kernel functions.
	\newblock {\em The Annals of Statistics}, pages 453--466.
	
	\bibitem[Robins et~al., 2000]{robins2000marginal}
	Robins, J.~M., Hernan, M.~A., and Brumback, B. (2000).
	\newblock Marginal structural models and causal inference in epidemiology.
	
	\bibitem[Rubin, 2006]{rubin2006causal}
	Rubin, D.~B. (2006).
	\newblock Causal inference through potential outcomes and principal
	stratification: application to studies with ``censoring" due to death.
	\newblock {\em Statistical Science}, 21(3):299--309.
	
	\bibitem[Stensrud et~al., 2017]{stensrud2017exploring}
	Stensrud, M.~J., Valberg, M., R{\o}ysland, K., and Aalen, O.~O. (2017).
	\newblock Exploring selection bias by causal frailty models.
	\newblock {\em Epidemiology}, 28(3):379--386.
	
	\bibitem[Tian et~al., 2005]{tian2005cox}
	Tian, L., Zucker, D., and Wei, L. (2005).
	\newblock On the cox model with time-varying regression coefficients.
	\newblock {\em Journal of the American statistical Association},
	100(469):172--183.
	
	\bibitem[Wasserman, 2006]{wasserman2006all}
	Wasserman, L. (2006).
	\newblock {\em All of nonparametric statistics}.
	\newblock Springer Science \& Business Media.
	
	\bibitem[Zehavi and Nevo, 2021]{zehavi2021matching}
	Zehavi, T. and Nevo, D. (2021).
	\newblock A matching framework for truncation by death problems.
	\newblock {\em arXiv preprint arXiv:2110.10186}.
	
	\bibitem[Zhang and Rubin, 2003]{zhang2003estimation}
	Zhang, J.~L. and Rubin, D.~B. (2003).
	\newblock Estimation of causal effects via principal stratification when some
	outcomes are truncated by “death”.
	\newblock {\em Journal of Educational and Behavioral Statistics},
	28(4):353--368.
	
	\bibitem[Zucker et~al., 1990]{zucker1990nonparametric}
	Zucker, D.~M., Karr, A.~F., et~al. (1990).
	\newblock Nonparametric survival analysis with time-dependent covariate
	effects: a penalized partial likelihood approach.
	\newblock {\em The Annals of Statistics}, 18(1):329--353.
	
\end{thebibliography}



\clearpage
\appendix
\begin{onehalfspacing}
\counterwithin{figure}{section}
\counterwithin{table}{section}
\counterwithin{equation}{section}
\renewcommand\thesection{\Alph{section}}
\section*{Description of contents in the Supplementary Materials}
\begin{itemize}
	\item Section \ref{SM:IdentProofs} provides proofs of the identification formulas (2) and (3) in the main text. In Section \ref{SM:prop1} we provide the $HR^C(t)$ identification formula proof under randomization. In Section \ref{SM:prop2} we provide $HR^C(t)$ identification formula proof in the presence of confounders.
	\item Section \ref{SM:formulas_for_varphi} provides formulas and their proofs for the function $\varphi(\Lambda^{A=1}(t),\Lambda^{A=0}(t),\theta)$ under different frailty distributions.
	\item Section \ref{app:details} provides the explicit form of the boundary kernel function, our proposed MSE estimation for the weighted scenario and some details on the initial parameters of the kernel estimation procedure.
	\item Section \ref{SM:simulations} provides technical details 
	about the simulations scenarios. In Section \ref{SM:sceanrio_Ia} we show that the hazard model in Scenario (Ia) results in the Cox model (4), in the main text, for the marginal hazard $\lambda^{A=a}(t)$ with a time-varying $HR^C(t)$. In Section \ref{SM:sceanrio_Ib} we show that the hazard model in Scenario (Ib) results in constant $HR^C(t)$. In Section \ref{SM:sceanrio_II} we show that the the hazard model in Scenario (II) results in constant $HR^C(t)$. In Section \ref{SM:additional_results} we provide complementary simulation results as discussed in Section 6 of the main text. For Scenarios (Ia) and (Ib) we present additional results under Kendall's $\tau\in (0.1,0.3,0.5,0.7)$, sample sizes $n=500,1000$ and censoring rates of $CR=20\%, 40\%$. We also present the empirical coverage rates for these settings. For Scenario (II) we present additional results of the mean estimates along with the empirical standard deviation of the estimates, the empirical coverage rates and the ratio between EMP.SD and EST.SE. In addition we present the relative bias of the different estimation methods under Kendall's $\tau=0.5$. 
	\item Section \ref{SM:Imvigor211} presents the Cox-based estimation along with the empirical standard deviation of the estimates for the Imvigor211 dataset. 
	\item Section \ref{SM:DIVAT} presents additional results for the DIVAT data set including: logistic regression for the calculating the weights, covariates balance, histogram of the obtained weights, the Kaplan-Meier plot of the DIVAT dataset and the results from the weighted Cox regression. 
\end{itemize}

\newpage
\section{Identification proofs}
\label{SM:IdentProofs}
In this Section, we provide proofs for the $HR^C(t)$ identification formula under randomization (Proposition 4.1) and the approximated identification formula in the presence of confounders. We will first show that $HR^{C}(t)=\frac{\psi^{A=1}(t)}{\psi^{A=0}(t)}$. Than we will show that $\psi^{A=a}(t)=\lambda^{A=a}(t)\gamma(\Lambda^{A=a}(t),\theta)$. Finally, we will show that under randomization $\lambda^{A=a}(t)$ can be identified by $\lambda(t|A=a)=\frac{E(dN(t|A=a))}{E(Y(t|A=a))}$, and in the presence of confounders by $\lambda_{IPTW}(t|A=a)=\frac{E\left[w(\bZ)dN(t|A=a)\right]}{E\left[w(\bZ)Y(t|A=a)\right]}$. For convenience of presentation, we omit $C=\infty$ in $T^{A=a,C=\infty}$, and write it as $T^{A=a}$. 
\subsection{$HR^C(t)$ identification formula proof under randomization}
\label{SM:prop1}
Under SUTVA, randomization, and the frailty assumptions, the $HR^C(t)$ can  be written as 
\begin{subequations}
	\label{eq:HRC_frac}
	\begin{align}
	\nonumber
	&HR^{C}(t)\\\nonumber
	&=\frac{\lim\limits_{dt \to 0}(dt)^{-1}\Pr\left[t\leq T^{A=1} < t +dt| T^{A=1}\geq t, T^{A=0}\geq t\right]}{\lim\limits_{dt \to 0}(dt)^{-1}\Pr\left[t\leq T^{A=0}< t+dt| T^{A=1}\geq t, T^{A=0}\geq t\right]}
	\\ \label{eq1:l.t.p}
	&=\frac{\int_{0}^{\infty}\lim\limits_{dt \to 0}(dt)^{-1}\Pr\left[t\leq T^{A=1} <t+dt| T^{A=1}\geq t, T^{A=0}\geq t, V=v\right]f(v|T^{A=1}\geq t, T^{A=0}\geq t)dv}{\int_{0}^{\infty}\lim\limits_{dt \to 0}(dt)^{-1}\Pr\left[t\leq T^{A=0} < t+dt| T^{A=1}\geq t, T^{A=0}\geq t, V=v\right]f(v|T^{A=1}\geq t, T^{A=0}\geq t)dv}
	\\ \label{eq1:ind}
	&=\frac{\int_{0}^{\infty}\lim\limits_{dt \to 0}(dt)^{-1}\Pr\left[t\leq T^{A=1} <t+dt| T^{A=1}\geq t, V=v\right] f(v|T^{A=1}\geq t, T^{A=0}\geq t)dv}{\int_{0}^{\infty}\lim\limits_{dt \to 0}(dt)^{-1}\Pr\left[t\leq T^{A=0} <t+dt| T^{A=0}\geq t, V=v\right] f(v|T^{A=1}\geq t, T^{A=0}\geq t)dv}
	\\\label{eq1:def}
	&=\frac{\int_{0}^{\infty}\lambda^{A=1}(t;v)f(v|T^{A=1}\geq t, T^{A=0}\geq t)dv}{\int_{0}^{\infty}\lambda^{A=0}(t;v)f(v|T^{A=1}\geq t, T^{A=0}\geq t)dv}
	\\\label{eq1:mul}
	&=\frac{\int_{0}^{\infty}v\psi^{A=1}(t)f(v|T^{A=1}\geq t, T^{A=0}\geq t)dv}{\int_{0}^{\infty}v\psi^{A=0}(t)f(v|T^{A=1}\geq t, T^{A=0}\geq t)dv}
	\\ \nonumber 
	&=\frac{\psi^{A=1}(t)\int_{0}^{\infty}vf(v|T^{A=1}\geq t, T^{A=0}\geq t)dv}{\psi^{A=0}(t)\int_{0}^{\infty}vf(v|T^{A=1}\geq t, T^{A=0}\geq t)dv}
	\\ \nonumber 
	&=\frac{\psi^{A=1}(t)}{\psi^{A=0}(t)}.
	\end{align}
\end{subequations}
In \eqref{eq1:l.t.p} we used the  Law of Total Probability and then the Monotone Convergence Theorem to swap the integral and the limit. In \eqref{eq1:ind} we used the assumption that given $V$, $T^{A=0}$ and $T^{A=1}$ are independent. In \eqref{eq1:def} we used the definition of the conditional hazard rate saying that $\lambda^{A=a}(t;v)=\lim\limits_{dt \to 0}(dt)^{-1}\Pr\left[t\leq T^{A=a} <t+dt| T^{A=1}\geq t, V=v\right]$. In \eqref{eq1:mul} we substituted the multiplicative frailty model  $\lambda^{A=a}(t;v)=v\psi^{A=a}(t)$. 

Next, we will show that $\psi^{A=a}(t)$ can be written as $\lambda^{A=a}(t)\tilde{\varphi}(\Lambda^{A=a}(t),\theta)$. Let $\Psi^{A=a}(t)=\int_{0}^{t}\psi^{A=a}(u)du$ and $S^{A=a}(t)=\Pr(T^{A=a} > t)$. Denote also $\Phi_V\left(u\right)=E\left(\exp(-Vu)\right)$ for the Laplace transform of $V$. Because $E_V(S^{A=a}(t|V))=S^{A=a}(t)$, and recalling the relationship between the survival function and the cumulative hazard function, we may write $\Phi_V(\Psi^{A=a}(t))=\exp(-\Lambda^{A=a}(t))$, and therefore $\Psi^{A=a}(t)=\Phi_V^{-1}[e^{-\Lambda^{A=a}(t)}]$. Thus, we may write $\psi^{A=a}(t)$ as
\begin{subequations}
	\nonumber
	\begin{align}
	\psi^{A=a}(t)=\frac{\partial}{\partial t}\Psi^{A=a}(t)&=\frac{\partial}{\partial t}\Phi_V^{-1}[\exp(-\Lambda^{A=a}(t))]\\
	&=\frac{\partial}{\partial e^{-\Lambda^{A=a}(t)}}\Phi_V^{-1}[\exp(-\Lambda^{A=a}(t))]\frac{\partial}{\partial t}\exp(-\Lambda^{A=a}(t))\\
	&=-\lambda^{A=a}(t)\frac{\partial}{\partial e^{-\Lambda^{A=a}(t)}}\Phi_V^{-1}[\exp(-\Lambda^{A=a}(t))]\exp(-\Lambda^{A=a}(t))\\
	&=\lambda^{A=a}(t)\tilde{\varphi}(\Lambda^{A=a}(t),\theta),
	\end{align}
\end{subequations}
where $\tilde{\varphi}(\Lambda^{A=a}(t),\theta)=-\frac{\partial}{\partial e^{-\Lambda^{A=a}(t)}}\Phi_V^{-1}[\exp(-\Lambda^{A=a}(t))]\exp(-\Lambda^{A=a}(t))$.

By substituting $\psi^{A=a}(t)=\lambda^{A=a}(t)\tilde{\varphi}(\Lambda^{A=a}(t),\theta)$ back into \eqref{eq:HRC_frac} we may write
\begin{equation}\label{HRC_final_prop}
\begin{split}
HR^{C}(t)&=\frac{\lambda^{A=1}(t)\tilde{\varphi}(\Lambda^{A=1}(t),\theta)}{\lambda^{A=0}(t)\tilde{\varphi}(\Lambda^{A=0}(t),\theta)}\\
&=\frac{\lambda^{A=1}(t)}{\lambda^{A=0}(t)}\varphi\left(\Lambda^{A=1}(t),\Lambda^{A=0}(t),\theta\right), 
\end{split}
\end{equation}
where $\varphi(\Lambda^{A=1}(t),\Lambda^{A=0}(t),\theta)=\frac{\lambda^{A=1}(t)\tilde{\varphi}(\Lambda^{A=1}(t),\theta)}{\lambda^{A=0}(t)\tilde{\varphi}(\Lambda^{A=0}(t),\theta)}$ is a function depending on $\Lambda^{A=1}(t)$, $\Lambda^{A=0}(t)$ and $\theta$, and may take a closed formula depending on the specific parametric family distribution of $V$. For example, we show in Section \ref{app:proof_gamma_frailty} that when $V\sim Gamma(\theta^{-1},\theta^{-1})$, we have $\varphi(\Lambda^{A=1}(t),\Lambda^{A=0}(t),\theta)=\exp \big\{\theta \left[\Lambda^{A=1}(t)-\Lambda^{A=0}(t)\right]\big\}$.

Now, in the final stage of the proof, we show that we can identify $\lambda^{A=a}(t)$ as $\lambda(t|A=a)=\frac{E[dN(t|a)]}{E[Y(t|a)]}$. The independent censoring assumption can be written as (see Chapter 1 in \cite{aalen2008survival})
\begin{equation*}
\Pr\left(t\leq X^{A=a} <t+dt,\delta^{A=a}=1 | X^{A=a}\geq t\right) = \Pr\left(t\leq T^{A=a,c=\infty} <t+dt| T^{A=a,c=\infty}\geq t\right).
\end{equation*}
Therefore, the hazard rate in terms of potential outcomes can be written as 
\begin{align}
\label{eq:hazard.prob.CP}
\begin{split}
\lambda^{A=a}(t)=& \lim\limits_{dt \to 0}(dt)^{-1}\Pr\left[t\leq T^{A=a,c=\infty} <t+dt| T^{A=a,c=\infty}\geq t\right] \\
=& \lim\limits_{dt \to 0}(dt)^{-1}\Pr\left[t\leq X^{A=a} <t+dt,\delta^{A=a}=1 | X^{A=a}\geq t\right] \\
=&\frac{ \lim\limits_{dt \to 0}(dt)^{-1}\Pr\left[t\leq X^{A=a} <t+dt,\delta^{A=a}=1 \right]}{\Pr\left(X^{A=a}\geq t\right)}\\
=& \frac{\Pr[dN^{A=a}(t)=1]}{\Pr\left(X^{A=a}\geq t\right)}.
\end{split}
\end{align}
Focusing on the numerator $\Pr[dN^{A=a}(t)=1]$, under randomization and SUTVA we have
\begin{align*}
\Pr\left[dN^{A=a}(t)=1 \right]&=\Pr[dN(t)=1 |A=a] \\
&=\frac{\Pr\left[dN(t)=1 ,A=a \right]}{\Pr(A=a)}\\
&= \frac{E[dN(t|A=a)]}{Pr(A=a)},
\end{align*}
where the final line follows from the definition of $dN(t|A=a)=dN(t)I(A=a)$. 
Similar arguments for  the denominator $\Pr\left(X^{A=a}\geq t\right)$, will give us 
\begin{align*}
\Pr\left(X^{A=a}\geq t\right)&=\Pr\left(X\geq t\right|A=a)\\
&=\frac{\Pr\left(X\geq t, A=a\right)}{\Pr(A=a)}\\
&=\frac{E[Y(t|A=a)]}{P(A=a)},
\end{align*}
where the final line follows from the definition of $Y(t|A=a)=I(X\geq t)I(A=a)$ and $\Pr(X\geq t)=E(Y(t))$. 
In total we get 
\begin{equation*}
\lambda^{A=a}(t)=\frac{E[dN(t|A=a)]/\Pr(A=a)}{E[Y(t|A=a)]/\Pr(A=a)} = \frac{E[dN(t|A=a)]}{E[Y(t|A=a)]}.
\end{equation*}
\newpage
\subsection{$HR^C(t)$ approximated identification formula proof  in the presence of confounders}
\label{SM:prop2}
Under SUTVA, conditional exchangeability, independent censoring,  the frailty assumptions, and under high censoring rate and/or low association between $T^{A=a,C=\infty}$ and $\bZ$, the $HR^C(t)$ can be written as 
\begin{subequations}
	\label{eq:HRC_w_obs}
	\begin{align}
	\nonumber
	&HR^{C}(t)\\\nonumber
	&=\frac{\lim\limits_{dt \to 0}(dt)^{-1}\Pr\left[t\leq T^{A=1} < t +dt| T^{A=1}\geq t, T^{A=0}\geq t\right]}{\lim\limits_{dt \to 0}(dt)^{-1}\Pr\left[t\leq T^{A=0}< t+dt| T^{A=1}\geq t, T^{A=0}\geq t\right]}
	\\ \label{subeq:HRC_w_obs_2}
	&=\frac{\int_{0}^{\infty}\int\lim\limits_{dt \to 0}(dt)^{-1}\Pr\left[t\leq T^{A=1} <t+dt| T^{A=1}\geq t, T^{A=0}\geq t,V=v, \bZ=\bz\right]f(v,\bz|T^{A=1}\geq t, T^{A=0}\geq t)d\bz dv}{\int_{0}^{\infty}\int\lim\limits_{dt \to 0}(dt)^{-1}\Pr\left[t\leq T^{A=0} < t+dt| T^{A=1}\geq t, T^{A=0}\geq t,V=v, \bZ=\bz\right]f(v,\bz|T^{A=1}\geq t, T^{A=0}\geq t)d\bz dv}
	\\\label{subeq:HRC_w_obs_3}
	&=\frac{\int_{0}^{\infty}\int\lim\limits_{dt \to 0}(dt)^{-1}\Pr\left[t\leq T^{A=1} <t+dt| T^{A=1}\geq t, V=v, \bZ=\bz\right]f(v,\bz|T^{A=1}\geq t, T^{A=0}\geq t)d\bz dv}{\int_{0}^{\infty}\int\lim\limits_{dt \to 0}(dt)^{-1}\Pr\left[t\leq T^{A=0} <t+dt| T^{A=0}\geq t, V=v, \bZ=\bz\right]f(v,\bz|T^{A=1}\geq t, T^{A=0}\geq t)d\bz dv}
	\\\label{subeq:HRC_w_obs_4}
	&=\frac{\int_{0}^{\infty}\int\lambda^{A=1}(t|v,\bz)f(v,\bz|T^{A=1}\geq t, T^{A=0}\geq t)d\bz dv}{\int_{0}^{\infty}\int\lambda^{A=0}(t|v,\bz)f(v,\bz|T^{A=1}\geq t, T^{A=0}\geq t)d\bz dv}
	\\ \label{subeq:HRC_w_obs_5}
	&=\frac{\int_{0}^{\infty}\int\lambda^{A=1}(t|v,\bz)f(\bz|T^{A=1}\geq t, v)\frac{f(v,\bz|T^{A=1}\geq t, T^{A=0}\geq t)}{f(\bz|T^{A=1}\geq t, v)}d\bz dv}{\int_{0}^{\infty}\int\lambda^{A=0}(t|v,\bz)f(\bz|T^{A=0}\geq t, v)\frac{f(v,\bz|T^{A=1}\geq t, T^{A=0}\geq t)}{f(\bz|T^{A=0}\geq t, v)}d\bz dv}
	\\ \label{subeq:HRC_w_obs_6}
	&=\frac{\int_{0}^{\infty}\int\lambda^{A=1}(t|v,\bz)f(\bz|T^{A=1}\geq t, v)\frac{\Pr(T^{A=0}\geq t|\bZ=\bz,V=v) \Pr(T^{A=1}\geq t|V=v)f(v)}{\Pr(T^{A=1}\geq t, T^{A=0}\geq t)}d\bz dv}{\int_{0}^{\infty}\int \lambda^{A=0}(t|v,\bz)f(\bz|T^{A=0}\geq t, v)\frac{\Pr(T^{A=0}\geq t|\bZ=\bz,V=v) \Pr(T^{A=1}\geq t|V=v)f(v)}{\Pr(T^{A=1}\geq t, T^{A=0}\geq t)}d\bz dv}\\\label{subeq:HRC_w_obs_7}
	&\approx\frac{\int_{0}^{ \infty}\left[\int\lambda^{A=1}(t|v,\bz)f(\bz|T^{A=1}\geq t, v)d\bz\right]\frac{\Pr(T^{A=0}\geq t|V=v) \Pr(T^{A=1}\geq t|V=v)f(v)}{\Pr(T^{A=1}\geq t, T^{A=0}\geq t)} dv}{\int_{0}^{\infty}\left[\int\lambda^{A=0}(t|v,\bz)f(\bz|T^{A=0}\geq t, v)d\bz\right]\frac{\Pr(T^{A=0}\geq t|V=v) \Pr(T^{A=1}\geq t|V=v)f(v)}{\Pr(T^{A=1}\geq t, T^{A=0}\geq t)} dv}
	\\ \label{subeq:HRC_w_obs_8}
	&= \frac{\int_{0}^{\infty}\lambda^{A=1}(t|v)\Pr(T^{A=0}\geq t|V=v)\Pr(T^{A=1}\geq t|V=v)f(v) dv}{\int_{0}^{ \infty}\lambda^{A=0}(t|v)\Pr(T^{A=1}\geq t|V=v) \Pr(T^{A=0}\geq t|V=v)f(v)dv}
	\\\label{subeq:HRC_w_obs_9}
	&=\frac{\int_{0}^{\infty }v\psi^{A=1}(t)\Pr(T^{A=0}\geq t|V=v)\Pr(T^{A=1}\geq t|V=v)f(v)dv}{\int_{0}^{ }v\psi^{A=0}(t)\Pr(T^{A=0}\geq t|V=v)\Pr(T^{A=1}\geq t|V=v)f(v)dv}
	\\ \nonumber 
	&=\frac{\psi^{A=1}(t)\int_{0}^{\infty}v\Pr(T^{A=0}\geq t|V=v)\Pr(T^{A=1}\geq t|V=v)f(v)dv}{\psi^{A=0}(t)\int_{0}^{\infty}v\Pr(T^{A=0}\geq t|V=v)\Pr(T^{A=1}\geq t|V=v)f(v)dv}
	\\\nonumber 
	&=\frac{\psi^{A=1}(t)}{\psi^{A=0}(t)}.
	\end{align}
\end{subequations}
In \eqref{subeq:HRC_w_obs_2} we used the Law of Total Probability and then the Monotone Convergence Theorem to swap the integral and the limit. In \eqref{subeq:HRC_w_obs_3} we used the assumption that Given $V$ and $\bZ$, $T^{A=0}$ and $T^{A=1}$ are independent. In \eqref{subeq:HRC_w_obs_4} we used the definition of the conditional hazard rate saying that $\lambda^{A=a}(t|v,\bz)=\lim\limits_{dt \to 0}dt^{-1}\Pr\left[t\leq T^{A=a} <t+dt| T^{A=1}\geq t, v,\bz\right]$. In \eqref{subeq:HRC_w_obs_5} we multiplied and divided the each integrand by $\Pr(T^{A=1}\geq t|z,v)$ and the denominator by $\Pr(T^{A=0}\geq t|z,v)$. In \eqref{subeq:HRC_w_obs_6} for $a=0,1$ we used Bayes' Theorem to write 
$$
\frac{f(v,\bz|T^{A=1}\geq t, T^{A=0}\geq t)}{f(\bz|T^{A=a}\geq t, v)}=\frac{\Pr(T^{A=1-a}\geq t|\bZ=\bz,V=v) \Pr(T^{A=a}\geq t|v)f(v)}{ \Pr(T^{A=1}\geq t, T^{A=0}\geq t)}.
$$ 
In \eqref{subeq:HRC_w_obs_7}, we used the fact that under high censoring rate and/or low association between $T^{A=a,C=\infty}$ and $\bZ$, $\Pr(T^{A=a}\geq t|z,v)\approx \Pr(T^{A=a}\geq t|v)$. In \eqref{subeq:HRC_w_obs_8} we used the Law of Total Probability and then the Monotone Convergence Theorem to write $\lambda^{A=a}(t|v)=\int_{}^{}\lambda^{A=a}(t|\bz,v)f(\bz|T^{A=a}\geq t, v)d\bz$. In \eqref{subeq:HRC_w_obs_9} we substituted the multiplicative frailty model $\lambda^{A=a}(t|v)=v\psi^{A=a}(t)$. 

Since \eqref{HRC_final_prop} depends on the multiplicity assumption of the frailty it is still holds even when confounders are present. Now, we still have to show that we can identify $\lambda^{A=a}(t)$ as $\lambda_{IPTW}(t|A=a)=\frac{E\left[w(\bZ)dN(t|A=a)\right]}{E\left[w(\bZ)Y(t|A=a)\right]}$, where $w(\bZ)=1/Pr(A=a|\bZ)$.
Under the independent censoring assumption we can write $\lambda^{A=a}(t)=\frac{\Pr\left[dN^{A=a}(t)=1 \right]}{\Pr\left(X^{A=a}\geq t\right)}$ (see Equation \eqref{eq:hazard.prob.CP}). Focusing on the numerator $\Pr\left[dN^{A=a}(t)=1 \right]$,  under SUTVA and conditional exchangeability we have
\begin{align*}
\Pr\left[dN^{A=a}(t)=1 \right]&=E_Z\left[\frac{\Pr\left(dN(t)=1 ,A=a|Z \right)}{\Pr(A=a|Z)}\right] \\
&=E_Z\left[E\left[\frac{dN(t|A=a) }{\Pr(A=a|Z)}\Bigg| Z\right]\right]\\
&=E\left[\frac{dN(t|A=a)}{\Pr(A=a|Z)}\right],
\end{align*}
where the final line follows from the definition of $dN(t|A=a)=dN(t)I(A=a)$ and the law of total expectation. Similar arguments for the denominator $\Pr\left(X^{A=a}\geq t\right)$, will give us 
\begin{align*}
\Pr\left(X^{A=a}\geq t\right)&=E_Z\left[\frac{\Pr\left(X\geq t, A=a|Z\right)}{\Pr(A=a|Z)}\right]\\
&= E_Z\left[E\left[\frac{I\left(X\geq t, A=a\right)}{\Pr(A=a|Z)}\Bigg| Z\right]\right]\\
&=E\left[\frac{Y(t|A=a)}{\Pr(A=a|Z)}\right]
\end{align*}
where the final line follows from the definition of $Y(t|A=a)=I(X\geq t)I(A=a)$ and $\Pr(X\geq t)=E(Y(t))$. and the law of total expectation. By defining $w(\bZ)=1/Pr(A=a|\bZ)$, in total we get 
\begin{equation*}
\lambda_{IPTW}(t|A=a)=\frac{E\left[w(\bZ)dN(t|A=a)\right]}{E\left[w(\bZ)Y(t|A=a)\right]}.
\end{equation*}

\newpage
\section{Formulas for the function $\varphi(\Lambda^{A=1}(t),\Lambda^{A=0}(t),\theta)$}
\label{SM:formulas_for_varphi}
In this section we provide closed-form formulas for $\varphi(\Lambda^{A=1}(t),\Lambda^{A=0}(t),\theta)$ under a select number of distributions for $V$. We use here the results from Section \ref{SM:prop1} that $HR^{C}(t)=\frac{\psi^{A=1}(t)}{\psi^{A=0}(t)}$ and that $\exp[-\Lambda^{A=a}(t)]=\Phi_V[\Psi^{A=a}(t)]$. Throughout the section, the distribution of $V$ always parameterized such that $V$ has mean 1 and variance $\theta$. Table \ref{SM:Tab:dists} summarizes our results while the subsections the follows provide the technical details. 
\begin{table}[h]\label{app:table_all_frailties}
	\caption{The Laplace transom $\Phi_V(u)$, $\psi^{A=a}(t)$, the causal hazard ratio ($HR^C(t)$) for the distributions: Gamma, inverse Gaussian (IG) and positive stable (PS)}
	\label{SM:Tab:dists}
	\centering
	\begin{tabular}{lllll}
		\hline
		Gamma  &  &  &  &  \\
		~~		$\Phi_V(u)=\frac{\frac{1}{\theta}^{\frac{1}{\theta}}}{(u+\frac{1}{\theta})^{\frac{1}{\theta}}}$ &  &  &  &  \\
		~~		$\psi^{A=a}(t)=\Lambda^{A=a}(t) \exp\big[\theta \Lambda^{A=a}(t)\big]$   &  &  &  &  \\
		~~		$HR^C(t)=\frac{\lambda^{A=1}(t)}{\lambda^{A=0}(t)} \exp \big\{\theta \left[\Lambda^{A=1}(t)-\Lambda^{A=0}(t)\right]\big\}$   &  &  &  &  \\
		\hline
		IG     &  &  &  &  \\
		~~$\Phi_V(u)=\exp \left(\frac{1}{\theta}\left(1-\sqrt{1+2\theta u }\right)\right)$&  &  &  &  \\
		~~$\psi^{A=a}(t)=\Lambda^{A=a}(t) \{1+\theta \Lambda^{A=a}(t)\}$   &  &  &  &  \\
		~~$HR^C(t)=\frac{\lambda^{A=1}(t)[1+\theta\Lambda^{A=1}(t)]}{\lambda^{A=0}(t)[1+\theta\Lambda^{A=0}(t)]}$   &  &  &  &  \\
		\hline
		PS &  &  &  &  \\
		~~$\Phi_V(u)=\exp\left(-u^\theta\right)$ &  &  &  &  \\
		~~$\psi^{A=a}(t)=\Lambda^{A=a}(t) \frac{1}{\theta}\{\lambda^{A=a}(t)^{1/\theta-1}\}$   &  &  &  &  \\
		~~$HR^C(t)=\frac{\Lambda^{A=1}(t)}{\Lambda^{A=0}(t)}\left(\frac{\Lambda^{A=1}(t)}{ \Lambda^{A=0}(t)}\right)^{1/\theta-1}$   &  &  &  &  \\
		\hline
	\end{tabular}
\end{table}
\subsection{Gamma-frailty}\label{app:proof_gamma_frailty}
For $V\sim Gamma(\theta^{-1},\theta^{-1})$, the Laplace transform is $\Phi_V(u)=\frac{\frac{1}{\theta}^{\frac{1}{\theta}}}{\big(u+\frac{1}{\theta}\big)^{\frac{1}{\theta}}}$. Following Section \ref{SM:prop1} we can write
\begin{equation*}
\exp[-\Lambda^{A=a}(t)]=\Phi_V[\Psi^{A=a}(t)]=\frac{\frac{1}{\theta}^{\frac{1}{\theta}}}{\big[\Psi^{A=a}(t)+\frac{1}{\theta}\big]^{\frac{1}{\theta}}}
\end{equation*}
and that $\Psi^{A=a}(t)=\frac{1-\exp\big[-\theta \Lambda^{A=a}(t)\big]}{\exp\big[-\theta \Lambda^{A=a}(t)\big]}$.
By taking derivative with respect to $t$ we get
\begin{equation*}
\psi^{A=a}(t)=\lambda^{A=a}(t) \exp\big[\theta\Lambda^{A=a}(t)\big]
\end{equation*}
and the causal hazard ratio is 
\begin{equation*}
HR^{C}(t)=\frac{\psi^{A=1}(t)}{\psi^{A=0}(t)}=\frac{\lambda^{A=1}(t) }{\lambda^{A=0}(t) } \exp \big\{\theta \left[\Lambda^{A=1}(t) -\Lambda^{A=0}(t) \right]\big\}.
\end{equation*}

\subsection{Inverse Gaussian (IG)-frailty}
For $V\sim IG(1,\theta^{-1})$, the Laplace transform is
$\Phi_V(u)=exp \left[\frac{1}{\theta}\left(1-\sqrt{1+2\theta u }\right)\right]$. Following Section \ref{SM:prop1} we can write
\begin{equation*}
exp[-\Lambda^{A=a}(t)]=\Phi_V[\Psi^{A=a}(t)]=exp \bigg\{\frac{1}{\theta}\left[1-\sqrt{1+2\theta \Psi^{A=a}(t) }\right]\bigg\}
\end{equation*}
and that $\Psi^{A=a}(t)=\frac{[1+\theta\Lambda^{A=a}(t)]^2-1}{2\theta}$. By taking derivative with respect to $t$ we get
\begin{equation*}
\psi^{A=a}(t)=\Lambda^{A=a}(t)\big[1+\theta\Lambda^{A=a}(t)\big]
\end{equation*}
and the causal hazard ratio is 
\begin{equation*}
HR^C(t)=\frac{\psi^{A=1}(t)}{\psi^{A=0}(t)}=\frac{\lambda^{A=1}(t)[1+\theta\lambda^{A=1}(t)]}{\lambda^{A=0}(t)[1+\theta\lambda^{A=0}(t)]}.
\end{equation*}

\subsection{Positive stable (PS)-frailty}
For $V$ distributed PS the Laplace transform is $\Phi_V(u)=\exp \left(-u^\theta\right)$. Following Section \ref{SM:prop1} we can write
\begin{equation*}
\exp[-\Lambda^{A=a}(t)]=\Phi_Z[\Psi^{A=a}(t)]=\exp \left[-\Psi^{A=a}(t)^\theta\right]\\
\end{equation*}
and that $\Psi^{A=a}(t)=\Lambda^{A=a}(t)^{(1/\theta)}$.
By taking derivative with respect to $t$ we get
\begin{equation*}
\psi^{A=a}(t)=\frac{1}{\theta}\lambda^{A=a}(t)\Lambda^{A=a}(t)^{\frac{1-\theta}{\theta}}
\end{equation*}
and the causal hazard ratio is 
\begin{equation*}
HR^C(t)=\frac{\psi^{A=1}(t)}{\psi^{A=0}(t)}=\frac{\lambda^{A=1}(t)}{\lambda^{A=0}(t)}\left[\frac{\Lambda^{A=1}(t)}{ \Lambda^{A=0}(t)}\right]^{1/\theta-1}
\end{equation*}

%

\newpage
\section{Details on the kernel function}\label{app:details}
In this section we present explicit form of the boundary kernel function, our proposed MSE estimation for the weighted scenario and some details on the initial parameters of the kernel estimation procedure.

Throughout the simulation studies and the analysis of the real data examples, we used the Epanechnikov boundary kernel \citep{muller1994hazard}. Let the finite interval $\left[beg,end\right]$ to be the support of the possible event times. Then, the kernel function $K_t(u)$ takes the form 
\begin{equation*}
K_t(u) =K_t(u;q)=
\begin{cases}
K_t\left(u;\frac{t-beg}{b(t)}\right) & beg<t<beg+b(t)\\
K_t(u;1) & beg+b(t)\leq t\leq end-b(t)\\
K_t\left(-u;\frac{end-t}{b(t)}\right) & end-b(t)<t<end
\end{cases},
\end{equation*}
where the explicit form of the kernel function $K_t(u;q)$ is
\begin{equation*}\label{eq:Epanechnikov_kernel}
K_t(u;q)=\frac{12}{(1+q)^{4}}(u+1)\left[u(1-2q)+(3q^2-2q+1)/2\right].
\end{equation*}

In the presence of confounders, we propose to estimate local mean squared error (MSE)
\begin{equation*}
\widehat{MSE}[t,b(t)]=\widehat{var}_{IPTW}[t,b(t)]+\widehat{bias}^2[t,b(t)],
\end{equation*}
where a modified estimator of the variance estimate defined as    
\begin{equation*}
\widehat{var}_{IPTW}\left[t,b(t)\right]=\frac{1}{b_a(t)\sum_{i=1}^{m_a}\hat{w}_i\left(z\right)}\int_{}^{}K^2_t\left(y\right) \frac{\hat{\lambda}^{IPTW}(t-b_a(t)y|A=a)}{\widetilde{L}_m^a\left(t-b_a(t)y\right)}dy
\end{equation*}
where $\widetilde{L}_m^a(t)=1-\frac{1}{\sum_{i=1}^{m_a}\hat{w}_i\left(z\right)+1}\sum_{i=1}^{m_a}\hat{w}_i\left(z\right)I\left(T^{(i)}_a\leq t ,\delta_i=1\right)$ is the weighted empirical survival function of the uncensored observations. The formulas for $\widehat{bias}$ is the same as the one in case of randomization and is given in \cite{muller1994hazard}. 

As indicated in previous work \citep{muller1994hazard,hess1999hazard}, in order to use the proposed algorithm the user need to provide several initial parameters. Most of the initial parameters we used is the recommended values by \cite{hess1999hazard} which are the default values in \texttt{muhaz} \textbf{R} package. The rest of them (number of points in the estimation grid and the minimal and maximal times of the estimation grid) were chosen in order to reduce the estimator's variance and are detailed in Section (6).

\newpage
\section{Simulations}\label{SM:simulations}
\subsection{Technical details about the simulation scenarios}
\subsubsection{Scenario (Ia) }\label{SM:sceanrio_Ia}
We show in this section that hazard model in Scenario (Ia) $V\exp\{\beta a+\theta \exp[\beta a]t\}$ results in the Cox model (4) for the marginal hazard $\lambda^{A=a}(t)$ with a time-varying $HR^C(t)$. 

Let $V\sim Gamma(\frac{1}{\theta},\frac{1}{\theta})$ with the corresponds (on the frailty) hazard rate of $\lambda^{A=a}(t|V)=V\exp\{\beta a+\theta \exp[\beta a]t\}$. It follows that $\phi^{A=a}(t)=\exp\{\beta a+\theta \exp[\beta a]t\}$ and $\Phi^{A=a}(t)=1/\theta \{\exp[\theta\exp(\beta a)t]-1\}$.
Following Section \ref{SM:prop1} and \ref{app:proof_gamma_frailty} we can write 
\begin{align*}
\Lambda^{A=a}(t)&=\frac{1}{\theta}log\left[\theta\left(\Phi^{A=a}(t)+\frac{1}{\theta}\right)\right]\\
&=\frac{1}{\theta}log\left[\theta\Phi^{A=a}(t)+1\right]\\
&=\frac{1}{\theta}log(\{\exp[\theta\exp(\beta a)t]-1\}+1)\\
&=t\exp(\beta a)\\
\end{align*}
By taking derivative with respect to $t$, we get $\lambda^{A=a}(t)=\exp(\beta a)$ which is a Cox model. Using \eqref{eq:HRC_frac} will give us $HR^C(t)=\frac{\exp\{\beta +\theta \exp[\beta ]t\}}{\exp\{\theta t\}}=\exp\{\beta+\theta t [\exp(\beta)-1)]\}$, a time varying $HR^C(t)$.

\subsubsection{Scenario (Ib)}
\label{SM:sceanrio_Ib}
In this section we show that the hazard model in Scenario (Ib) results in constant $HR^C(t)$ which equals to $0.5$. Under scenario (Ib), we have $\lambda^{A=a}(t|V)=V\exp[1.5t+\log(0.5) a]$. Therefore, $\phi^{A=a}=\exp[1.5t+\log(0.5) a]$. From equation (1) we have
\begin{equation}
HR^C(t)=\frac{\psi^{A=1}(t)}{\psi^{A=0}(t)}=\frac{\exp[1.5t+\log(0.5) a]}{\exp(1.5t)}=0.5
\end{equation}

\subsubsection{Scenario (II) }\label{SM:sceanrio_II}
In this section we show that the hazard model in Scenario (II) results in constant $HR^C(t)$  which equals to $0.5$.
Under scenario (Ib),  we have $\lambda^{A=a}(t|V,Z)=V\gamma \exp(\log(0.5) a+\beta_z Z)$. From equation (7e) we have 
\begin{equation*}
HR^C(t)=\frac{\int_{0}^{\infty }\int_{}^{}\lambda^{A=1}(t|v,\bz)f(v,\bz|T^{A=1}\geq t, T^{A=0}\geq t)d\bz dv}{\int_{0}^{ }\int_{}^{ }\lambda^{A=0}(t|v,\bz)f(v,\bz|T^{A=1}\geq t, T^{A=0}\geq t)d\bz dv}
\end{equation*}
Then, by substituting $\lambda^{A=a}(t|V,Z)=V\gamma \exp(\log(0.5) a+\beta_z Z)$ into (7e) we have 
\begin{align}
&=\frac{\int_{0}^{\infty }\int_{}^{}v\gamma \exp(\log(0.5) a+\beta_z z)f(v,\bz|T^{A=1}\geq t, T^{A=0}\geq t)dz dv}{\int_{0}^{ }\int_{}^{ }v\gamma \exp(\log(0.5) a+\beta_z z)f(v,\bz|T^{A=1}\geq t, T^{A=0}\geq t)dz dv}\\
&=\frac{\gamma\exp(\log(0.5))\int_{0}^{\infty}\int_{}^{}v\exp(\beta_z z)f(v,z|T^{A=1}\geq t, T^{A=0}\geq t)dzdv}{\gamma\int_{0}^{\infty}\int_{}^{}v\exp(\beta_z z)f(v,z|T^{A=1}\geq t, T^{A=0}\geq t)dzdv}\\
&=0.5
\end{align}

\newpage
\subsection{Additional simulation results }\label{SM:additional_results}
\begin{figure}
	\centering
	\includegraphics[width=12cm, height=8cm]{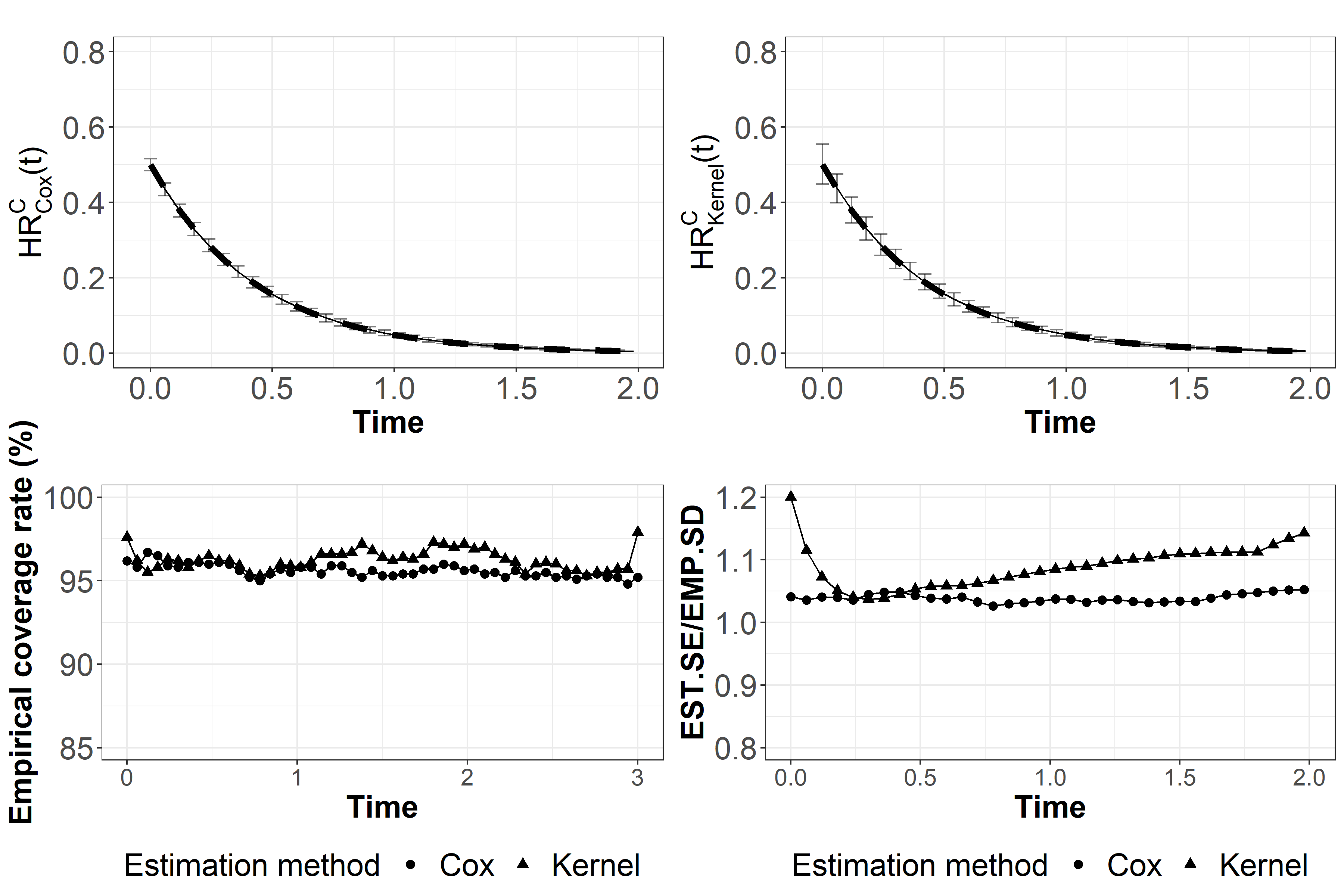}
	\caption{Performance of the Cox-based and kernel-based estimators in Scenario (Ia) under sample size of $n=5000$, 20\% censoring rate, and a Gamma frailty distribution with Kendall's $\tau=0.7$. Results are presented only until $t=2$. The upper row presents the mean estimated $HR^C(t)$ across the simulations, plus/minus one empirical standard deviation. The dashed line represents the true $HR^C(t)$. The bottom row presents the empirical coverage rate (left bottom corner) the ratio between the mean estimated standard error (EST.SE) and empirical standard deviation of the estimates (EST.SD) (right bottom corner). $\widehat{HR}_{Cox}^C(t)$: Cox-based estimator $\widehat{HR}_{kernel}^C(t)$: Kernel-based estimator.}
	\label{fig:IA_tau07_n5000}
	
\end{figure}

\begin{figure}	
	\centering
	\begin{subfigure}[t]{1\textwidth}
		\centering
		\includegraphics[width=12cm, height=8cm]{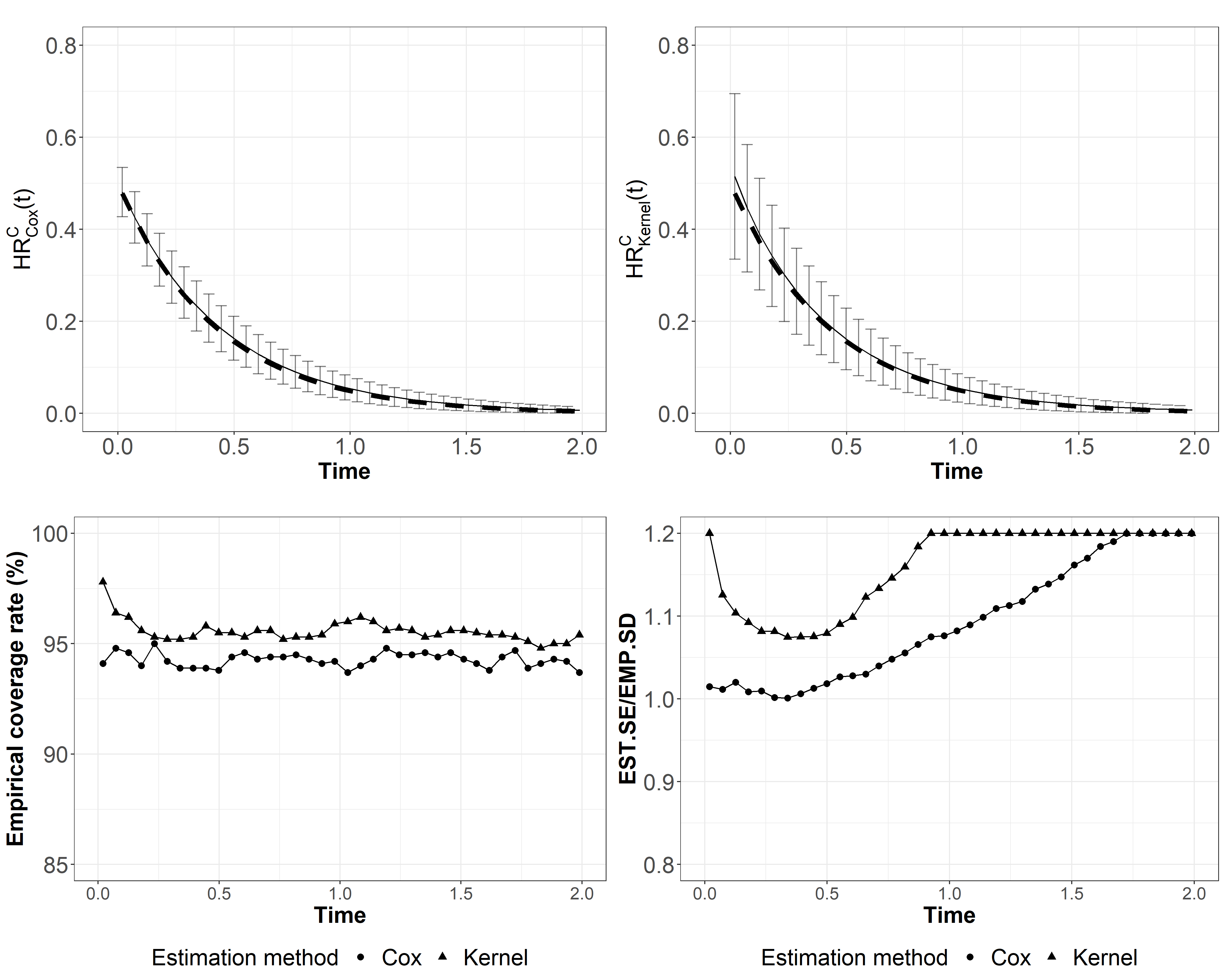}
		\caption{Sample size $n=500$}
	\end{subfigure}
	\vfill
	\begin{subfigure}[t]{1\textwidth}
		\centering
		\includegraphics[width=12cm, height=8cm]{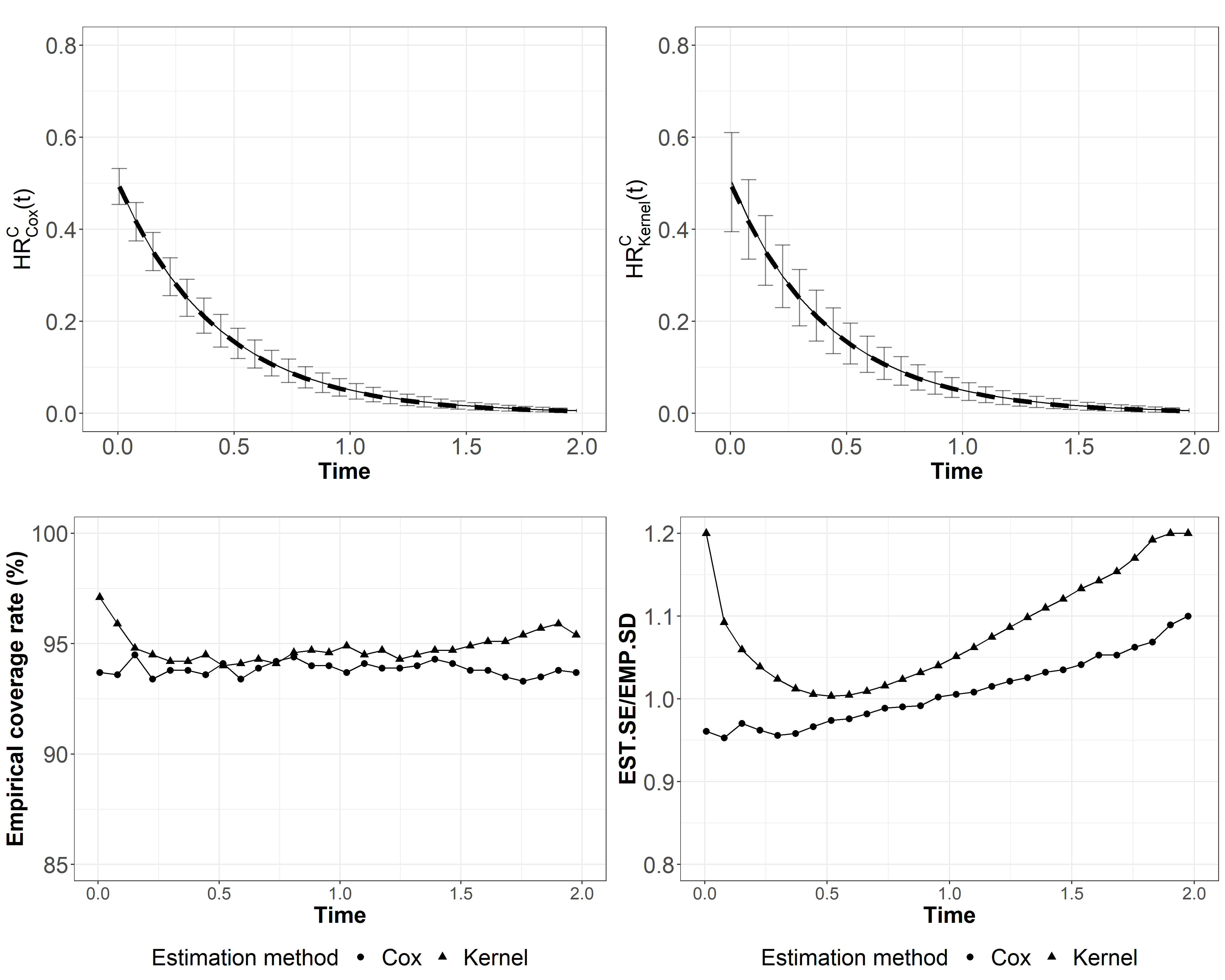}
		\caption{Sample size $n=1000$}
	\end{subfigure}
	\centering
\caption{Performance of the Cox-based and kernel-based estimators in Scenario (Ia) under sample sizes of $n=500, 1000$, 20\% censoring rate, and a Gamma frailty distribution with Kendall's $\tau=0.7$. Results are presented only until $t=2$. The upper row presents presents the mean estimated $HR^C(t)$ across the simulations, plus/minus one empirical standard deviation. The dashed line represents the true $HR^C(t)$. The bottom row presents the empirical coverage rate (left bottom corner) the ratio between the mean estimated standard error (EST.SE) and empirical standard deviation of the estimates (EST.SD) (right bottom corner). $\widehat{HR}_{Cox}^C(t)$: Cox-based estimator $\widehat{HR}_{kernel}^C(t)$: Kernel-based estimator.}
\label{fig:IA_07_smaller_sample_sizes}
\end{figure}

\begin{figure}
	\begin{subfigure}[t]{1\textwidth}
		\centering
		\includegraphics[width=12cm, height=8cm]{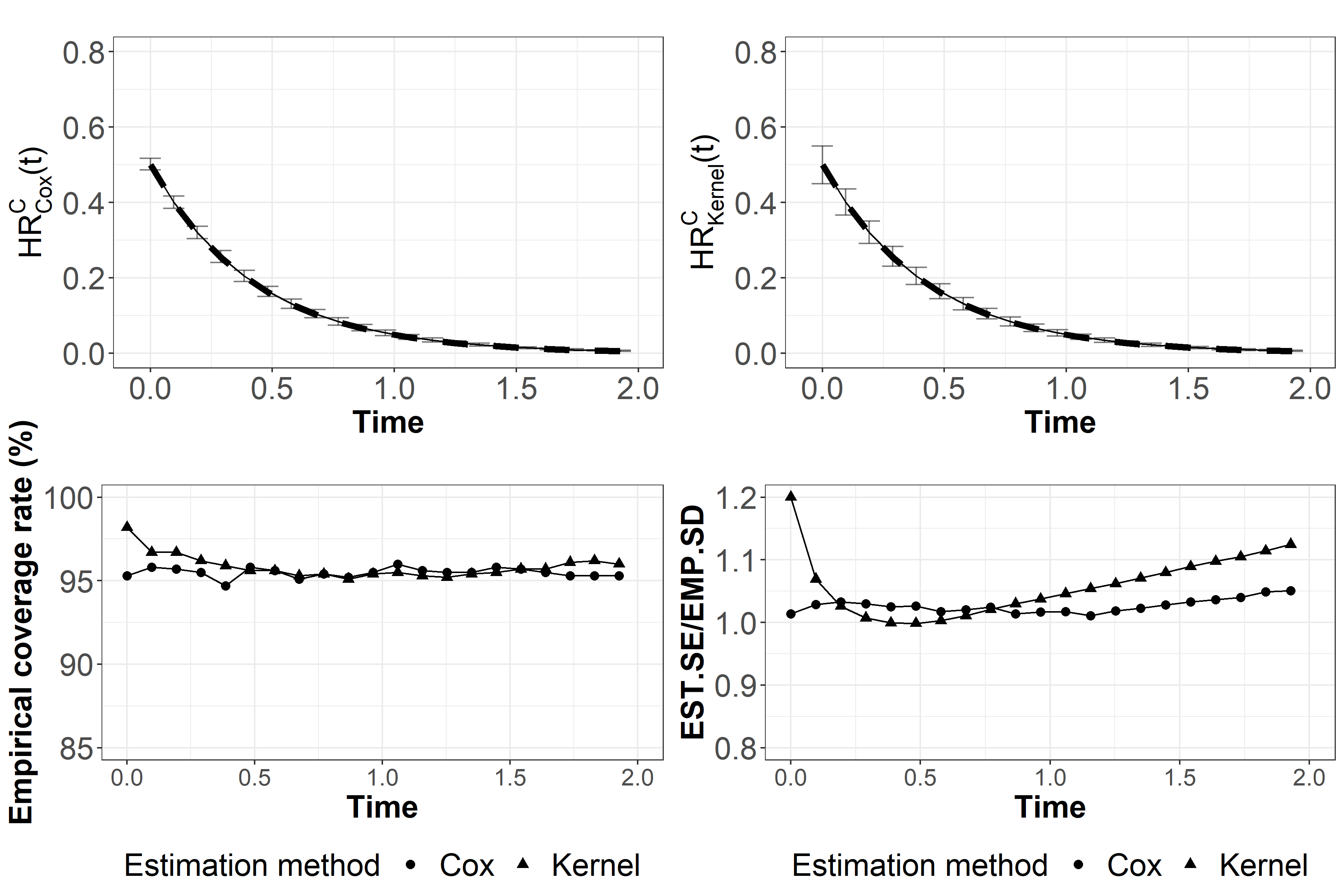}
		\caption{$CR=20\%$}
	\end{subfigure}
	\vfill
	\begin{subfigure}[t]{1\textwidth}
		\centering
		\includegraphics[width=12cm, height=8cm]{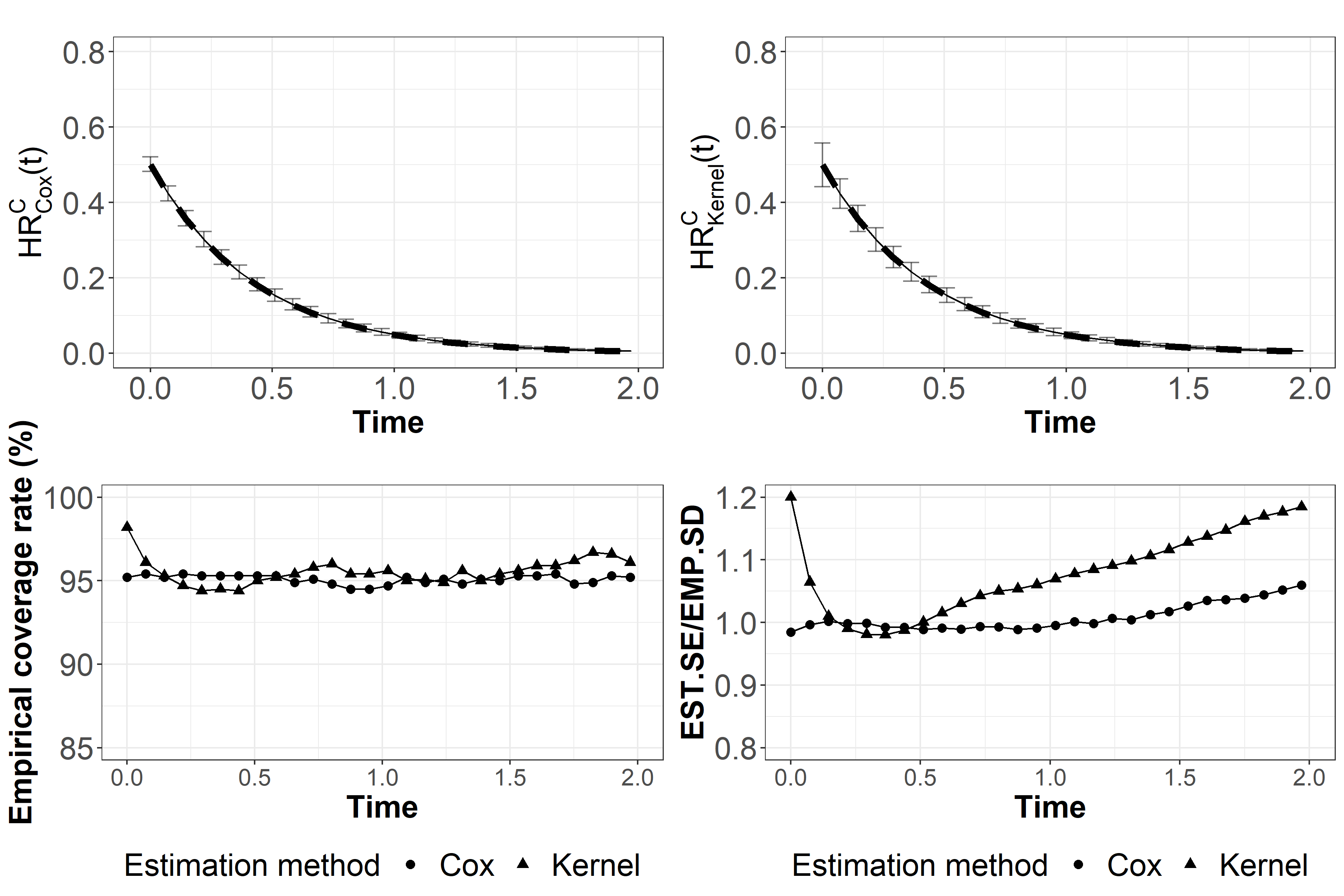}
		\caption{$CR=40\%$}
		\label{fig:IA_tau07_n500}
	\end{subfigure}
	\centering
	\caption{Performance of the Cox-based and kernel-based estimators in Scenario (Ia) under sample size of $n=5000$, censoring rates of 20\% and 40\%, and a Gamma frailty distribution with Kendall's $\tau=0.7$. Results are presented only until $t=2$. The upper row presents presents the mean estimated $HR^C(t)$ across the simulations, plus/minus one empirical standard deviation. The dashed line represents the true $HR^C(t)$. The bottom row presents the empirical coverage rate (left bottom corner) the ratio between the mean estimated standard error (EST.SE) and empirical standard deviation of the estimates (EST.SD) (right bottom corner). $\widehat{HR}_{Cox}^C(t)$: Cox-based estimator $\widehat{HR}_{kernel}^C(t)$: Kernel-based estimator.}
		\label{fig:IA_07_different_CR}
\end{figure}

\begin{figure}

	\centering
	\begin{subfigure}[t]{1\textwidth}
		\centering
		\includegraphics[width=12cm, height=8cm]{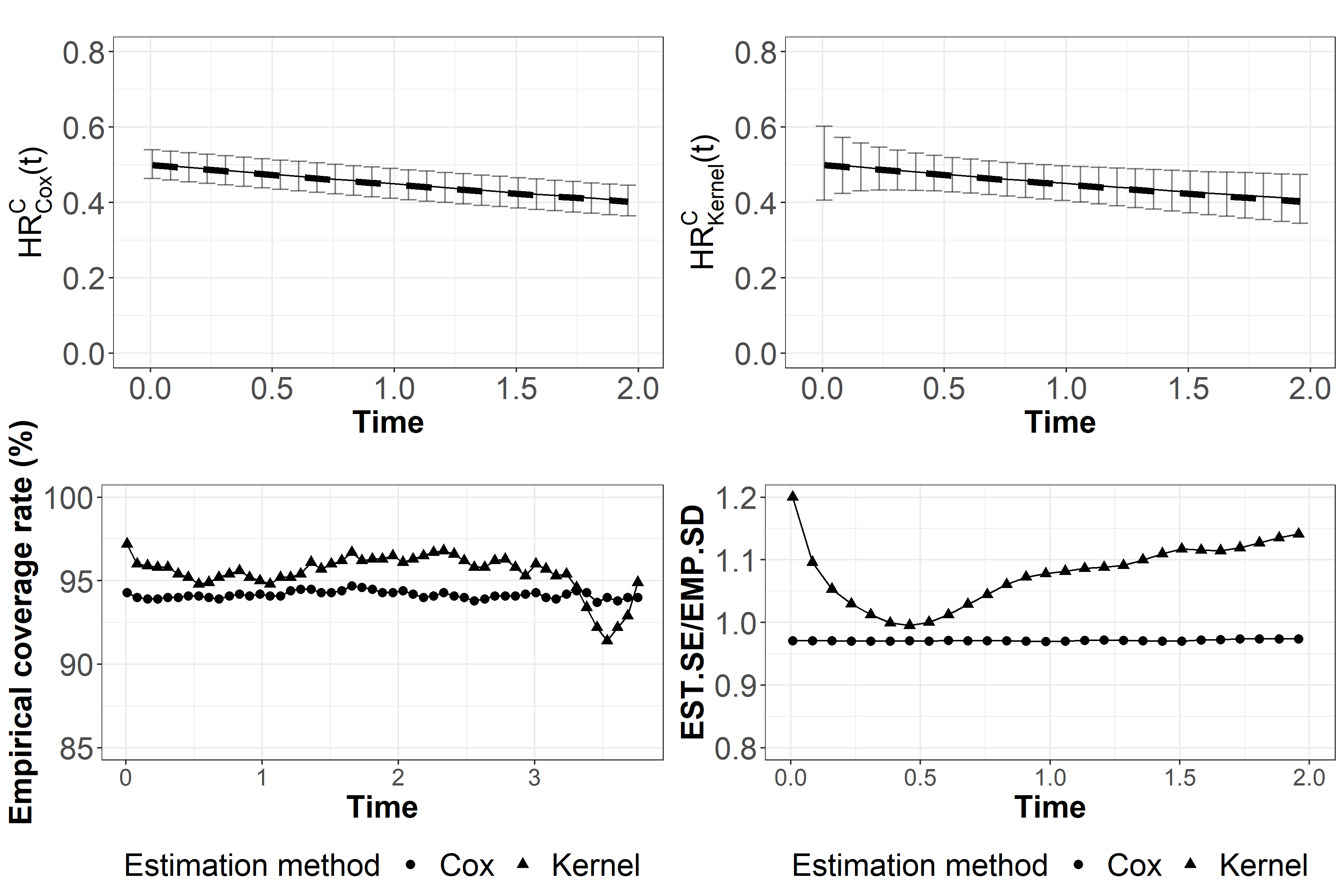}
		\caption{Kendall's $\tau=0.1$}
	\end{subfigure}
	\vfill
	\begin{subfigure}[t]{1\textwidth}
		\centering
		\includegraphics[width=12cm, height=8cm]{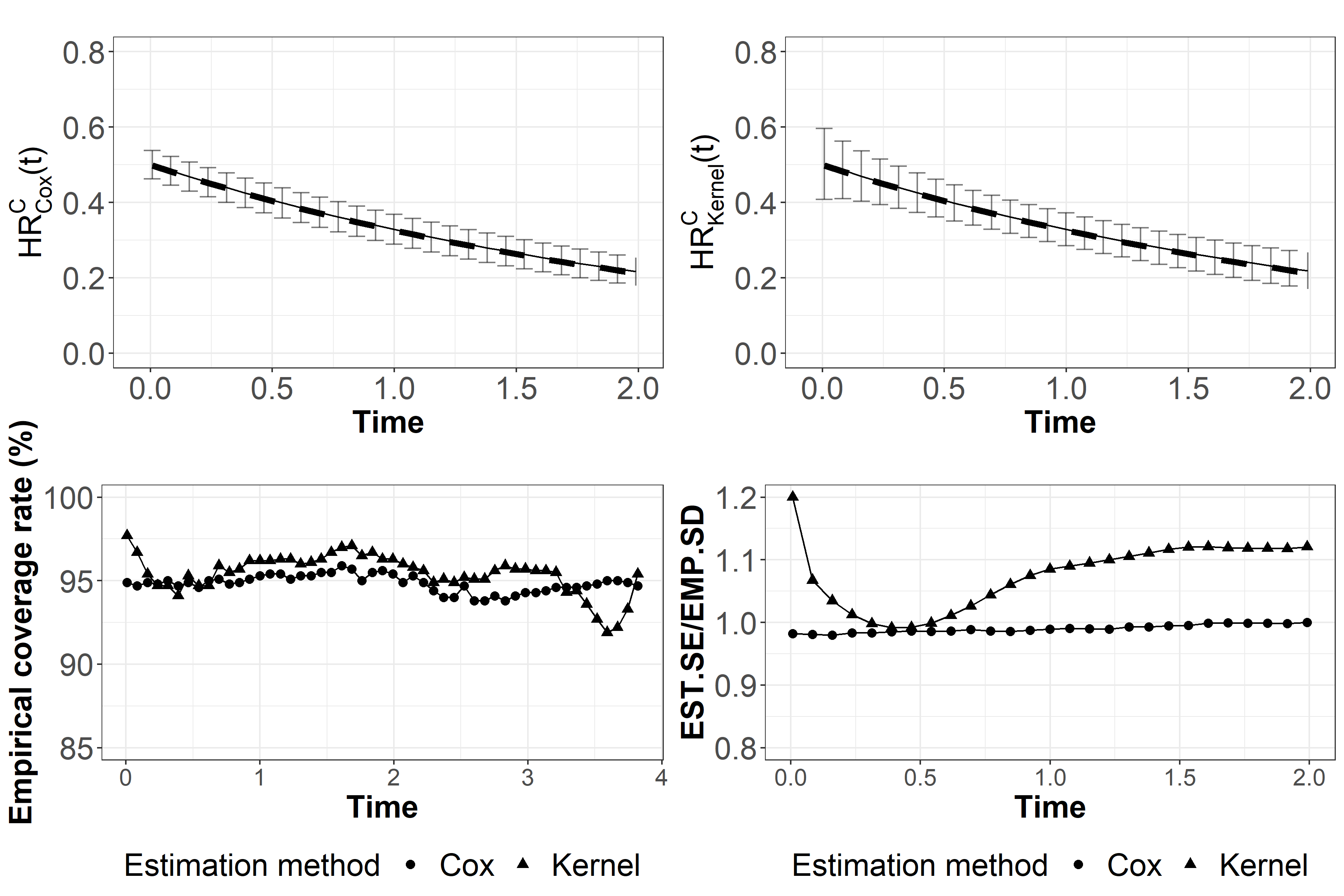}
		\caption{Kendall's $\tau=0.3$}
	\end{subfigure}
	\centering
\caption{
	Performance of the Cox-based and kernel-based estimators in Scenario (Ia) under sample size of $n=5000$, censoring rate of $CR=20\%$, and a Gamma frailty distribution with Kendall's $\tau=0.1, 0.3, 0.5$. Results are presented only until $t=2$. The upper row presents presents the mean estimated $HR^C(t)$ across the simulations, plus/minus one empirical standard deviation. The dashed line represents the true $HR^C(t)$. The bottom row presents the empirical coverage rate (left bottom corner) the ratio between the mean estimated standard error (EST.SE) and empirical standard deviation of the estimates (EST.SD) (right bottom corner). $\widehat{HR}_{Cox}^C(t)$: Cox-based estimator $\widehat{HR}_{kernel}^C(t)$: Kernel-based estimator.}
\label{fig:IA_07_smaller_different_taus}
\end{figure}

\begin{figure}

	\centering
	\begin{subfigure}[t]{1\textwidth}
		\centering
		\includegraphics[width=12cm, height=8cm]{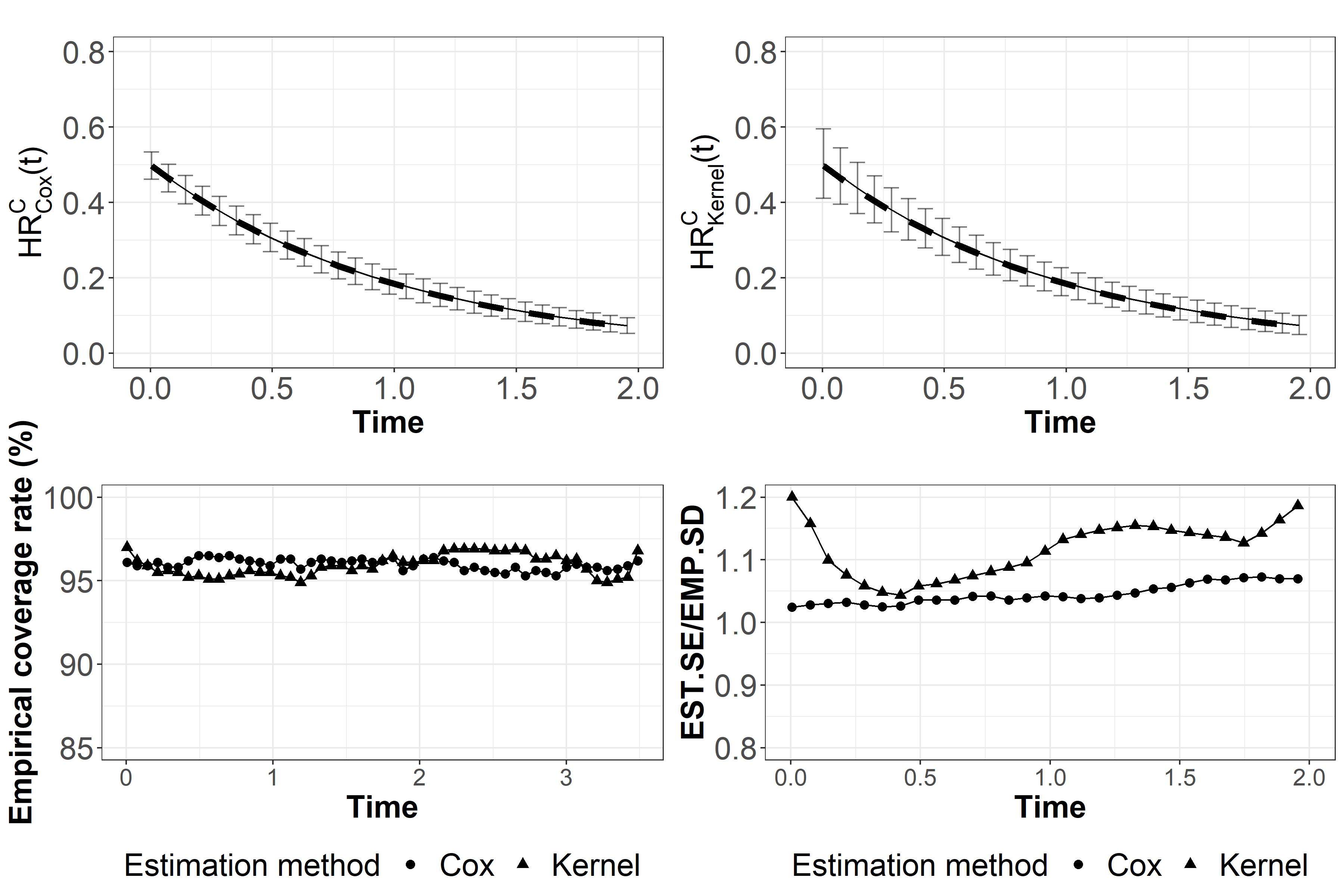}
	\end{subfigure}
	\caption{Performance of the Cox-based and kernel-based estimators in Scenario (Ia) under sample size of $n=5000$, censoring rate of $CR=20\%$, and a Gamma frailty distribution with Kendall's $\tau=0.5$. Results are presented only until $t=2$.  The upper row presents presents the mean estimated $HR^C(t)$ across the simulations, plus/minus one empirical standard deviation. The dashed line represents the true $HR^C(t)$. The bottom row presents the empirical coverage rate (left bottom corner) the ratio between the mean estimated standard error (EST.SE) and empirical standard deviation of the estimates (EST.SD) (right bottom corner). $\widehat{HR}_{Cox}^C(t)$: Cox-based estimator $\widehat{HR}_{kernel}^C(t)$: Kernel-based estimator.}
	\label{fig:IA_07_smaller_different_taus2}
\end{figure}

\begin{figure}

	\centering
	\includegraphics[width=12cm, height=8cm]{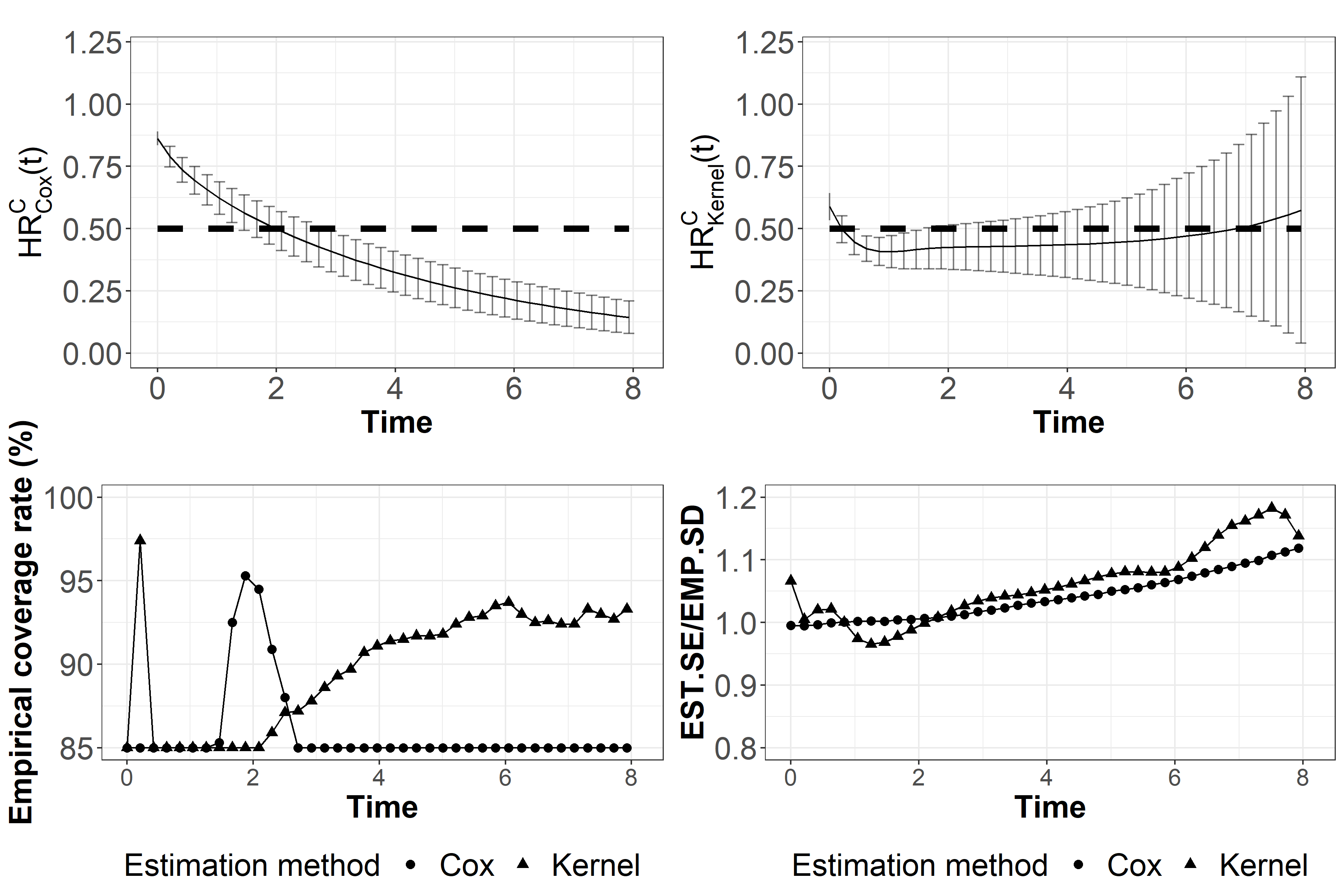}
		\centering
		\caption{Performance of the Cox-based and kernel-based estimators in Scenario (Ib) under sample size of $n=5000$, 20\% censoring rate, and a Gamma frailty distribution with Kendall's $\tau=0.7$. Results are presented only until $t=8$.  The upper row presents presents the mean estimated $HR^C(t)$ across the simulations, plus/minus one empirical standard deviation. The dashed line represents the true $HR^C(t)$. The bottom row presents the empirical coverage rate (left bottom corner) the ratio between the mean estimated standard error (EST.SE) and empirical standard deviation of the estimates (EST.SD) (right bottom corner). $\widehat{HR}_{Cox}^C(t)$: Cox-based estimator $\widehat{HR}_{kernel}^C(t)$: Kernel-based estimator.}
		\label{fig:IB_tau07_n5000}
\end{figure}

\begin{figure}

	\centering
	\begin{subfigure}[t]{1\textwidth}
		\centering
		\includegraphics[width=12cm, height=8cm]{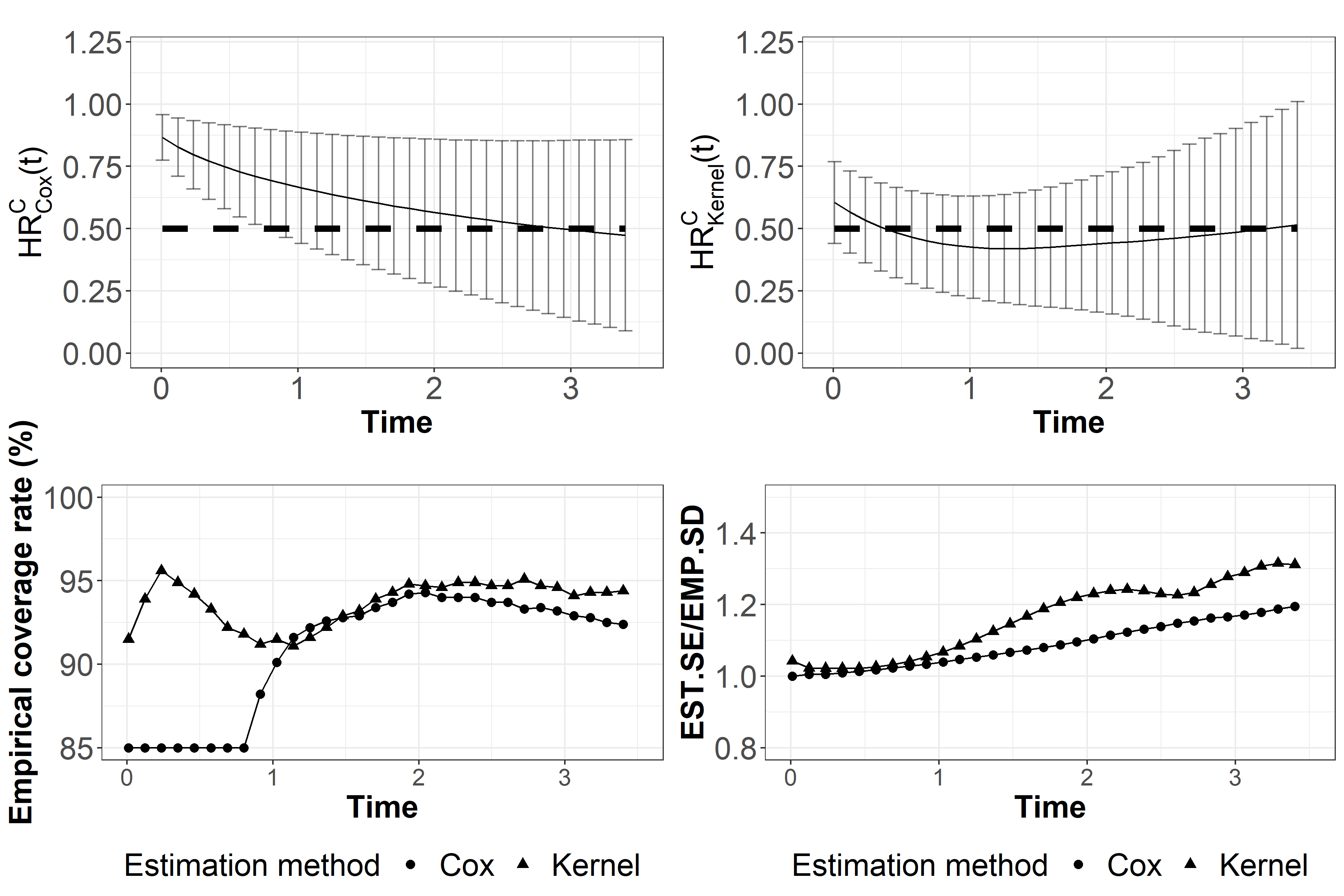}
		\caption{Sample size $n=500$}
	\end{subfigure}
	\vfill
	\begin{subfigure}[t]{1\textwidth}
		\centering
		\includegraphics[width=12cm, height=8cm]{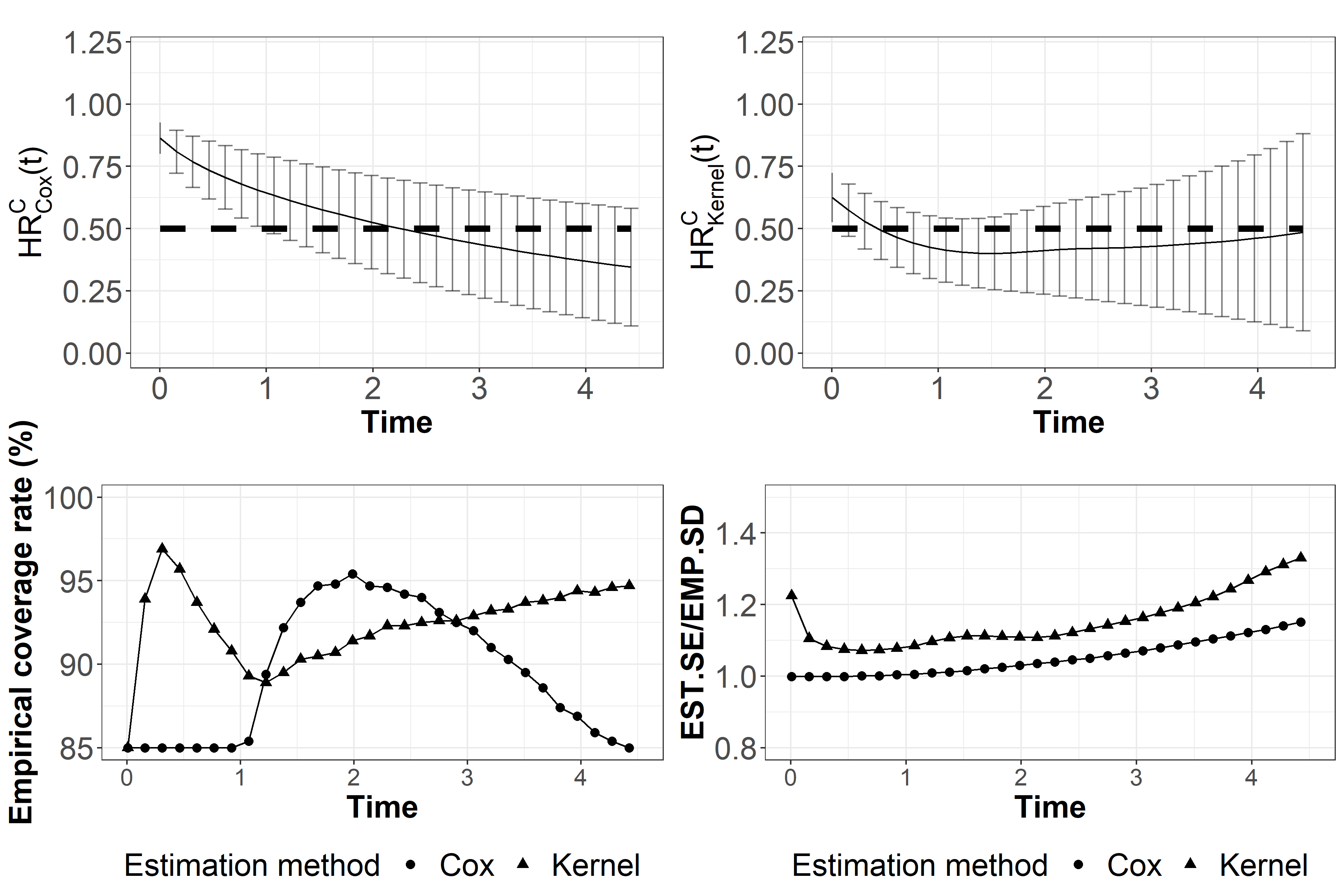}
		\caption{Sample size $n=1000$}
	\end{subfigure}
	\centering
	\caption{Performance of the Cox-based and kernel-based estimators in Scenario (Ib) under sample size of $n=500,1000$, 20\% censoring rate, and a Gamma frailty distribution with Kendall's $\tau=0.7$. Results are presented only until $t=8$. The upper row presents presents the mean estimated $HR^C(t)$ across the simulations, plus/minus one empirical standard deviation. The dashed line represents the true $HR^C(t)$. The bottom row presents the empirical coverage rate (left bottom corner) the ratio between the mean estimated standard error (EST.SE) and empirical standard deviation of the estimates (EST.SD) (right bottom corner). $\widehat{HR}_{Cox}^C(t)$: Cox-based estimator $\widehat{HR}_{kernel}^C(t)$: Kernel-based estimator.}
		\label{fig:IB_07_smaller_sample_sizes}
\end{figure}

\begin{figure}

	\centering
	\begin{subfigure}[t]{1\textwidth}
		\centering
		\includegraphics[width=12cm, height=8cm]{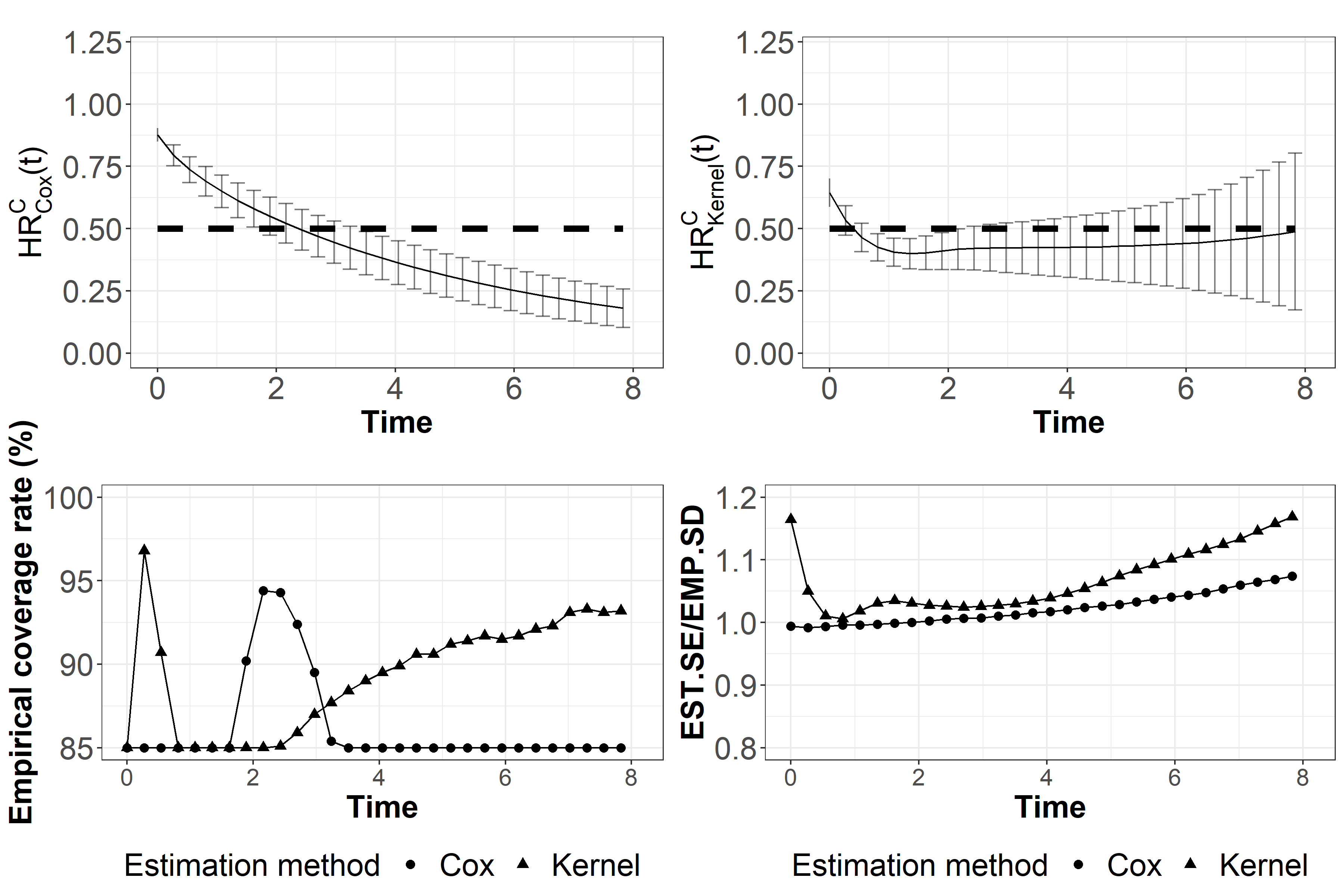}
		\caption{$CR=20\%$}
	\end{subfigure}
	\vfill
	\begin{subfigure}[t]{1\textwidth}
		\centering
		\includegraphics[width=12cm, height=8cm]{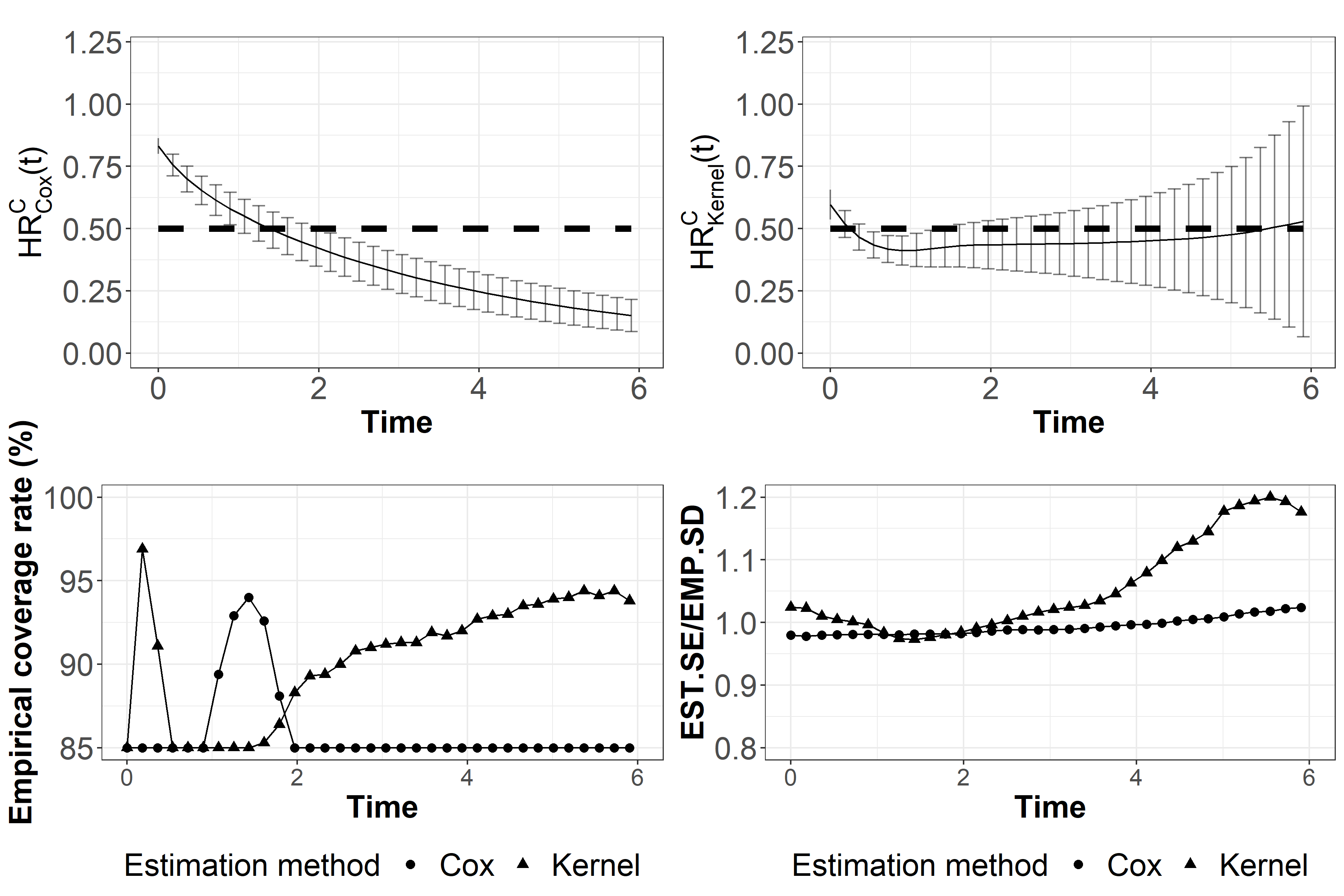}
		\caption{$CR=40\%$}
	\end{subfigure}
	\centering
	\caption{Performance of the Cox-based and kernel-based estimators in Scenario (Ib) under sample size of $n=5000$, censoring rate of $CR=20\%, 40\%$, and a Gamma frailty distribution with Kendall's $\tau=0.7$. Results are presented only until $t=8$. The upper row presents presents the mean estimated $HR^C(t)$ across the simulations, plus/minus one empirical standard deviation. The dashed line represents the true $HR^C(t)$. The bottom row presents the empirical coverage rate (left bottom corner) the ratio between the mean estimated standard error (EST.SE) and empirical standard deviation of the estimates (EST.SD) (right bottom corner). $\widehat{HR}_{Cox}^C(t)$: Cox-based estimator $\widehat{HR}_{kernel}^C(t)$: Kernel-based estimator.}
\label{fig:IB_07_different_CR}
\end{figure}

\begin{figure}

	\centering
	\begin{subfigure}[t]{1\textwidth}
		\centering
		\includegraphics[width=12cm, height=8cm]{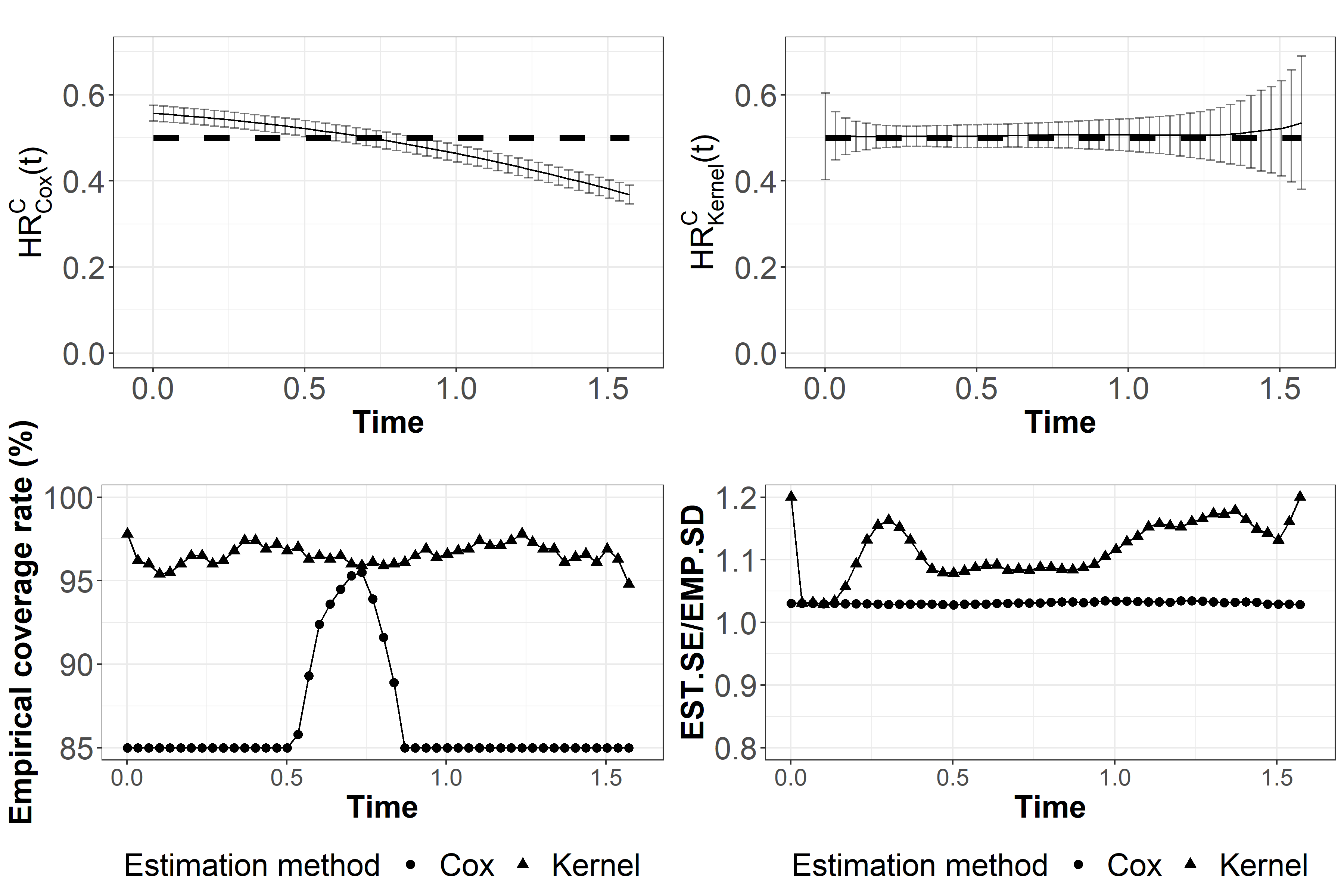}
		\caption{Kendall's $\tau=0.1$}
	\end{subfigure}
	\vfill
	\begin{subfigure}[t]{1\textwidth}
		\centering
		\includegraphics[width=12cm, height=8cm]{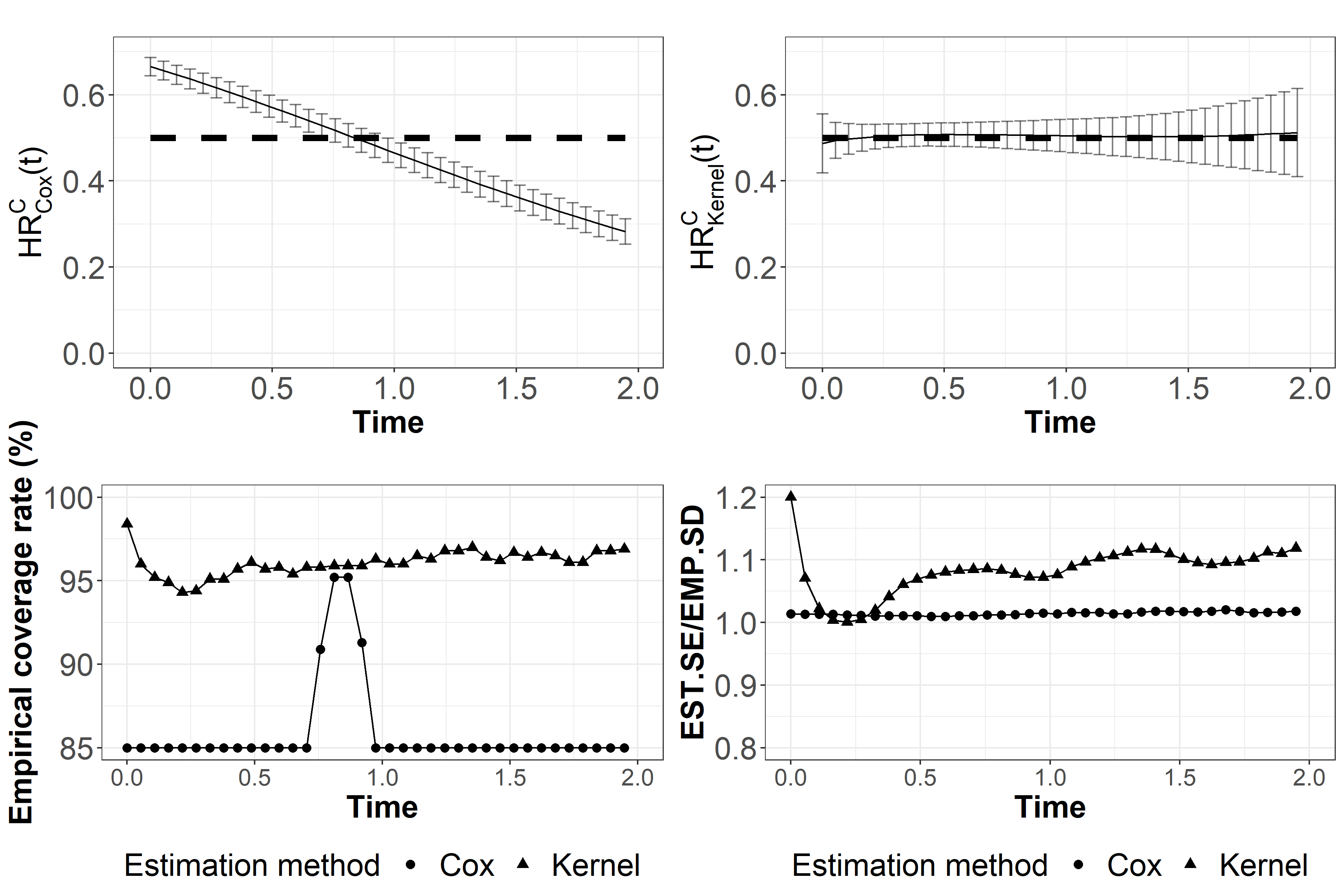}
		\caption{Kendall's $\tau=0.3$}
	\end{subfigure}
	\centering
	\caption{Performance of the Cox-based and kernel-based estimators in Scenario (Ib) under sample size of $n=5000$, censoring rate of $CR=20\%$, and a Gamma frailty distribution with Kendall's $\tau=0.1, 0.3, 0.5$. Results are presented only until $t=8$. The upper row presents presents the mean estimated $HR^C(t)$ across the simulations, plus/minus one empirical standard deviation. The dashed line represents the true $HR^C(t)$. The bottom row presents the empirical coverage rate (left bottom corner) the ratio between the mean estimated standard error (EST.SE) and empirical standard deviation of the estimates (EST.SD) (right bottom corner). $\widehat{HR}_{Cox}^C(t)$: Cox-based estimator $\widehat{HR}_{kernel}^C(t)$: Kernel-based estimator.}
		\label{fig:IB_07_smaller_different_taus}
\end{figure}

\begin{figure}

	\centering
	\begin{subfigure}[t]{1\textwidth}
		\centering
		\includegraphics[width=12cm, height=8cm]{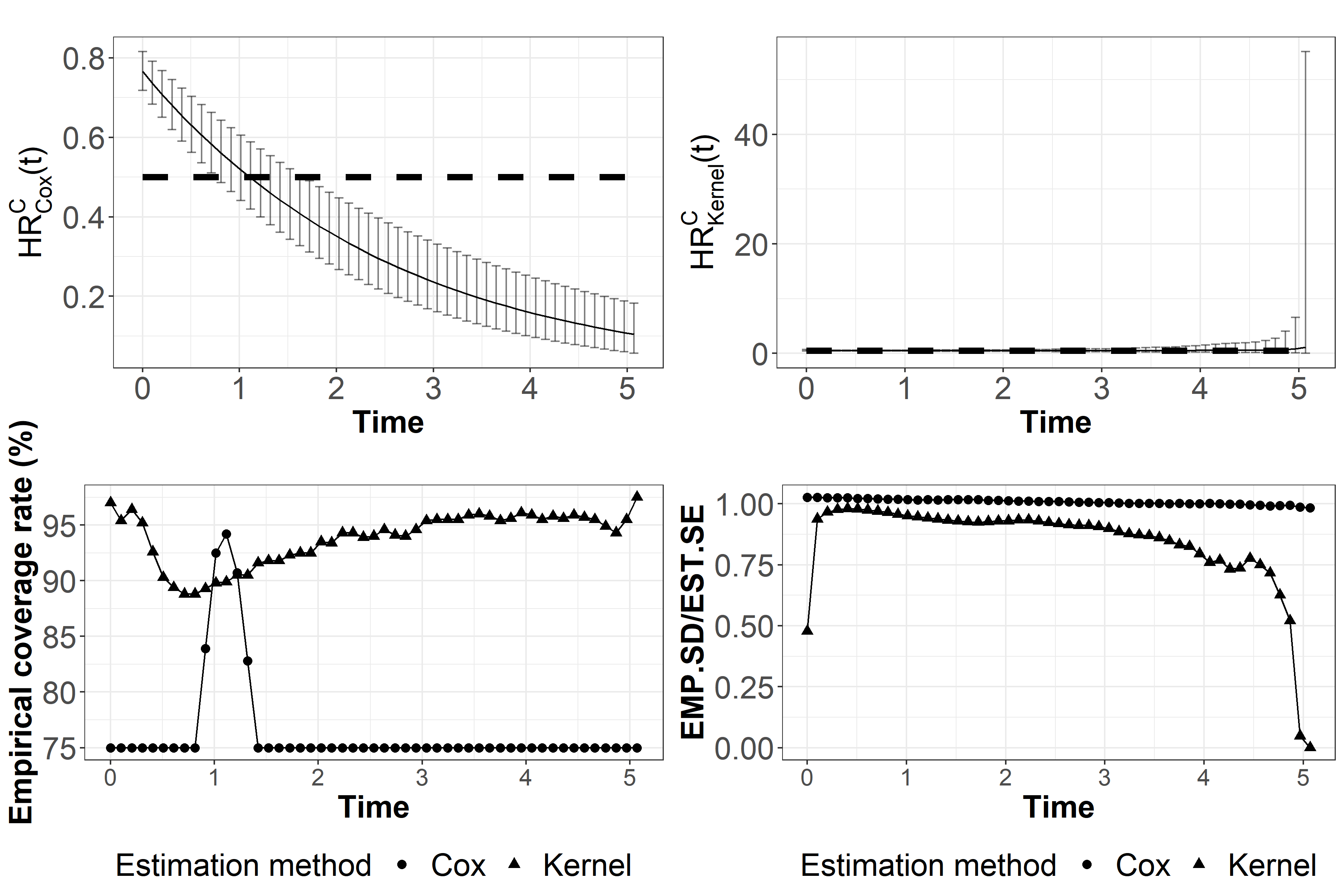}
	\end{subfigure}
	\centering
	\caption{Performance of the Cox-based and kernel-based estimators in Scenario (Ib) under sample size of $n=5000$, censoring rate of $CR=20\%$, and a Gamma frailty distribution with Kendall's $\tau=0.5$. Results are presented only until $t=8$. The upper row presents presents the mean estimated $HR^C(t)$ across the simulations, plus/minus one empirical standard deviation. The dashed line represents the true $HR^C(t)$. The bottom row presents the empirical coverage rate (left bottom corner) the ratio between the mean estimated standard error (EST.SE) and empirical standard deviation of the estimates (EST.SD) (right bottom corner). $\widehat{HR}_{Cox}^C(t)$: Cox-based estimator $\widehat{HR}_{kernel}^C(t)$: Kernel-based estimator.}
	\label{fig:IB_07_smaller_different_taus2}
\end{figure}

\begin{figure}
		
	\includegraphics[scale=0.3]{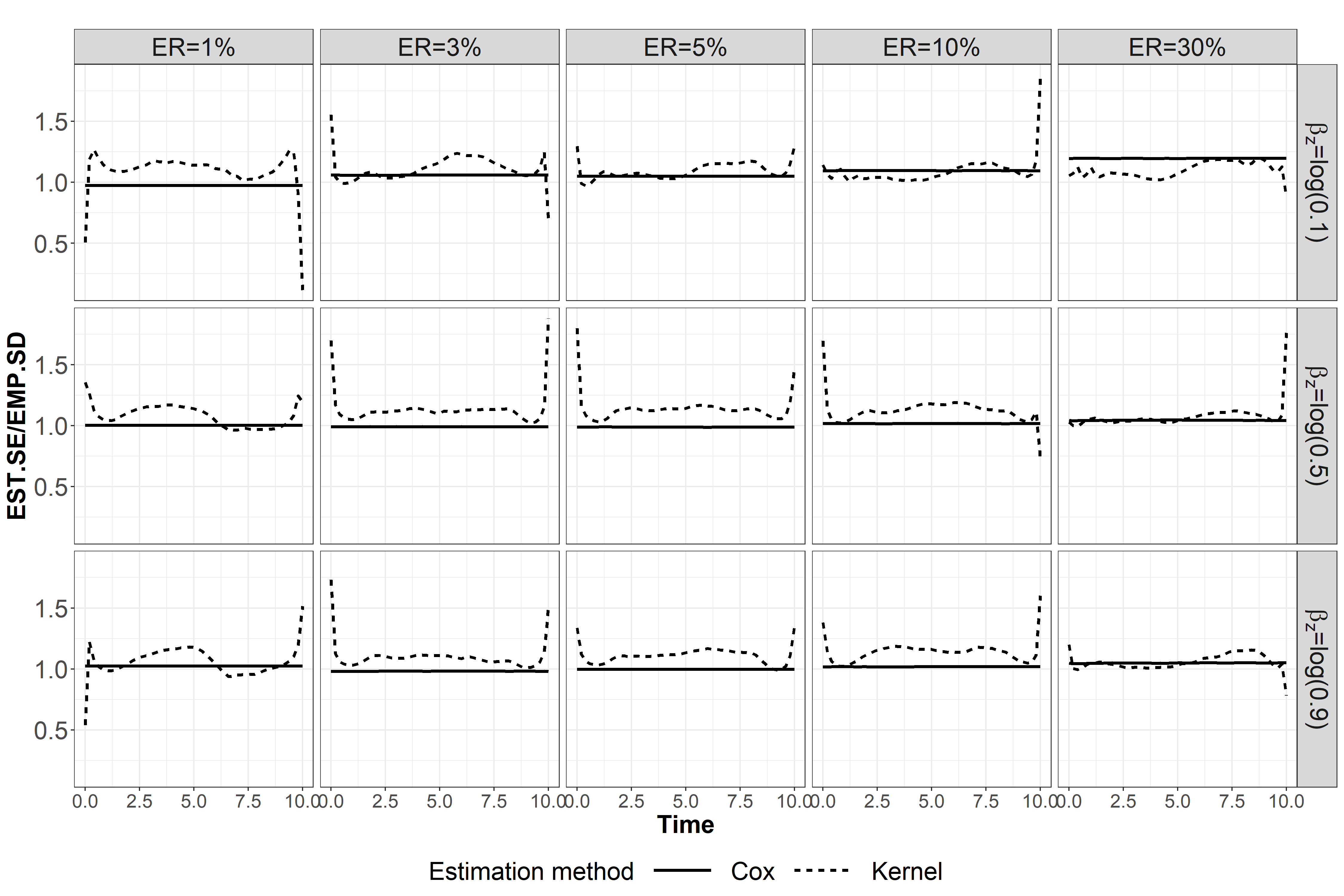}
	\centering
	\caption{The ratio between the EST.SE and EMP.SD ( EST.SE: mean estimated standard error, EMP.SD: empirical standard deviation of the estimates) of the Cox-based and kernel-based estimators in Scenario (II). Results presented under sample size of $n=50000$, Gamma frailty distribution and Kendall's $\tau$ of 0.7 between potential event times. ER: Event rate.}
	\label{fig:sim.results.II.SD}
\end{figure}

\begin{figure}

	\includegraphics[scale=0.3]{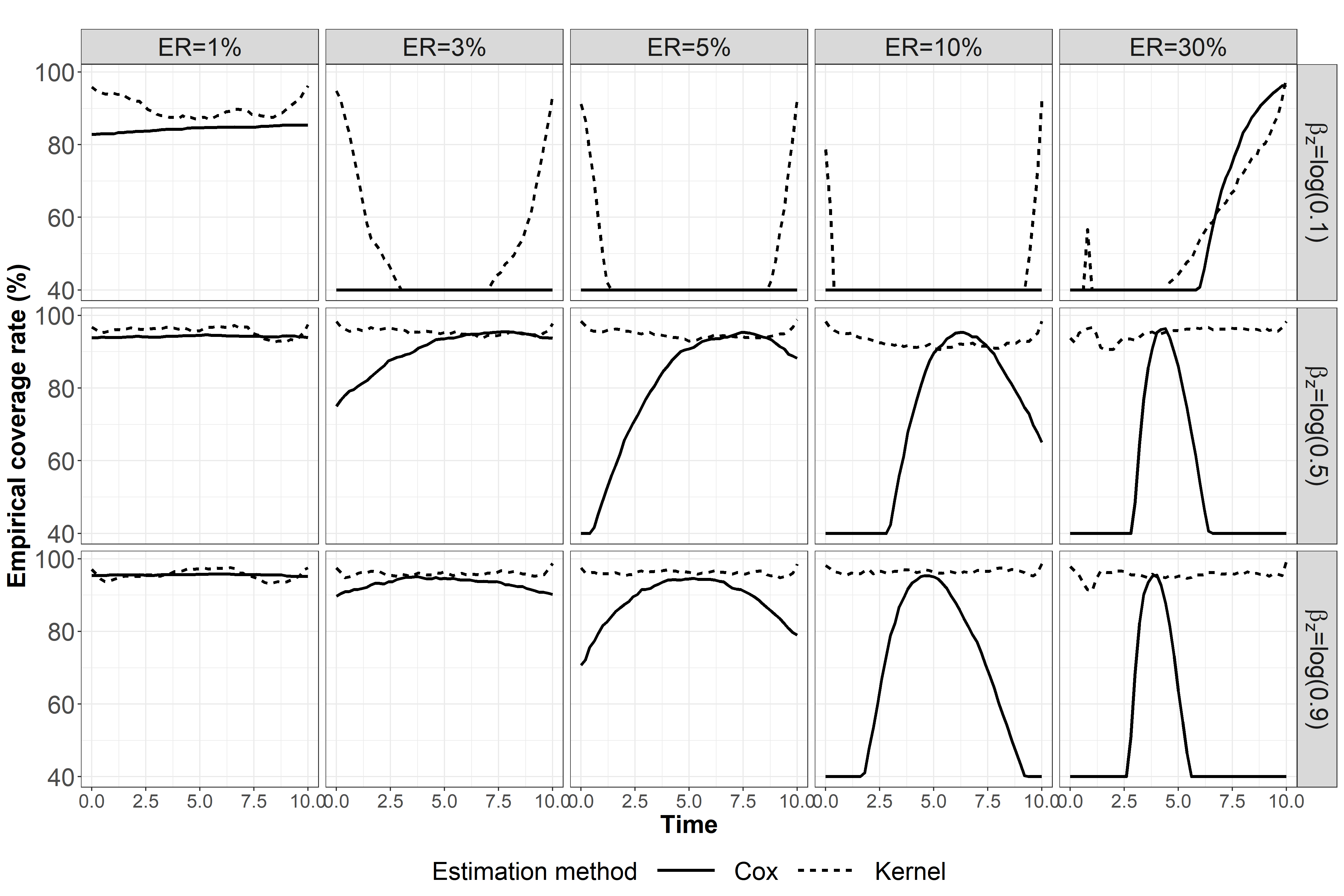}
	\centering
	\caption{The empirical coverage of the Cox-based and kernel-based estimators in Scenario (II). Results presented under sample size of $n=50000$, Gamma frailty distribution and Kendall's $\tau$ of 0.7 between potential event times. ER: Event rate.}
	\label{fig:sim.results.II.EC}	
\end{figure}

\begin{figure}[h]	
	\includegraphics[scale=0.3]{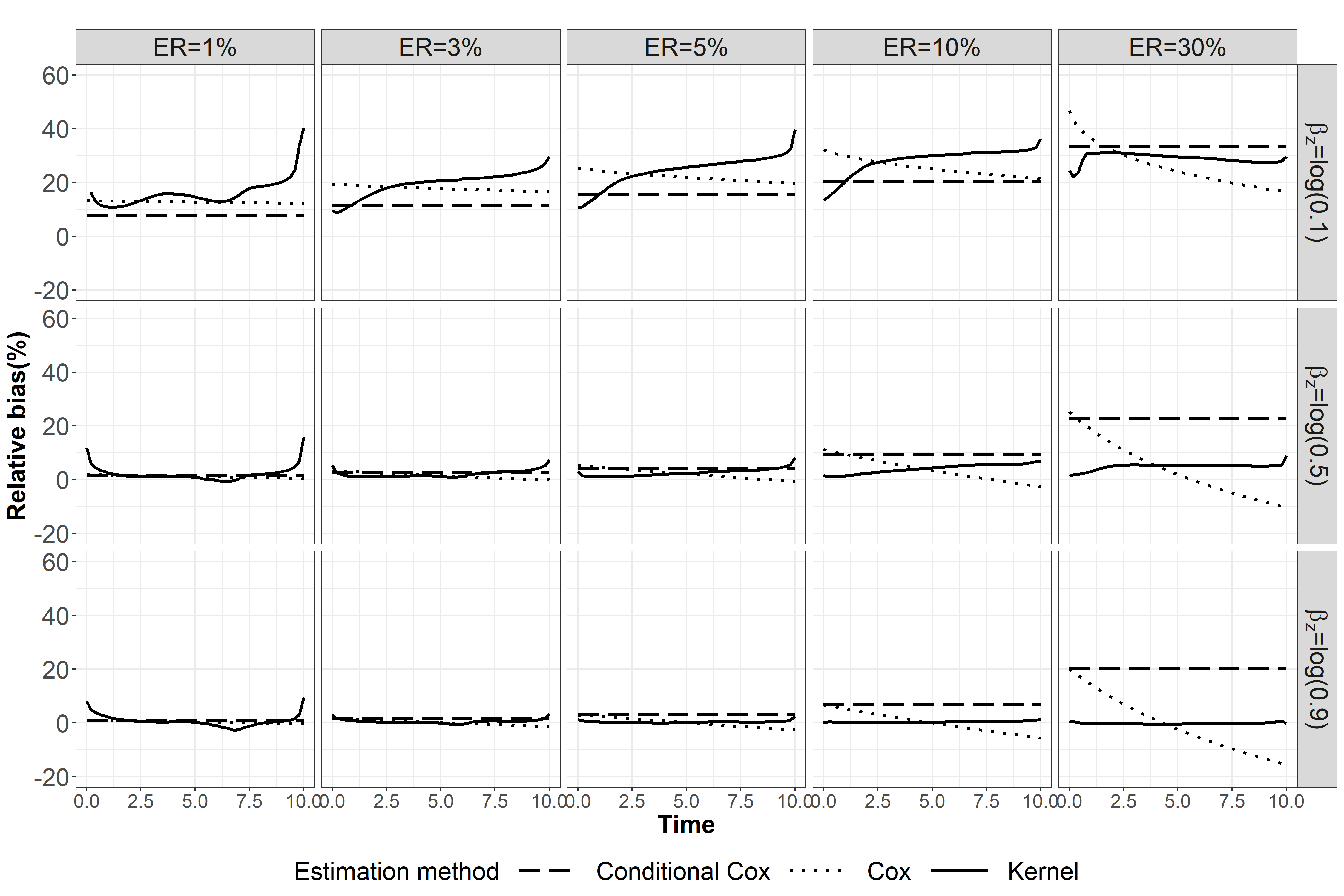}
	\centering
	\caption{Relative bias of Cox-based and kernel-based $HR^C(t)$ estimators in Scenario (II) under sample size of $n=50,000$, Gamma frailty distribution and Kendall's $\tau$ of 0.5 between potential event times. ER: Event rate.}
	\label{fig:sim.results.II.RB.05}
\end{figure}

\begin{figure}
	
	\includegraphics[scale=0.3]{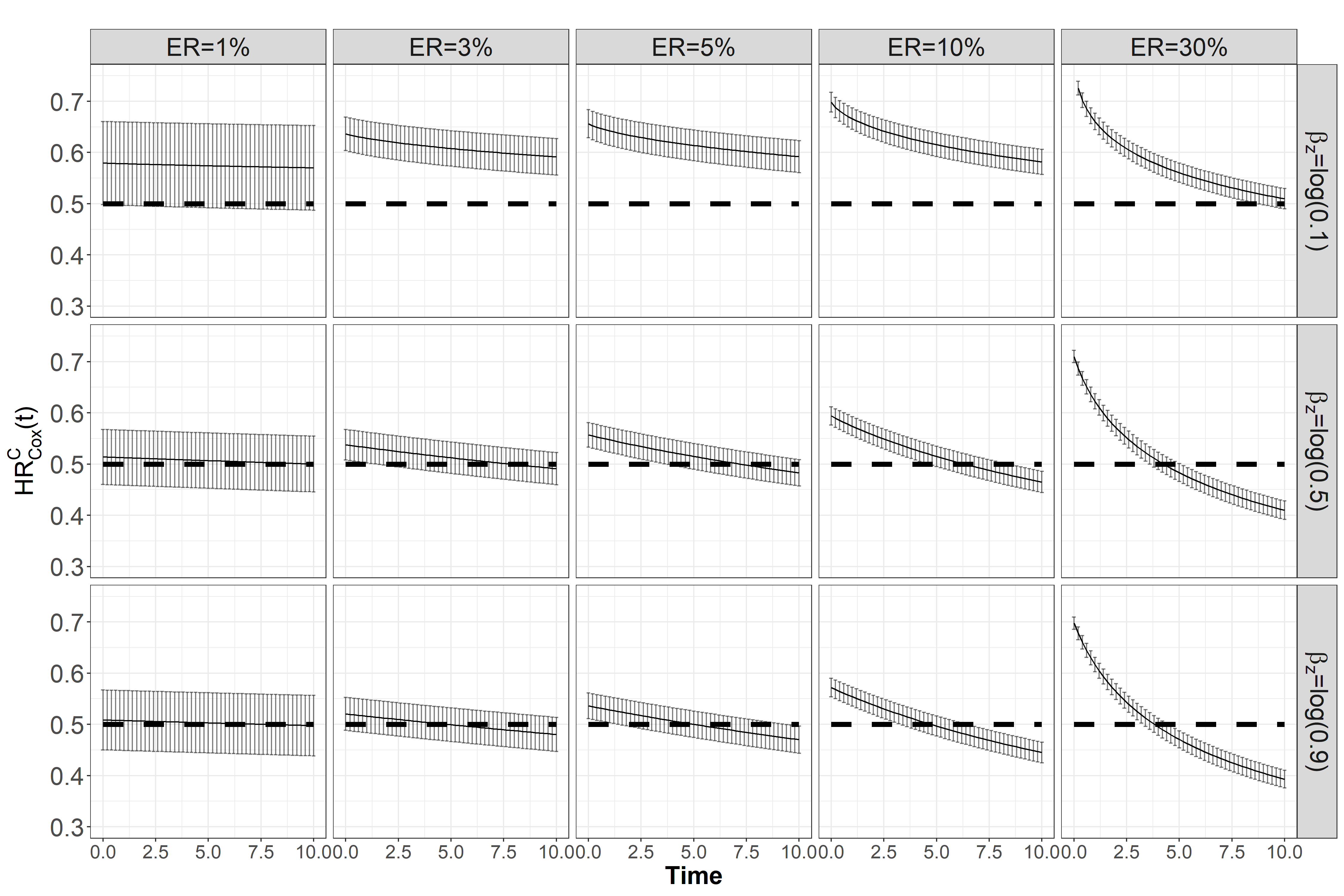}
	\centering
	
	\caption{Mean Cox-based $HR^C(t)$ estimates in Scenario (II) across the simulations, along with  plus/minus one empirical standard deviation. The dashed line represents the true $HR^C(t)$. Results presented under sample size of $n=50000$, Gamma frailty distribution and Kendall's $\tau$ of 0.7 between potential event times. ER: Event rate.}
	\label{fig:sim.results.II.Cox}
\end{figure}

\begin{figure}
	
	\includegraphics[scale=0.3]{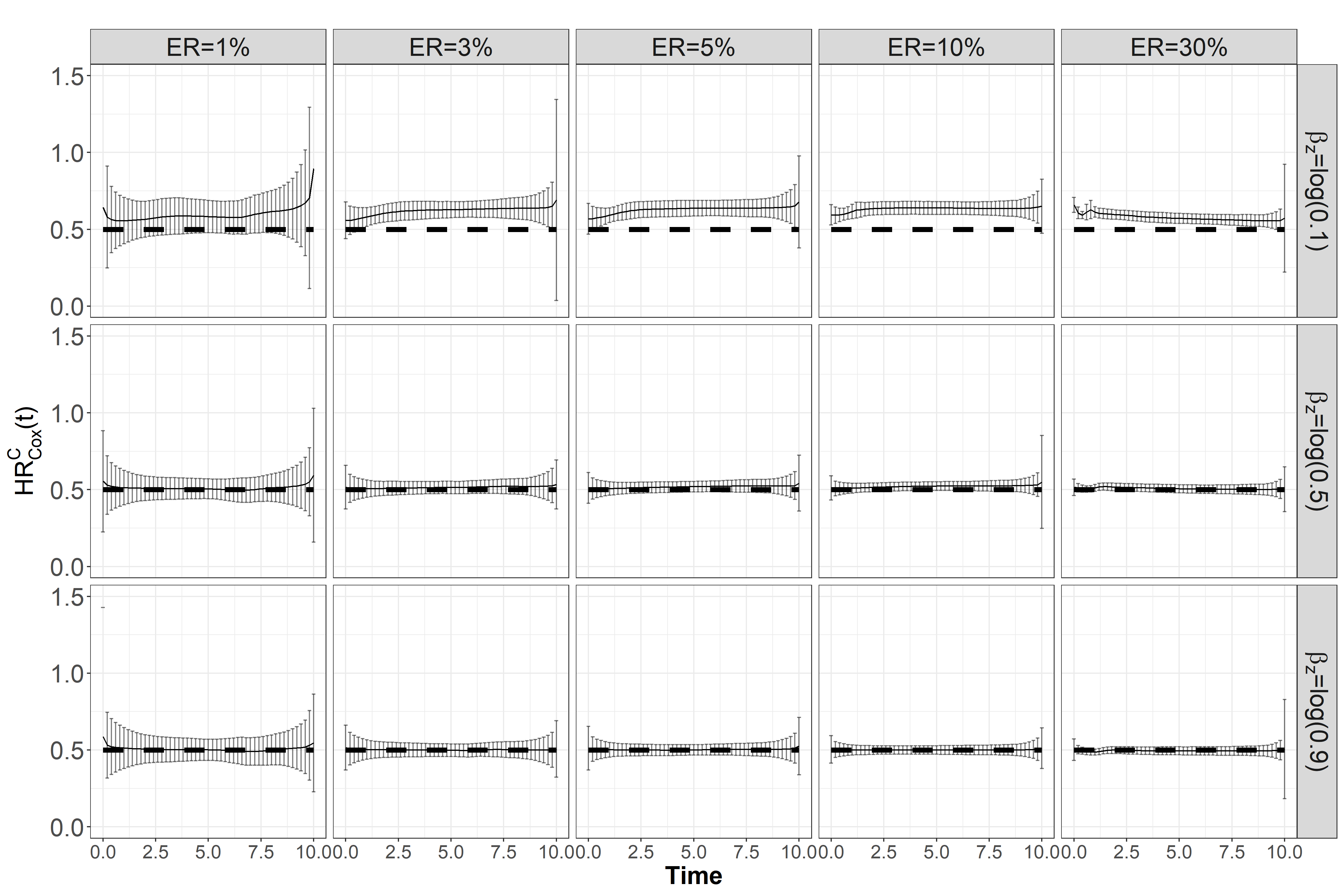}
	\centering
	
	\caption{Mean kernel-based $HR^C(t)$ estimates in Scenario (II) across the simulations, along with plus/minus one empirical standard deviation. The dashed line represents the true $HR^C(t)$. Results presented under sample size of $n=50000$, Gamma frailty distribution and Kendall's $\tau$ of 0.7 between potential event times. ER: Event rate.}
	\label{fig:sim.results.II.Kernel}
\end{figure}

\begin{figure}
	
	\includegraphics[scale=0.3]{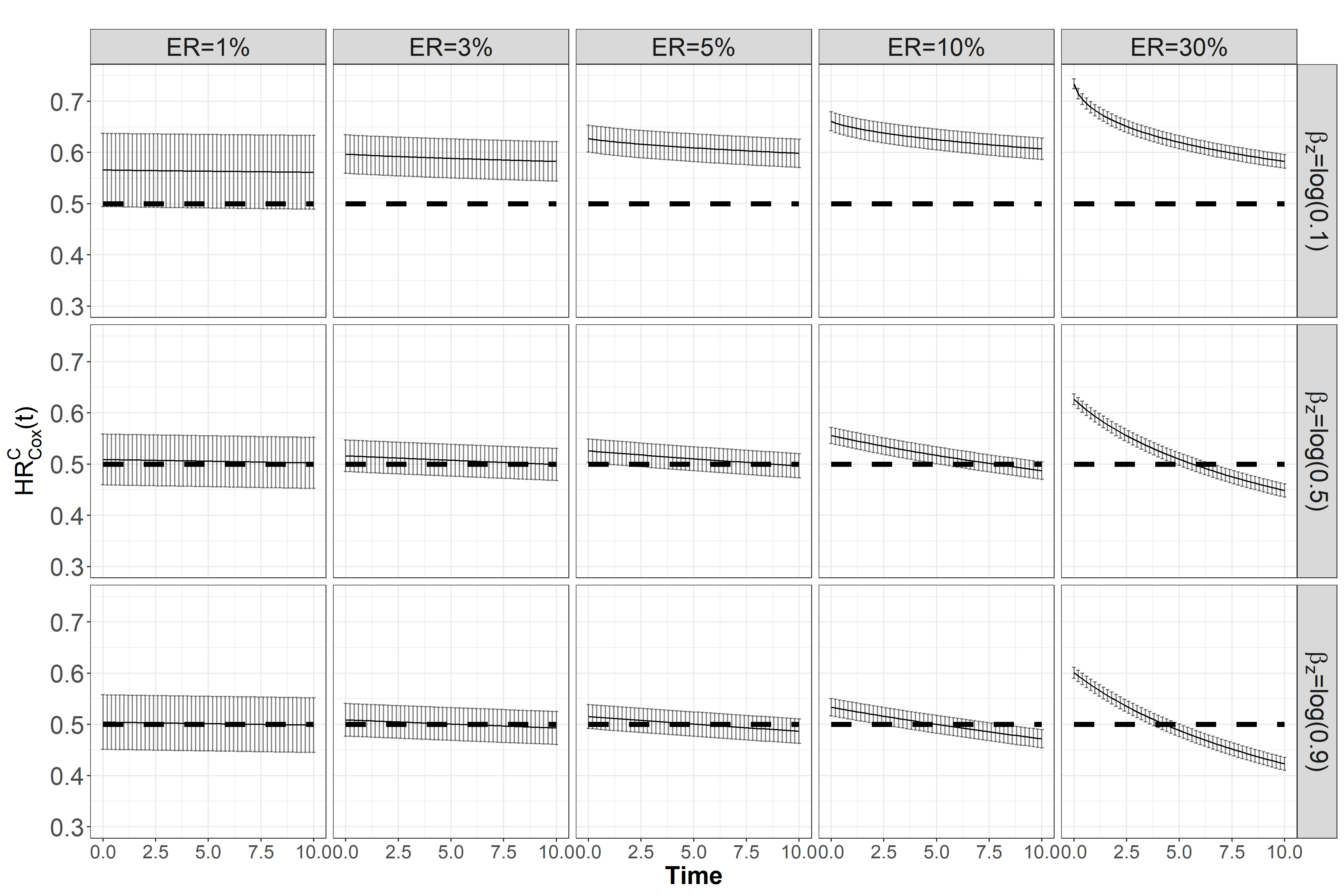}
	\centering
	
	\caption{Mean Cox-based $HR^C(t)$ estimates in Scenario (II) across the simulations, along with plus/minus one empirical standard deviation. The dashed line represents the true $HR^C(t)$. Results presented under sample size of $n=50000$, Gamma frailty distribution and Kendall's $\tau$ of 0.5 between potential event times. ER: Event rate.}
	\label{fig:sim.results.II.Cox.05}
\end{figure}

\begin{figure}

	\includegraphics[scale=0.3]{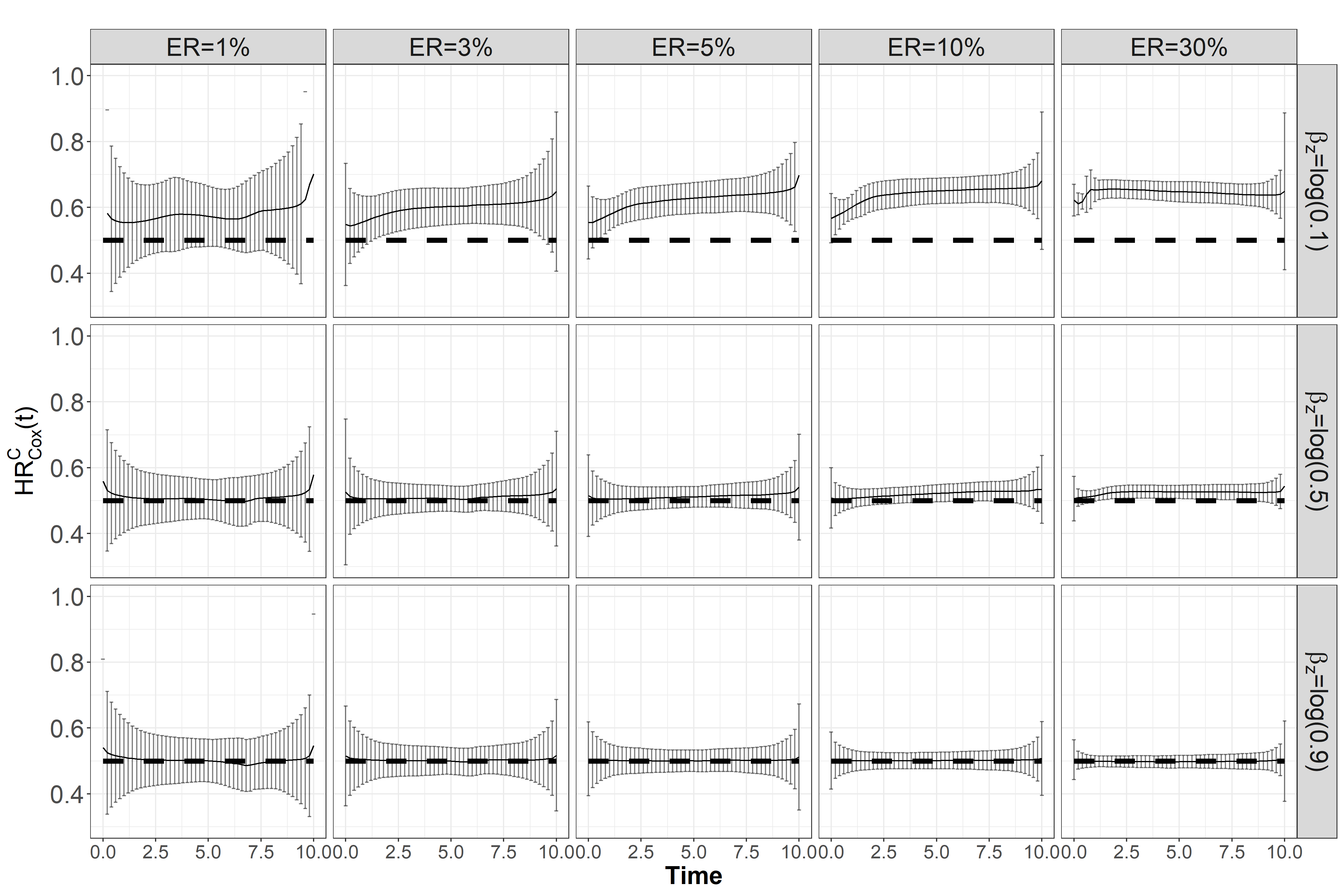}
	\centering

	\caption{Mean kernel-based $HR^C(t)$ estimates in Scenario (II) across the simulations, along with  plus/minus one empirical standard deviation. The dashed line represents the true $HR^C(t)$. Results presented under sample size of $n=50000$, Gamma frailty distribution and Kendall's $\tau$ of 0.5 between potential event times. ER: Event rate.}
		\label{fig:sim.results.II.Kernel.05}
\end{figure}

This section provides complementary simulation results as discussed in Section 6. We present the following
\begin{itemize}
	\item Figures \ref{fig:IA_tau07_n5000} to \ref{fig:IA_07_smaller_different_taus2} presents additional results for the Scenario (Ia). Figure \ref{fig:IA_tau07_n5000} presents more results under sample size of $n=5000$, 20\% censoring rate, and a Gamma frailty distribution with Kendall's $\tau=0.7$. Figure \ref{fig:IA_07_smaller_sample_sizes} presents results under sample sizes $n=500,1000$, 20\% censoring rate, and a Gamma frailty distribution with Kendall's $\tau=0.7$. Figure \ref{fig:IA_07_different_CR} presents results under sample sizes $n=5000$, 20\% and $40\%$ censoring rate, and a Gamma frailty distribution with Kendall's $\tau=0.7$. Figure \ref{fig:IA_07_smaller_different_taus} and \ref{fig:IA_07_smaller_different_taus2} presents results under sample sizes $n=5000$, 20\% censoring rate, and a Gamma frailty distribution with Kendall's $\tau=0.1,0.3,0.5$.
	\item Figures \ref{fig:IB_tau07_n5000} to \ref{fig:IB_07_smaller_different_taus2} presents additional results for the Scenario (Ib). Figure \ref{fig:IB_tau07_n5000} presents more results under sample size of $n=5000$, 20\% censoring rate, and a Gamma frailty distribution with Kendall's $\tau=0.7$. Figure \ref{fig:IB_07_smaller_sample_sizes} presents results under sample sizes $n=500,1000$, 20\% censoring rate, and a Gamma frailty distribution with Kendall's $\tau=0.7$. Figure \ref{fig:IB_07_different_CR} presents results under sample sizes $n=5000$, 20\% and $40\%$ censoring rate, and a Gamma frailty distribution with Kendall's $\tau=0.7$. Figure \ref{fig:IB_07_smaller_different_taus} and \ref{fig:IB_07_smaller_different_taus2} presents results under sample sizes $n=5000$, 20\% censoring rate, and a Gamma frailty distribution with Kendall's $\tau=0.1,0.3,0.5$.
	\item Figures \ref{fig:sim.results.II.SD} to \ref{fig:sim.results.II.Kernel.05} present additional simulations results in Scenario (II). Figure \ref{fig:sim.results.II.SD} presents the ratio between EST.SE and EMP.SD under and a Gamma frailty distribution with Kendall's $\tau=0.7$. Figure \ref{fig:sim.results.II.EC} presents the empirical coverage rate under and a Gamma frailty distribution with Kendall's $\tau=0.7$. Figure \ref{fig:sim.results.II.RB.05} presents the relative bias of the estimators under and a Gamma frailty distribution with Kendall's $\tau=0.5$. Figure \ref{fig:sim.results.II.Cox} and \ref{fig:sim.results.II.Kernel} presents the mean estimates along with plus/minus one empirical standard deviation for the Cox-based and Kernel-based $HR^C(t)$ estimates under and a Gamma frailty distribution with Kendall's $\tau=0.7$, respectively. \ref{fig:sim.results.II.Cox.05} and \ref{fig:sim.results.II.Kernel.05} presents the same results but under and a Gamma frailty distribution with Kendall's $\tau=0.5$.
	
\end{itemize}


\newpage


\newpage
\clearpage
\section{Imvigor211 data example- More results}\label{SM:Imvigor211}
In this Section we present the Cox-based estimator for the $HR^C(t)$.
\subsection{Cox-based estimator}\label{SM:Imvigor211-cox}
\begin{figure}[h]

	\includegraphics[scale=0.3]{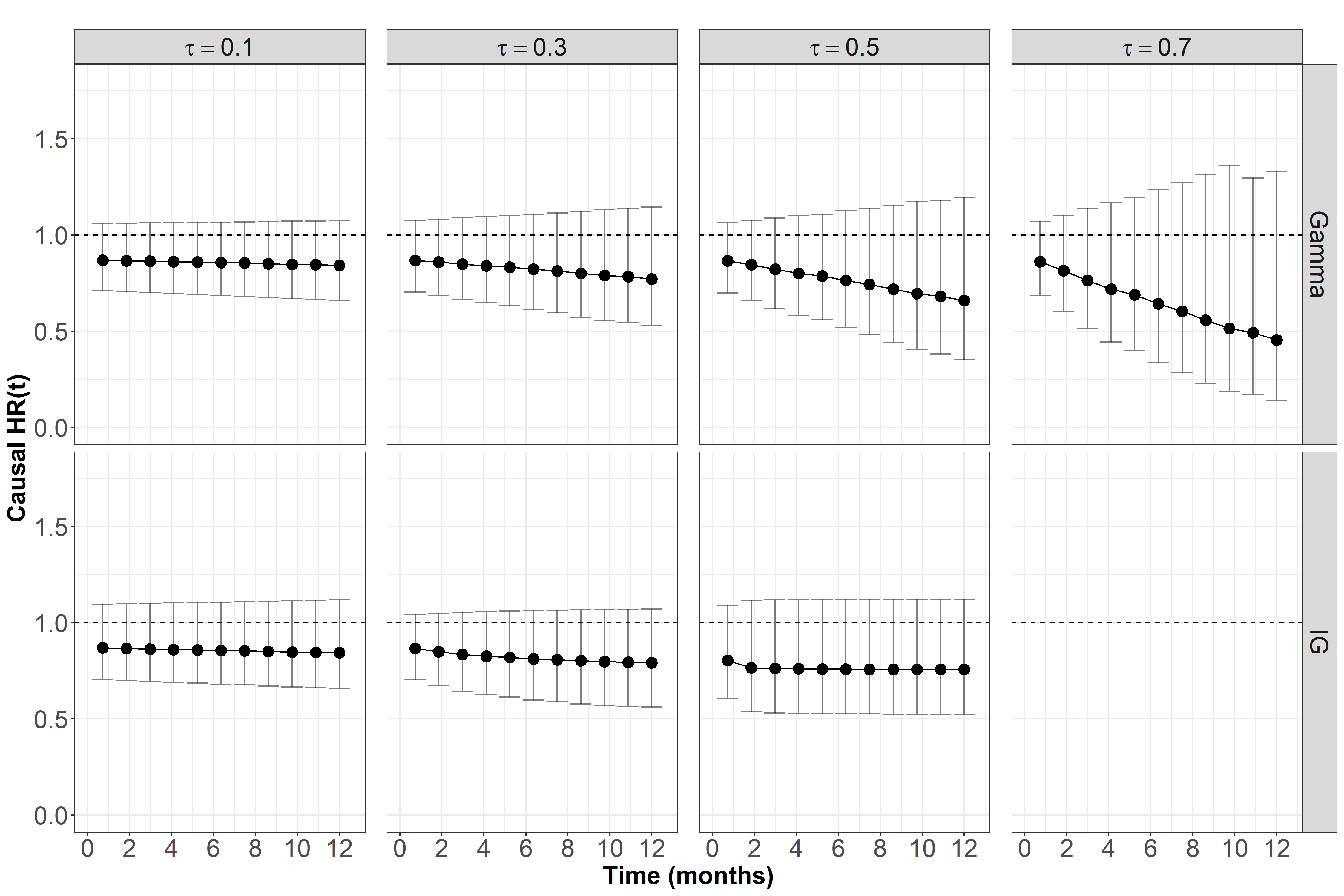}
	\centering
	\label{fig:Atezo.HRC.Cox}
		\caption{Cox-based estimation along with plus/minus one standard error for the $HR^C(t)$ in the IMvigor211 study, under different frailty distributions and  Kendall's $\tau$ values. The horizontal dashed line in these figures represents $HR^C(t)=1$.}	
\end{figure}
The Cox-based estimator of the $HR^C(t)$ is below one under all frailty distributions and the possible Kendall's $\tau$. At time 0 months, the $HR^C(0)=0.84$ and then it decreases monotonically over time under the Gamma distribution and remains approximately constant under the IG distribution. It suggests that Atezo treatment, when compared to Chemo, has a protective effect for every $t$ on patients who would have survived up to time $t$ regardless of the their treatment effect. 

\newpage
\clearpage
\section{DIVAT data example- More results}\label{SM:DIVAT}
\begin{figure}[h]
	
	\includegraphics[scale=0.3]{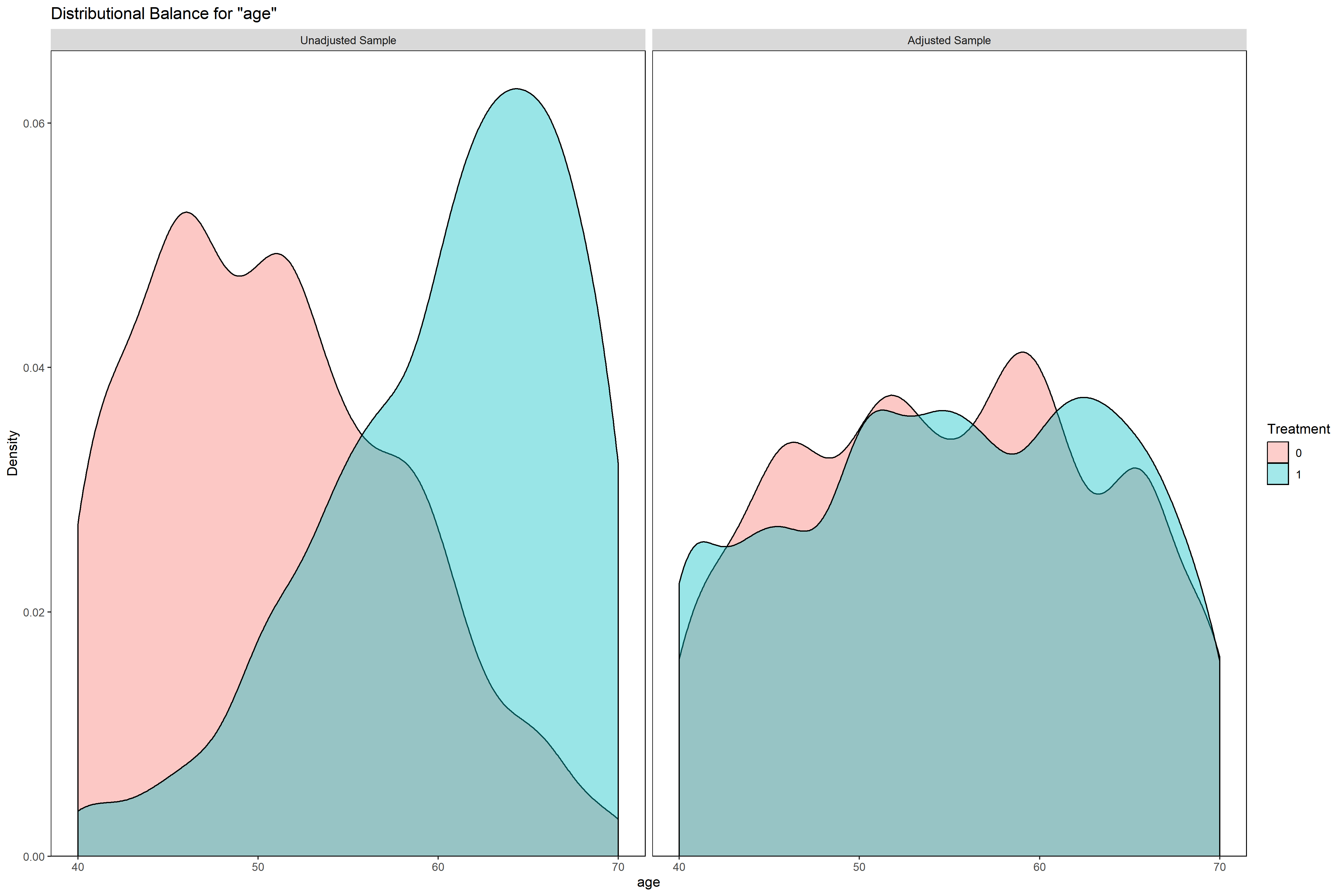}
	\centering
	
	\caption{Distributional balance of age covariate}
	\label{fig:age_balance}	
\end{figure}

\begin{figure}[h]

	\includegraphics[scale=0.3]{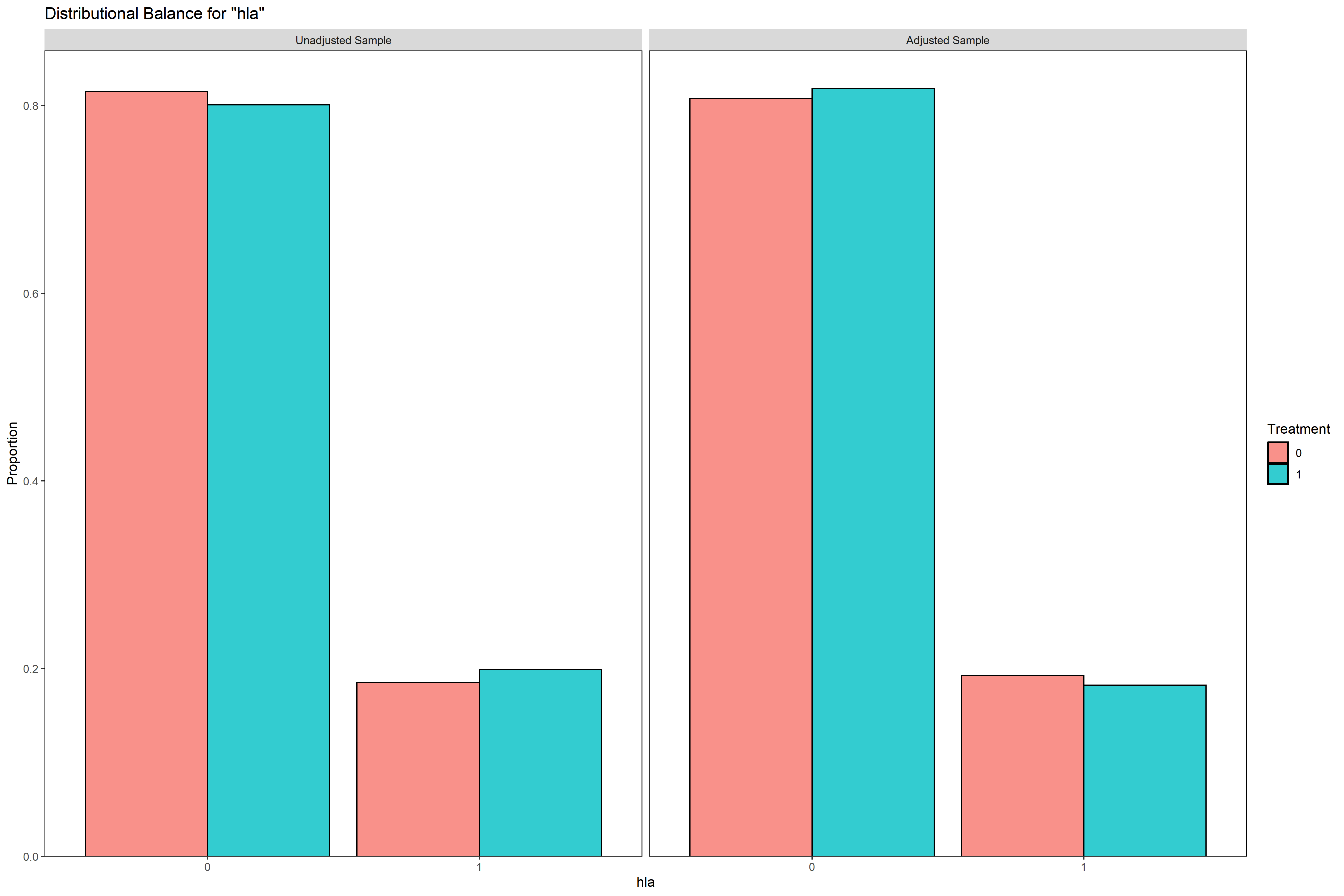}
	\centering

		\caption{Distributional balance of hla covariate}
			\label{fig:hla_balance}	
\end{figure}

\begin{figure}[h]

	\includegraphics[scale=0.3]{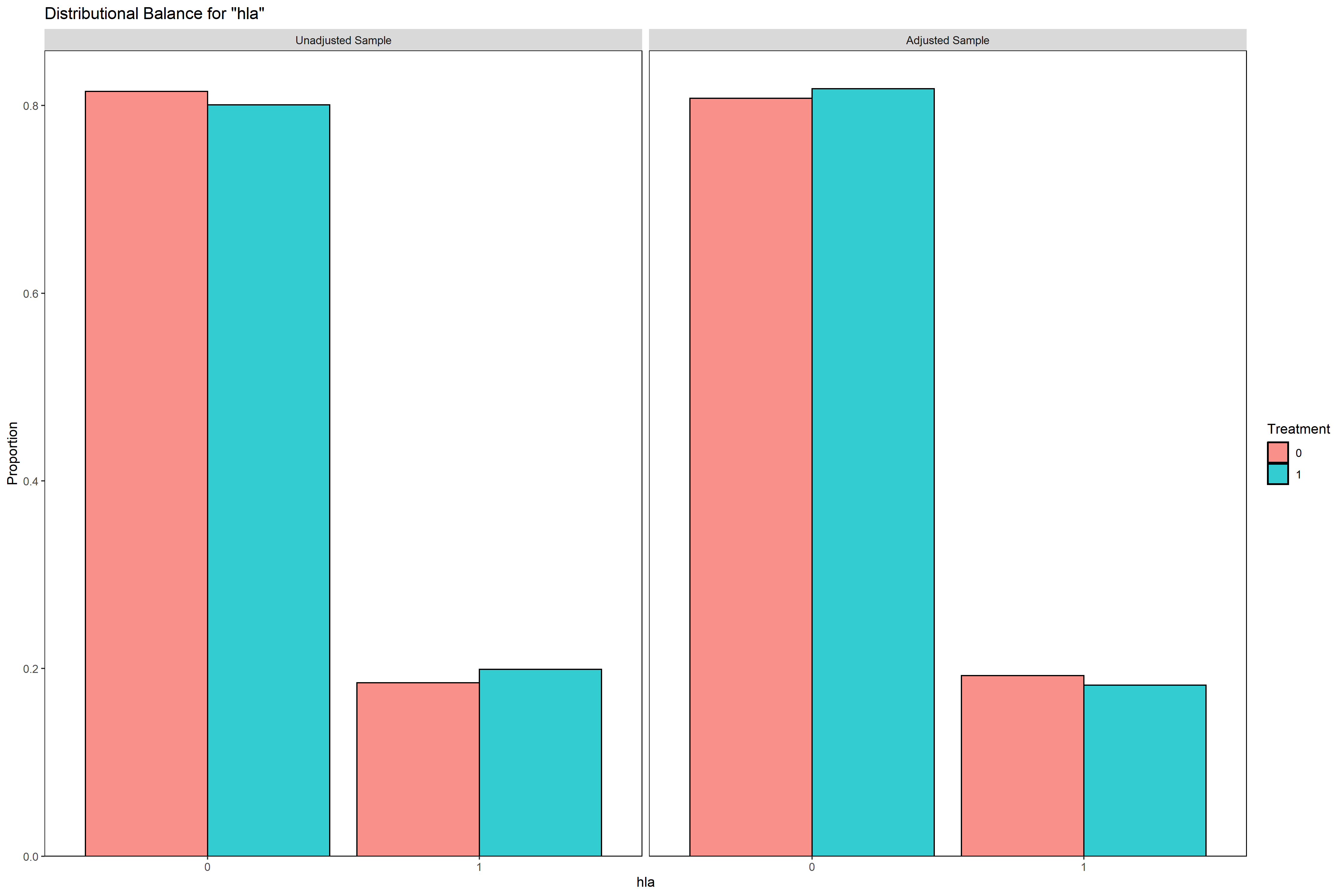}
	\centering

		\caption{Distributional balance of re-transplant covariate}	
			\label{fig:re-transplant_balance}
\end{figure}

\begin{figure}[h]

	\includegraphics[scale=0.3]{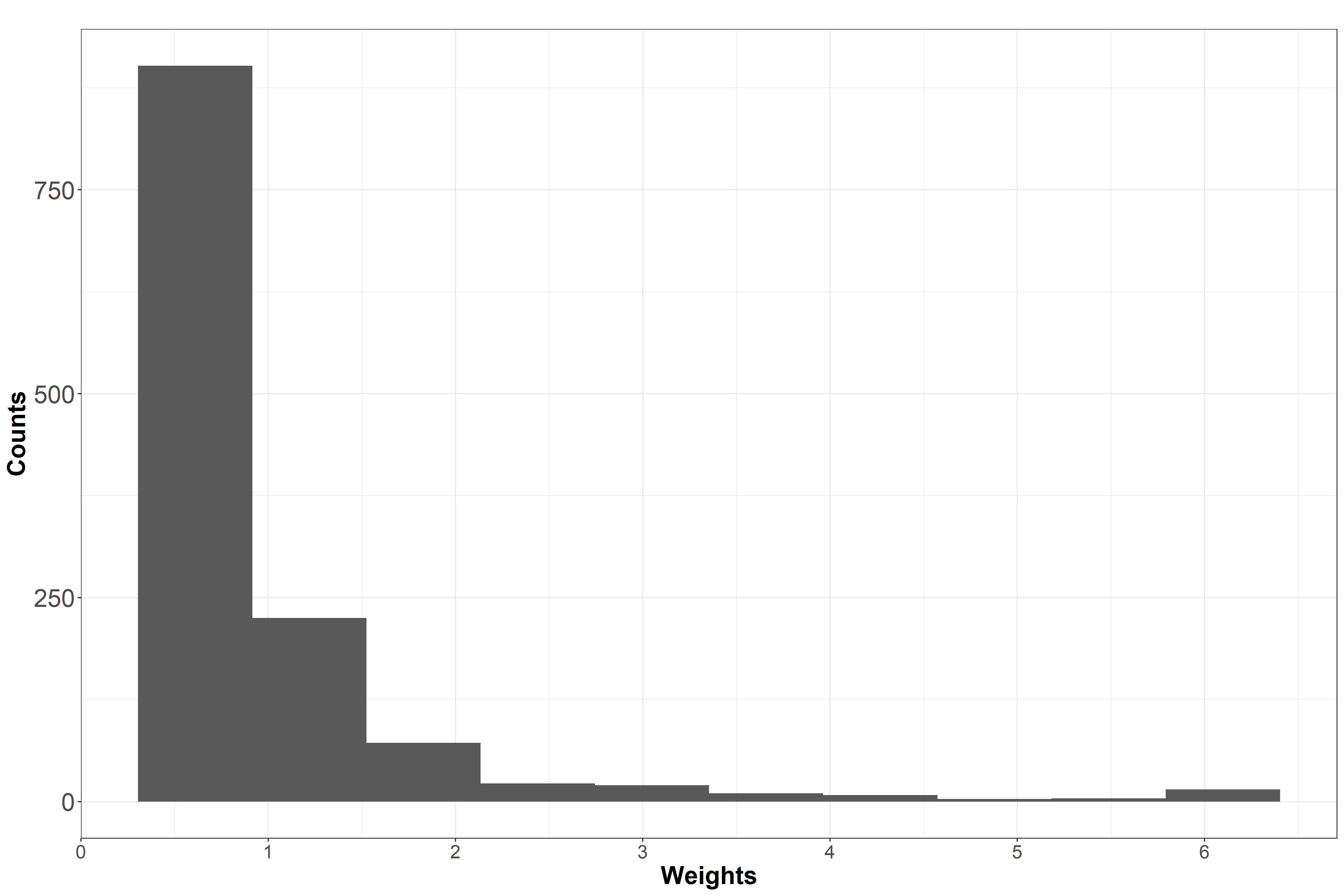}
	\centering

		\caption{Histogram of the obtained weights after truncation at the 99th percentile}	
			\label{fig:histogram of weights}
\end{figure}

\begin{figure}[h]
	\includegraphics[scale = 0.5]{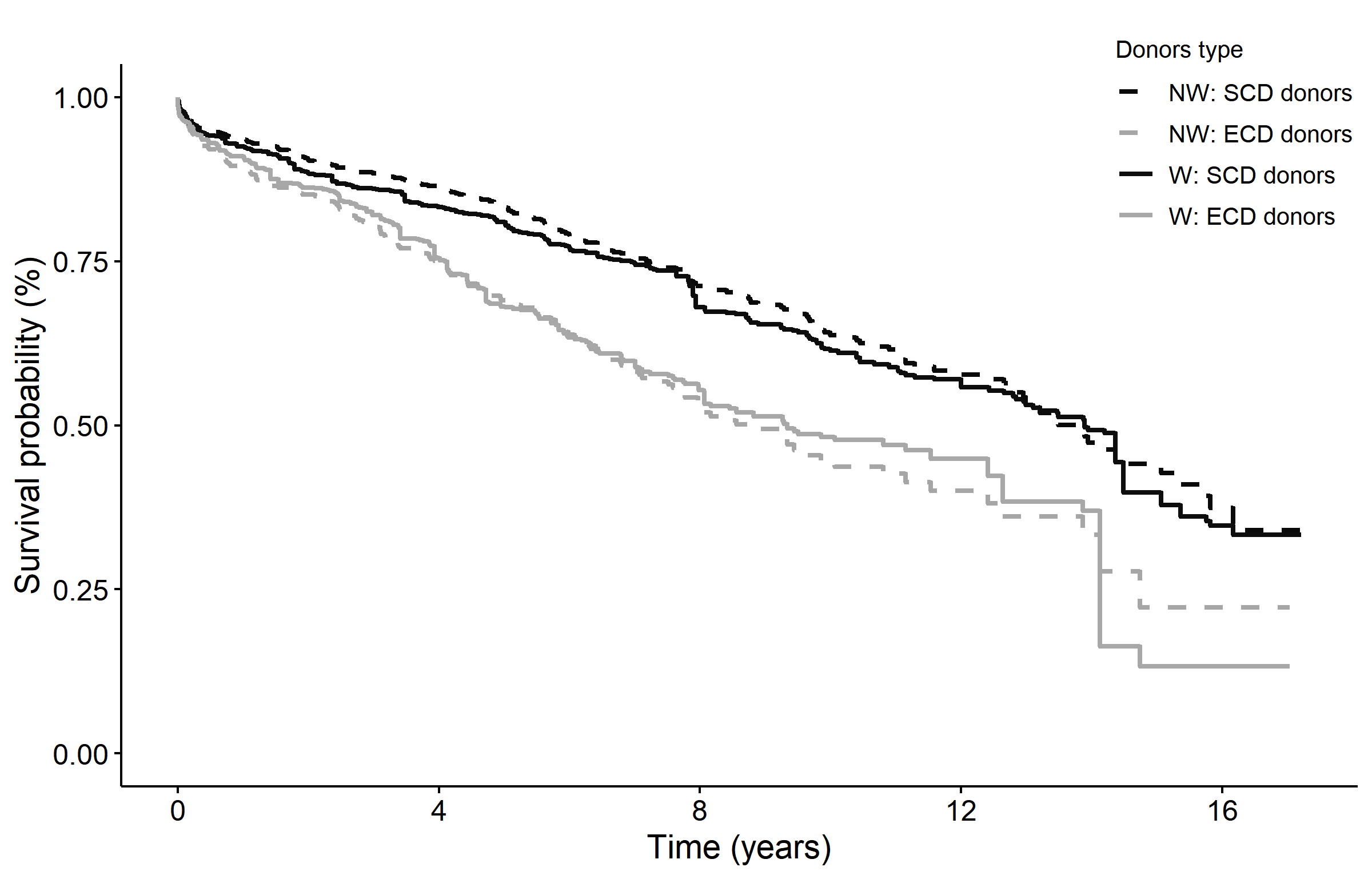}
	\centering
		\caption{Kaplan-Meier curves for the DIVAT study (IPTW and regular curves)}	
			\label{fig:KP_DIVAT}
\end{figure}

In this section, we present more results for the DIVAT dataset. In section \ref{sec:log.reg} we present the results from the  logistic regression for $\pi(\bZ)$. In Section \ref{sec:Covariates balance} we present a covariates balance table with the standardized mean difference. We also present in Figures \ref{fig:age_balance}--\ref{fig:re-transplant_balance} comparison of the covariates distribution before and after applying the calculated weights. In Figure \ref{fig:histogram of weights} we present the histogram of the obtained weights we used in analyzing the DIVAT dataset. In figure \ref{fig:KP_DIVAT} we present the Kaplan-Meier plot for the DIVAT dataset. We present the non-weighted and the weighted curves. In Figure \ref{tab:con.cox.table} we present the results from applying Cox regression including the donor criterion type and the confounders. 

\subsection{DIVAT: Logistic regression for the calculating the weights}\label{sec:log.reg}
\begin{table}[h]
	\centering
	\caption{Logistic regression results, OR: odds ratio, 2.50\%: lower confidence interval, 97.50\%: upper confidence interval, p.v: p-value}
	\begin{tabular}{lllll}
		\toprule
		& \multicolumn{1}{l}{OR} & 2.50\% & 97.50\% & \multicolumn{1}{l}{p.v} \\
		\midrule
		(Intercept) & 0.00  & 0.00  & 0.00  & 0.00 \\
		age   & 1.20  & 1.18  & 1.23  & 0.00 \\
		HLA   & 0.96  & 0.68  & 1.37  & 0.84 \\
		re-transplant & 1.29  & 0.92  & 1.83  & 0.14 \\
		\bottomrule
	\end{tabular}%
	\label{tab:log.table.divat}%
\end{table}%
\newpage
\subsection{DIVAT: Covariates balance}\label{sec:Covariates balance}
\begin{table}[h]
	\centering
	\caption{Covariates balance in DIVAT data-set. SMD: Standardized mean difference}
	\begin{tabular}{lllllll}
		& \multicolumn{3}{c}{Before adjusting weights} & \multicolumn{3}{c}{After adjusting weights} \\
		\midrule
		& SCD   & ECD   & SMD   & SCD   & ECD   & SMD \\
		\midrule
		age   & 50.84 & 60.61 & 1.15  & 55.04 & 55.16 & 0.01 \\
		HLA   & 0.18  & 0.20  & 0.01  & 0.19  & 0.18  & -0.01 \\
		retransplant & 0.24  & 0.20  & -0.04 & 0.24  & 0.27  & 0.03 \\
		\bottomrule
	\end{tabular}%
	\label{tab:balance.table.divat}%
\end{table}%

\label{lastpage}

\clearpage
\newpage
\subsection{Histogram of the obtained weights}
\newpage
\subsection{DIVAT: Kaplan-Meier plot}
\subsection{DIVAT:Cox's regression results}
\begin{table}[h]
	\centering
	\caption{Cox's regression results, HR: hazard ratio, 2.50\%: lower confidence interval, 97.50\%: upper confidence interval, p.v: p-value}
	\begin{tabular}{lllll}
		\toprule
		& HR    & 2.50\% & 97.50\% & p.v \\
		\midrule
		treatment & 1.59  & 1.26  & 2.01  & <0.01 \\
		age   & 1.01  & 1.00  & 1.03  & 0.05 \\
		hla   & 1.16  & 0.91  & 1.48  & 0.22 \\
		retransplant & 1.40  & 1.10  & 1.78  & 0.01 \\
		\bottomrule
	\end{tabular}%
	\label{tab:con.cox.table}%
\end{table}%

\end{onehalfspacing}

|

\label{lastpage}

\end{document}